# Overall thermomechanical properties of layered materials for energy devices applications

A. Bacigalupo*[1], L. Morini[1], and A. Piccolroaz[2]

[1] IMT School for Advanced Studies, Lucca, Italy
[2] Department of Civil, Environmental and Mechanical Engineering, University of Trento, Italy

August 2, 2016

**Abstract**

This paper is concerned with the analysis of effective thermomechanical properties of multi-layered materials of interest for solid oxide fuel cells (SOFC) and lithium ions batteries fabrication. The recently developed asymptotic homogenization procedure is applied in order to express the overall thermoelastic constants of the first order equivalent continuum in terms of microfluctuations functions, and these functions are obtained by the solution of the corresponding recursive *cell problems*. The effects of thermal stresses on periodic multi-layered thermoelastic composite reproducing the characteristics of solid oxide fuel cells (SOFC-like) are studied assuming periodic body forces and heat sources, and the solution derived by means of the asymptotic homogenization approach is compared with the results obtained by finite elements analysis of the associate heterogeneous material.

## 1 Introduction

Solid oxide fuel cells (SOFC) and lithium ions batteries are two of the most performing and promising battery devices which can play an important role in realizing efficient small-scale power generation systems providing renewable energy for industrial applications. Due to the high temperatures which can be reached in operative scenarios (Pitakthapanaphong and Busso, 2005), the components of such batteries are subject to severe thermomechanical stresses which can cause damage and crack formation, compromising the performance of the devices in terms of power generation and energy conversion efficiency (Atkinson and Sun, 2007; Delette et al., 2013). Modelling the thermomechanical properties of SOFCs devices and lithium ions batteries represent a crucial issue in order to predict these phenomena and then to ensure the successful manufacture and the reliability of the systems.

*Corresponding author. Tel.: +39 0583 4326613, email address: andrea.bacigalupo@imtlucca.it

Both SOFCs and lithium ions batteries are characterized by a multi-layered configuration possessing many phases of composite materials, where the elementary cell is represented by the anode-electrolyte-cathode system. Moreover, in many operative situations solid oxide fuel cells are organized in stacks where several anode-electrolyte-cathode systems are separated by metallic interconnections. Since the macroscopic behaviour of these multi-layered structures is influenced by phenomena occurring at scale-lengths characteristic of the microscopic constituents, which is small compared to the macroscopic dimension (i.e. structural size), multiscale modelling of SOFCs and lithium ions batteries implies challenging numerical computations which require very fine mesh of finite elements and then strong computational resources (Richardson et al., 2012; Hajimolana et al., 2011). Homogenization techniques represent an useful and advantageous method for providing a rigorous and synthetic description of the effects of the microscopic phases on the overall properties of the materials. The application of these approaches makes possible to avoid the challenging numerical computations required by computational modelling of heterogeneous media, and are particularly suitable for periodic composite media, such as multi-layered battery devices. Several homogenization techniques have been proposed for studying overall properties of composite materials, such as the asymptotic (see for example Sanchez-Palencia (1974a,b, 1986), Bensoussan et al. (1978); Bakhvalov and Panasenko (1984); Gambin and Kroner (1989); Allaire (1992); Boutin and Auriault (1993); Meguid and Kalamkarov (1994); Boutin (1996); Andrianov et al. (2008); Tran et al. (2012)), the variational-asymptotic methods (see for example Smyshlyaev and Cherednichenko (2000); Peerlings and Fleck (2004); Smyshlyaev (2009); Bacigalupo (2014); Bacigalupo and Gambarotta (2014b)) and the computational approaches (see for example Forest and Sab (1998); Forest (2002); Kouznetsova et al. (2002, 2004); Kaczmarczyk et al. (2008); Forest and Trinh (2011); Bacigalupo and Gambarotta (2010, 2011); De Bellis and Addessi (2011); Addessi et al. (2013); Bacca et al. (2013a,b,c)).

The principal aim of this article is to provide exact closed-form expressions to estimate the overall thermoelastic and heat conduction tensors of multi-layered battery devices avoiding the challenging computations required by standard numerical modelling of the heterogeneous structures (Bove and Ubertini, 2008). With this purpose, an ideal periodic multi-layered thermoelastic composite material reproducing the planar geometry of an idealized battery device is introduced (see Fig 1). The thermoelastic and heat conduction tensors of the first order continuum equivalent to the introduced multi-layered battery-like thermoelastic composite are derived applying the asymptotic homogenization approach recently developed by Bacigalupo et al. (2016) for studying heterogeneous media in presence of thermodiffusive phenomena. Following the rigorous procedure developed in Bakhvalov and Panasenko (1984); Smyshlyaev and Cherednichenko (2000); Bacigalupo and Gambarotta (2013, 2014a,b, 2012) and Bacigalupo (2014) for composite elastic media with periodic microstructures and generalized by Bacigalupo et al. (2016) to the case of thermodiffusive materials, the fields equation for the homogenized first order thermoelastic continuum equivalent to the multi-layered battery devices are derived, and exact expressions for the overall thermoelastic constants of this equivalent medium are obtained. These expressions are used to determine analytically the components of the overall elastic, thermoelastic and heat conduction tensors corresponding to a tri-phase layered thermoelastic composite of interests for SOFCs devices fabrication. The thermoelastic constants of the three phases are assumed to possess values typical of the constituents of real SOFCs devices, evaluated by means of accurate experimental techniques and homogenization methods and accounting for the microstructure, such as the porosity, of the electrolyte and the electrodes. of The fields equation of the first order equivalent thermoelastic media are solved considering periodic heat sources, which localized and unlocalized profiles are representative for modelling some thermal effects detected in real situations. The solution of the homogenized field equations is compared with the numerical results obtained by the heterogeneous model assuming periodic body force and heat and mass sources acting on the considered three-phase layered composite.

The article is organized as follows: in Section 2 the geometry of the idealized periodic thermoelastic battery-like material is illustrated, and the corresponding constitutive relations and balance equations are introduced. The developed multi-scale asymptotic homogenization technique is described in Section 3, based on down-scaling relations correlating the microscopic fields to the macroscopic displacements and temperature. The unknown perturbation functions describing the effects of the material heterogeneities are defined as solutions of the corresponding non-homogeneous cell problems. In the same Section, the fields equations and explicit expressions for the components of the elastic, thermoelastic and heat conduction tensors of the equivalent first order homogeneous continuum are derived. In Section 4, these results are used for studying overall properties of three-phase layered thermoelastic composites of interests for SOFCs devices fabrication, represented by an an idealized cathode-electrolyte-anode-interconnection system. Finally, a critical discussion about the obtained results is reported together with conclusions and future perspectives in Section 5.

# 2 Multiscale modelling of periodic thermoelastic composites

Many energy battery devices such as lithium ions batteries and solid oxide fuel cells (SOFC) are characterized by multi-layered structures (Nakajo et al., 2012; Dev et al., 2014; Ellis et al., 2012). In order to develop a general approach for estimating effective thermomechanical properties of both lithium ions batteries and solid oxide fuel cells, we introduce a periodic multi-layered thermoelastic composite media reproducing the planar geometry of an idealized battery device as shown in Fig. 1.

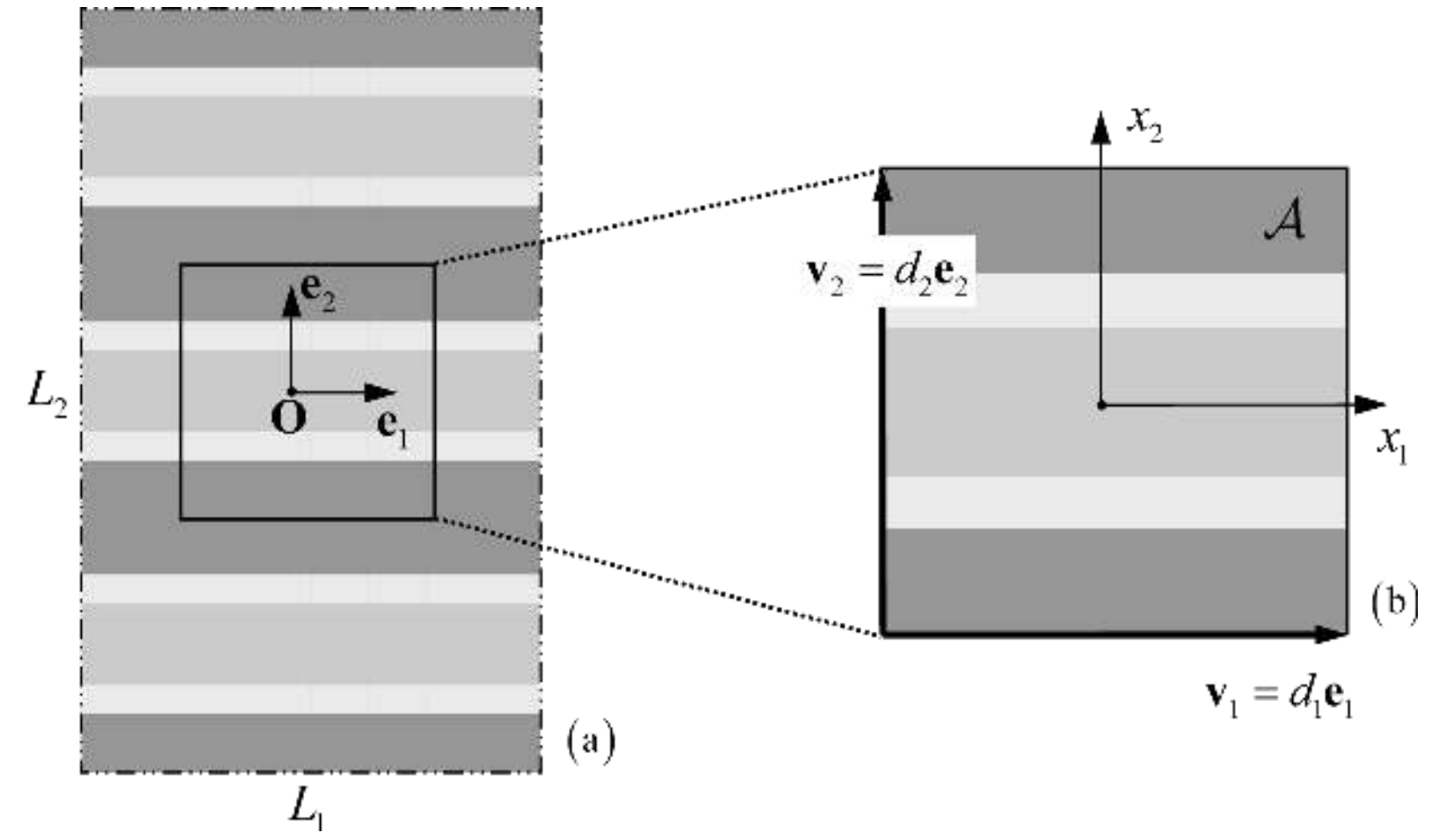

Figure 1: (a) Heterogeneous material – Periodic domain $L$ ; (b) Periodic cell $\mathcal{A}$ and periodicity vectors.

The constituent elements of the medium are modelled as a linear thermoelastic Cauchy continuum subject to small strains. The material point is identified by position vector $\boldsymbol{x} = x_1\boldsymbol{e}_1 + x_2\boldsymbol{e}_2$ referred to a system of coordinates with origin at point $O$ and orthogonal base $\{\boldsymbol{e}_1, \boldsymbol{e}_2\}$. Figure 1b shows the periodic cell $\mathcal{A} = [0, \varepsilon] \times [0, \delta\varepsilon]$ with characteristic size $\varepsilon$. The entire periodic medium can be obtained spanning the cell $\mathcal{A}$ by the two orthogonal vectors $\boldsymbol{v}_1 = d_1\boldsymbol{e}_1 = \varepsilon\boldsymbol{e}_1, \boldsymbol{v}_2 = d_2\boldsymbol{e}_2 = \delta\varepsilon\boldsymbol{e}_2$. $\mathcal{A}$ represents the elementary cell period of the elasticity tensor $\mathbb{C}^{(m,\varepsilon)}(\boldsymbol{x})$, the heat conduction tensor $\boldsymbol{K}^{(m,\varepsilon)}(\boldsymbol{x})$ and the thermal dilatation tensor $\boldsymbol{\alpha}^{(m,\varepsilon)}(\boldsymbol{x})$, which are defined as follows

$$\mathbb{C}^{(m,\varepsilon)}(\boldsymbol{x} + \boldsymbol{v}_i) = \mathbb{C}^{(m,\varepsilon)}(\boldsymbol{x}), \quad i = 1, 2, \quad \forall \boldsymbol{x} \in \mathcal{A}. \tag{1}$$

$$\boldsymbol{K}^{(m,\varepsilon)}(\boldsymbol{x} + \boldsymbol{v}_i) = \boldsymbol{K}^{(m,\varepsilon)}(\boldsymbol{x}), \quad \boldsymbol{\alpha}^{(m,\varepsilon)}(\boldsymbol{x} + \boldsymbol{v}_i) = \boldsymbol{\alpha}^{(m,\varepsilon)}(\boldsymbol{x}), \quad i = 1, 2, \quad \forall \boldsymbol{x} \in \mathcal{A}. \tag{2}$$

The tensors (1) and (2) are commonly referred to as $\mathcal{A}-$periodic functions.

The system is subject to body forces $\boldsymbol{b}(\boldsymbol{x})$ and heat sources $r(\boldsymbol{x})$ which are assumed to be $\mathcal{L}-$periodic with period $\mathcal{L} = [0, L] \times [0, \delta L]$ and to have vanishing mean values on $\mathcal{L}$. Since $L$ is a large multiple of $\varepsilon$, then $\mathcal{L}$ can be assumed to be a representative portion of the overall body. This means that $\boldsymbol{b}(\boldsymbol{x})$ and $r(\boldsymbol{x})$ possess a period much greater than the microstructural size $\varepsilon$.

A non-dimensional unit cell $\mathcal{Q} = [0, 1] \times [0, \delta]$ that reproduces the periodic microstructure by rescaling with the small parameter $\varepsilon$ is introduced (Bacigalupo, 2014). Two distinct scales are represented by the macroscopic (slow) variables $\boldsymbol{x} \in \mathcal{A}$ and the microscopic (fast) variable $\boldsymbol{\xi} = \boldsymbol{x}/\varepsilon \in \mathcal{Q}$ (see for example Bakhvalov and Panasenko (1984) and Smyshlyaev and Cherednichenko (2000)). The constitutive tensors (1), and (2) are functions of the microscopic variable, whereas the body forces and heat sources depend on the slow macroscopic variable. Consequently, the mapping of both the elasticity and thermodiffusive tensors may be defined on $\mathcal{Q}$ as follows: $\mathbb{C}^{(m,\varepsilon)}(\boldsymbol{x}) = \mathbb{C}^m(\boldsymbol{\xi} = \boldsymbol{x}/\varepsilon)$, $\boldsymbol{K}^{(m,\varepsilon)}(\boldsymbol{x}) = \boldsymbol{K}^m(\boldsymbol{\xi} = \boldsymbol{x}/\varepsilon)$, $\boldsymbol{\alpha}^{(m,\varepsilon)}(\boldsymbol{x}) = \boldsymbol{\alpha}^m(\boldsymbol{\xi} = \boldsymbol{x}/\varepsilon)$, respectively.

The relevant microscopic fields are the micro-displacement $\boldsymbol{u}(\boldsymbol{x})$, and the microscopic temperature $\theta(\boldsymbol{x}) = T(\boldsymbol{x}) - T_0$ evaluated with respect to the natural state ($T = T_0$). The micro-stress $\boldsymbol{\sigma}(\boldsymbol{x})$ and the microscopic heat flux $\boldsymbol{q}(\boldsymbol{x})$ are defined by the following constitutive relations:

$$\boldsymbol{\sigma}(\boldsymbol{x}) = \mathbb{C}^m \left(\frac{\boldsymbol{x}}{\varepsilon}\right) \boldsymbol{\varepsilon}(\boldsymbol{x}) - \boldsymbol{\alpha}^m \left(\frac{\boldsymbol{x}}{\varepsilon}\right) \theta(\boldsymbol{x}), \tag{3}$$

$$\boldsymbol{q}(\boldsymbol{x}) = -\boldsymbol{K}^m \left(\frac{\boldsymbol{x}}{\varepsilon}\right) \nabla\theta(\boldsymbol{x}), \tag{4}$$

where $\boldsymbol{\varepsilon}(\boldsymbol{x}) = \text{sym}\nabla\boldsymbol{u}(\boldsymbol{x})$ is the micro-strain tensor which is assumed to be zero at the fundamental state of the system. The micro-stresses (3) and the microscopic heat flux (4) satisfy the local balance equations on the domain $\mathcal{A}$

$$\nabla \cdot \boldsymbol{\sigma}(\boldsymbol{x}) + \boldsymbol{b}(\boldsymbol{x}) = \boldsymbol{0}, \tag{5}$$

$$\nabla \cdot \boldsymbol{q}(\boldsymbol{x}) - r(\boldsymbol{x}) = 0, \tag{6}$$

Substituting expressions (3) and (4) in equations (5) and (6), and remembering the symmetry of the elasticity tensor, the resulting set of partial differential equations is written in the form

$$\nabla \cdot \left(\mathbb{C}^m \left(\frac{\boldsymbol{x}}{\varepsilon}\right) \nabla\boldsymbol{u}(\boldsymbol{x})\right) - \nabla \cdot \left(\boldsymbol{\alpha}^m \left(\frac{\boldsymbol{x}}{\varepsilon}\right) \theta(\boldsymbol{x})\right) + \boldsymbol{b}(\boldsymbol{x}) = \boldsymbol{0} \tag{7}$$

$$\nabla \cdot \left(\boldsymbol{K}^m \left(\frac{\boldsymbol{x}}{\varepsilon}\right) \nabla\theta(\boldsymbol{x})\right) + r(\boldsymbol{x}) = 0, \tag{8}$$

The micro-displacement and microscopic temperature may be seen in the form $\boldsymbol{u}(\boldsymbol{x}, \boldsymbol{\xi} = \boldsymbol{x}/\varepsilon)$, $\theta(\boldsymbol{x}, \boldsymbol{\xi} = \boldsymbol{x}/\varepsilon)$ as functions of both the slow and the fast variable. The solution of microscopic fields equations (7) and (8) is computationally very expensive and provides too detailed results to be of practical use, so that it is convenient to replace the heterogeneous model with an equivalent homogeneous one to obtain equations whose coefficients are not rapidly oscillating while their solutions are close to those of the original equations.

Further in the paper, the overall elastic moduli, thermal expansion and heat conductivity tensors of a first order (Cauchy) homogeneous thermoelastic continuum equivalent to multi-layered material reported in Fig.1 are derived applying the asymptotic homogenization approach recently developed for periodic thermodiffusive composites by Bacigalupo et al. (2016). Assuming that the size of the microstructure $\varepsilon$ is sufficiently small with respect to the structural size $L$, the overall thermoelastic constants of the homogeneous continuum are expressed in terms of geometrical, mechanical, and thermal diffusive properties of the microstructure by means of an asymptotic expansion for the microscopic fields. The asymptotic expansion is performed in terms of the parameter $\varepsilon$ that keeps the dependence on the slow variable $\boldsymbol{x}$ separate from the fast one $\boldsymbol{\xi} = \boldsymbol{x}/\varepsilon$ such that two distinct scales are represented.

Let us define the macroscopic physical quantities characterizing the first order homogenized continuum equivalent to the periodic material reported in Fig. 1. The macro-displacement $\boldsymbol{U}(\boldsymbol{x})$ of component $U_i$, and the macroscopic temperature $\Theta(\boldsymbol{x})$ are defined at a point $\boldsymbol{x}$ in the reference $(\boldsymbol{e}_i,\ i = 1, 2)$. The displacement gradient is given by $\nabla \boldsymbol{U}(\boldsymbol{x}) = \frac{\partial U_i}{\partial x_j} \boldsymbol{e}_i \otimes \boldsymbol{e}_j = H_{ij} \boldsymbol{e}_i \otimes \boldsymbol{e}_j = \boldsymbol{H}(\boldsymbol{x})$,and then the macroscopic strain is $\boldsymbol{E}(\boldsymbol{x}) = \mathrm{sym}\nabla \boldsymbol{U}(\boldsymbol{x})$. The macro-stresses $\boldsymbol{\Sigma}(\boldsymbol{x})$ associate to $\boldsymbol{E}(\boldsymbol{x})$ are defined as $\boldsymbol{\Sigma}(\boldsymbol{x}) = \Sigma_{ij} \boldsymbol{e}_i \otimes \boldsymbol{e}_j$ with $\Sigma_{ij} = \Sigma_{ji}$, and the macroscopic heat flux is given by $\boldsymbol{Q}(\boldsymbol{x}) = Q_i \boldsymbol{e}_i$.

# 3 Asymptotic homogenization approach to thermoelastic composites

In this Section, explicit expressions for the elasticity, thermal expansion and heat conductivity tensors of the homogeneous first order continuum equivalent to a thermoelastic composite material with periodic microstructure are derived by means of asymptotic homogenization approach developed in Bacigalupo et al. (2016).

## 3.1 Multiscale analysis of micro-displacement fields

According to Bakhvalov and Panasenko (1984); Smyshlyaev and Cherednichenko (2000); Bacigalupo and Gambarotta (2014b); Bacigalupo (2014) and Bacigalupo et al. (2016) the microscopic displacement and temperature fields are represented through an asymptotic expansion with respect to the parameter $\varepsilon$, and the following *down-scaling* relations are derived

$$u_k\left(\boldsymbol{x}, \boldsymbol{\xi} = \frac{\boldsymbol{x}}{\varepsilon}\right) = U_k(\boldsymbol{x}) + \varepsilon \left( N^{(1)}_{kpq_1}(\boldsymbol{\xi}) \frac{\partial U_p(\boldsymbol{x})}{\partial x_{q_1}} + \tilde{N}^{(1)}_k(\boldsymbol{\xi}) \Theta(\boldsymbol{x}) \right)_{\boldsymbol{\xi}=\boldsymbol{x}/\varepsilon} + \\ + \varepsilon^2 \left( N^{(2)}_{kpq_1q_2}(\boldsymbol{\xi}) \frac{\partial^2 U_p(\boldsymbol{x})}{\partial x_{q_1} \partial x_{q_2}} + \tilde{N}^{(2)}_{kq_1}(\boldsymbol{\xi}) \frac{\partial \Theta(\boldsymbol{x})}{\partial x_{q_1}} \right)_{\boldsymbol{\xi}=\boldsymbol{x}/\varepsilon} + \mathcal{O}(\varepsilon^3), \tag{9}$$

$$\theta\left(\boldsymbol{x}, \boldsymbol{\xi} = \frac{\boldsymbol{x}}{\varepsilon}\right) = \Theta(\boldsymbol{x}) + \varepsilon \left( M^{(1)}_{q_1}(\boldsymbol{\xi}) \frac{\partial \Theta(\boldsymbol{x})}{\partial x_{q_1}} \right)_{\boldsymbol{\xi}=\boldsymbol{x}/\varepsilon} + \varepsilon^2 \left( M^{(2)}_{q_1q_2}(\boldsymbol{\xi}) \frac{\partial^2 \Theta(\boldsymbol{x})}{\partial x_{q_1} \partial x_{q_2}} \right)_{\boldsymbol{\xi}=\boldsymbol{x}/\varepsilon} + \mathcal{O}(\varepsilon^3). \tag{10}$$

Note that due to their dependence on the slow space variable $\boldsymbol{x}$, the macroscopic fields $U_k$ and $\Theta$ are $\mathcal{L}-$periodic functions. $N^{(1)}_{kpq_1}$ and $N^{(2)}_{kpq_1q_2}$ are the mechanical fluctuations functions, $M^{(1)}_{q_1}$ and $M^{(2)}_{q_1q_2}$ are the thermal fluctuations functions, $\tilde{N}^{(1)}_{k}$ and $\tilde{N}^{(2)}_{kq_1}$ denote the additional fluctuations functions corresponding to the contribution of the temperature to local displacement. All these perturbation functions depend on the fast space variable $\boldsymbol{\xi} = \boldsymbol{x}/\varepsilon$, and are $\mathcal{Q}-$periodic with zero mean value over $\mathcal{Q}$, namely $\left\langle N^{(1)}_{kpq_1}\right\rangle = 0$, $\left\langle N^{(2)}_{kpq_1q_2}\right\rangle = 0$, $\left\langle \tilde{N}^{(1)}_{k}\right\rangle = 0$, $\left\langle \tilde{N}^{(2)}_{kq_1}\right\rangle = 0$, $\left\langle M^{(1)}_{q_1}\right\rangle = 0$ and $\left\langle M^{(2)}_{q_1q_2}\right\rangle = 0$ where $\langle\cdot\rangle$ denotes the averaging operator over $\mathcal{Q}$ (normalization conditions).

The *down-scaling* relations (9) and (10) can be substituted into the microscopic fields equations (7) and (8), and using the property $\frac{\partial}{\partial x_j} f(\boldsymbol{x}, \boldsymbol{\xi} = \frac{\boldsymbol{x}}{\varepsilon}) = \left(\frac{\partial f}{\partial x_j} + \frac{1}{\varepsilon}\frac{\partial f}{\partial \xi_j}\right)_{\boldsymbol{\xi}=\boldsymbol{x}/\varepsilon} = \left(\frac{\partial f}{\partial x_j} + \frac{f_{,j}}{\varepsilon}\right)_{\boldsymbol{\xi}=\boldsymbol{x}/\varepsilon}$, two fields equations of infinite order in which the unknowns are the macroscopic quantities $U_k(\mathbf{x})$ and $\Theta(\mathbf{x})$ are obtained (see Bacigalupo and Gambarotta (2014b) and Bacigalupo (2014) for details). According to Bacigalupo et al. (2016), the fields equations of the equivalent first order thermoelastic continuum can be obtained considering only the $\varepsilon^{-1}$ and $\varepsilon^{0}$ terms of the sequence of PDEs derived by the asymptotic procedure. In order to obtain a set of PDEs with constant coefficients, the fluctuations functions must satisfy non-homogeneous equations commonly known as *cell problems.* At the order $\varepsilon^{-1}$, the following equation are derived from Navier's equation (7):

$$\begin{aligned} &\left(C^{\varepsilon}_{ijkl} N^{(1)}_{kpq_1,l}\right)_{,j} + C^{\varepsilon}_{ijpq_1,j} = n^{(1)}_{ipq_1}, \\ &\left(C^{\varepsilon}_{ijkl} \tilde{N}^{(1)}_{k,l}\right)_{,j} - \alpha^{\varepsilon}_{ij,j} = \tilde{n}^{(1)}_{i}, \end{aligned} \tag{11}$$

whereas the heat conduction equation (8) yields

$$\left(K^{\varepsilon}_{ij} M^{(1)}_{q_1,j}\right)_{,i} + K^{\varepsilon}_{iq_1,i} = m^{(1)}_{q_1}, \tag{12}$$

where as a consequence of the $\mathcal{Q}-$periodicity of $C^{\varepsilon}_{ijpq_1}, \alpha^{\varepsilon}_{ij}$ and $K^{\varepsilon}_{iq_1}$ it can be easily verified that

$$n^{(1)}_{ipq_1} = \langle C^{\varepsilon}_{ijpq_1,j}\rangle = 0, \quad \tilde{n}^{(1)}_{i} = -\langle \alpha^{\varepsilon}_{ij,j}\rangle = 0, \quad m^{(1)}_{q_1} = \langle K^{\varepsilon}_{iq_1,i}\rangle = 0. \tag{13}$$

At the order $\varepsilon^{0}$, the following *cell problems* are derived from equation (7):

$$\begin{aligned} &\left(C^{\varepsilon}_{ijkl} N^{(2)}_{kpq_1q_2,l}\right)_{,j} + \frac{1}{2}\Bigg[\left(C^{\varepsilon}_{ijkq_2} N^{(1)}_{kpq_1}\right)_{,j} + C^{\varepsilon}_{iq_2pq_1} + C^{\varepsilon}_{iq_2kl} N^{(1)}_{kpq_1,l} \\ &+ \left(C^{\varepsilon}_{ijkq_1} N^{(1)}_{kpq_2}\right)_{,j} + C^{\varepsilon}_{iq_1pq_2} + C^{\varepsilon}_{iq_1kl} N^{(1)}_{kpq_2,l}\Bigg] = n^{(2)}_{ipq_1q_2}, \\ &\left(C^{\varepsilon}_{ijkl} \tilde{N}^{(2)}_{kq_1,l}\right)_{,j} + \left(C^{\varepsilon}_{ijkq_1} \tilde{N}^{(1)}_{k}\right)_{,j} + C^{\varepsilon}_{iq_1kl}\tilde{N}^{(1)}_{k,l} - \alpha^{\varepsilon}_{iq_1} - \left(\alpha^{\varepsilon}_{ij} M^{(1)}_{q_1}\right)_{,j} = \tilde{n}^{(2)}_{iq_1}, \end{aligned} \tag{14}$$

at the same order, the heat conduction equation (8) yields

$$\begin{aligned} &\left(K^{\varepsilon}_{ij} M^{(2)}_{q_1q_2,j}\right)_{,i} + \frac{1}{2}\Bigg[\left(K^{\varepsilon}_{iq_1} M^{(1)}_{q_2}\right)_{,i} + K^{\varepsilon}_{q_2q_1} + K^{\varepsilon}_{q_1j} M^{(1)}_{q_2,j} \\ &+ \left(K^{\varepsilon}_{iq_2} M^{(1)}_{q_1}\right)_{,i} + K^{\varepsilon}_{q_1q_2} + K^{\varepsilon}_{q_2j} M^{(1)}_{q_1,j}\Bigg] = m^{(2)}_{q_1q_2}, \end{aligned} \tag{15}$$

where:

$$n^{(2)}_{ipq_1q_2} = \frac{1}{2}\left\langle C^{\varepsilon}_{iq_2pq_1} + C^{\varepsilon}_{iq_2kl} N^{(1)}_{kpq_1,l} + C^{\varepsilon}_{iq_1pq_2} + C^{\varepsilon}_{iq_1kl} N^{(1)}_{kpq_2,l}\right\rangle,$$

$$\tilde{n}^{(2)}_{iq_1} = \left\langle C^{\varepsilon}_{iq_1kl}\tilde{N}^{(1)}_{k,l} - \alpha^{\varepsilon}_{iq_1}\right\rangle, \quad m^{(2)}_{q_1q_2} = \frac{1}{2}\left\langle K^{\varepsilon}_{q_2q_1} + K^{\varepsilon}_{q_1j}M^{(1)}_{q_2,j} + K^{\varepsilon}_{q_1q_2} + K^{\varepsilon}_{q_2j}M^{(1)}_{q_1,j}\right\rangle. \tag{16}$$

The perturbation functions characterizing the *down-scaling* relations (9) and (10) are obtained by the solution of the previously defined cells problems, derived by imposing the normalization conditions.

According to Bakhvalov and Panasenko (1984) and Smyshlyaev and Cherednichenko (2000), the constants (13) and (16) are determined by imposing that the non-homogeneous terms in equations (11), (12), (14) and (15) (associated to the auxiliary body forces (Bacigalupo, 2014) and heat sources) possess vanishing mean values over the unit cell $\mathcal{Q}$. This implies the $\mathcal{Q}-$periodicity of the perturbations functions $N^{(1)}_{kpq_1}$, $N^{(2)}_{kpq_1q_2}$, $M^{(1)}_{q_1}$, $M^{(2)}_{q_1q_2}$, $\tilde{N}^{(1)}_{k}$ and $\tilde{N}^{(2)}_{kq_1}$ and then the continuity and regularity of the microscopic fields (micro-displacements and micro-temperature) at the interface between adjacent cells are guaranteed. Using the cell problems (11), (12), (14) and (15) together with the constants definitions (13) and (16) into the asymptotic expansion of the microscopic fields equations (9) and (10) and truncating the asymptotic expansion at the order $\varepsilon^0$, the following averaged equations are derived

$$n^{(2)}_{ipq_1q_2}\frac{\partial^2 U_p}{\partial x_{q_1}\partial x_{q_2}} + \tilde{n}^{(2)}_{iq_1}\frac{\partial \Theta}{\partial x_{q_1}} + \mathcal{O}(\varepsilon) + b_i = 0 \tag{17}$$

$$m^{(2)}_{q_1q_2}\frac{\partial^2 \Theta}{\partial x_{q_1}\partial x_{q_2}} + \mathcal{O}(\varepsilon) + r = 0. \tag{18}$$

It is important to note that the solution of equations (17) and (18) requires that the following normalization condition is satisfied:

$$\frac{1}{\delta L^2}\int_{\mathcal{L}} U_p(\mathbf{x})d\mathbf{x} = 0, \quad \frac{1}{\delta L^2}\int_{\mathcal{L}} \Theta(\mathbf{x})d\mathbf{x} = 0, \tag{19}$$

where the $\mathcal{L}-$periodic domain is the same defined in previous Section as $\mathcal{L} = [0, L] \times [0, \delta L]$.

In the next Section, using the symmetry properties of the tensors (16) together with the ellipticity of the fields equations (17) and (18), the coefficients $n^{(2)}_{ipq_1q_2}$, $\tilde{n}^{(2)}_{iq_1}$, $m^{(2)}_{q_1q_2}$ are related to the overall elastic and thermodiffusive constants of the media $C_{iq_1pq_2}$, $\alpha_{iq_1}$ and $K_{q_1q_2}$, and the homogenized field equations of the first order continuum equivalent to a thermoelastic composite material with periodic microstructure are derived from the (17) and (18) (see Bacigalupo et al. (2016)).

## 3.2 Overall properties of equivalent homogeneous continuum

The field equations of an homogeneous first order continuum in presence of thermodiffusion are given by

$$C_{iq_1pq_2}\frac{\partial^2 U_p}{\partial x_{q_1}\partial x_{q_2}} - \alpha_{iq_1}\frac{\partial \Theta}{\partial x_{q_1}} + b_i = 0, \tag{20}$$

$$K_{q_1q_2}\frac{\partial^2 \Theta}{\partial x_{q_1}\partial x_{q_2}} + r = 0, \tag{21}$$

where $C_{iq_1pq_2}$ are the components of the overall elastic tensor, $\alpha_{iq_1}$ are the components of the overall thermal dilatation tensor and $K_{q_1q_2}$ denotes the components of the overall heat conduction tensor. Remembering the approximation, the macroscopic field equations (20) and (21) can be compared to the zero order terms of the averaged field equation for determining the overall properties of the thermodiffusive Cauchy continuum. In order to relate the coefficients $n^{(2)}_{ipq_1q_2}$, $\tilde{n}^{(2)}_{iq_1}$, $m^{(2)}_{q_1q_2}$ to the overall elastic and thermodiffusive constants of the media $C_{iq_1pq_2}$, $\alpha_{iq_1}$ and $K_{q_1q_2}$ the symmetries of the tensors of components $n^{(2)}_{ipq_1q_2}$, $\tilde{n}^{(2)}_{iq_1}$, $m^{(2)}_{q_1q_2}$, and the ellipticity of the fields equations (17) and (18) are required. A demonstration of these properties is reported in Bacigalupo et al. (2016). As a consequence of these properties, it can be observed that: $n^{(2)}_{ipq_1q_2} = \frac{1}{2}(C_{iq_1pq_2} + C_{iq_2pq_1})$, $\tilde{n}^{(2)}_{iq_1} = \alpha_{iq_1}$ and $m^{(2)}_{q_1q_2} = K_{q_1q_2}$. In particular, comparing the field equation (20) to (17), and remembering the relationship between $n^{(2)}_{ipq_1q_2}$ and $C_{iq_1pq_2}$, it is easy to note that due to the repetition of the indexes $q_1$ and $q_2$: $C_{iq_1pq_2}\frac{\partial^2 U_p}{\partial x_{q_1}\partial x_{q_2}} = n^{(2)}_{ipq_1q_2}\frac{\partial^2 U_p}{\partial x_{q_1}\partial x_{q_2}} = \frac{1}{2}(C_{iq_1pq_2} + C_{iq_2pq_1})\frac{\partial^2 U_p}{\partial x_{q_1}\partial x_{q_2}}$.

The overall elastic and thermodiffusive tensors, obtained in terms of fluctuations functions, and the components of microscopic elastic and thermodiffusive tensors, take the form (see Bacigalupo et al. (2016) for details):

$$C_{iq_1pq_2} = \frac{1}{4}\left\langle C^{\varepsilon}_{rjkl}\left(N^{(1)}_{riq_1,j} + \delta_{ri}\delta_{jq_1} + N^{(1)}_{rq_1i,j} + \delta_{rq_1}\delta_{ij}\right)\left(N^{(1)}_{kpq_2,l} + \delta_{kp}\delta_{q_2,l} + N^{(1)}_{kq_2p,l} + \delta_{kq_2}\delta_{lp}\right)\right\rangle,$$

$$\alpha_{iq_1} = \left\langle C^{\varepsilon}_{iq_1kl}\tilde{N}^{(1)}_{k,l} - \alpha^{\varepsilon}_{iq_1}\right\rangle,$$

$$K_{q_1q_2} = \left\langle K^{\varepsilon}_{ij}(M^{(1)}_{q_1,j} + \delta_{jq_1})(M^{(1)}_{q_2,i} + \delta_{iq_2})\right\rangle. \tag{22}$$

The components $C_{iq_1pq_2}$ and $K_{q_1q_2}$ of the overall constitutive tensors of the material coincide with those derived by asymptotic homogenization techniques applied to uncoupled static elastic (Bakhvalov and Panasenko, 1984; Smyshlyaev and Cherednichenko, 2000; Bacigalupo, 2014) and heat conduction problems (Zhang et al., 2007) in media with periodic microstructures. The components $\alpha_{iq_1}$ of the coupling thermoelastic tensor have been obtained by means of a consistent generalization of the down-scaling relations (9) and (10). These expressions relate the microscopic displacement field to the macroscopic displacements and temperature. The asymptotic homogenization procedure described in this Section is an extension of the general methods proposed by Bakhvalov and Panasenko (1984); Smyshlyaev and Cherednichenko (2000); Bacigalupo and Gambarotta (2014b) and Bacigalupo (2014) to the case of periodic thermoelastic materials, and it can be applied to study the effects of any generic periodic microstructure, including both two- and three-dimensional geometries without additional restrictions. This means that expressions (22) are valid for all thermoelastic composite media with periodic microstructure, and the characteristics of the microstructures are described by means of the fluctuations functions $N^{(1)}_{riq_j}$, $\tilde{N}^{(1)}_{k}$ and $M^{(1)}_{q_j}$. For layered media composed by an arbitrary number of phases of arbitrary thickness, such as for example the tri-phase thermoelastic laminate considered in next Section, the fluctuations functions can be determined analitically, whereas for more complex topologies of the microstructure, they must be estimated by means of numerical techniques.

# 4 Homogenization of multi-phase layered thermoelastic composites of interest for SOFC devices fabrication

The developed general homogenization procedure is now applied to the case of a three-phase thermoelastic composite which can be used to model the thermomechanical behaviour of energy devices possessing a layered structure, such as solid oxide fuel cells or lithium ions batteries. Exact analytical expressions for the perturbation functions and then the overall thermomechanical constants are derived. Two different examples of space-varying heat sources are considered. The analytical results derived by the solution the homogenized model are compared with those provided by the finite element analysis of the corresponding heterogeneous problem.

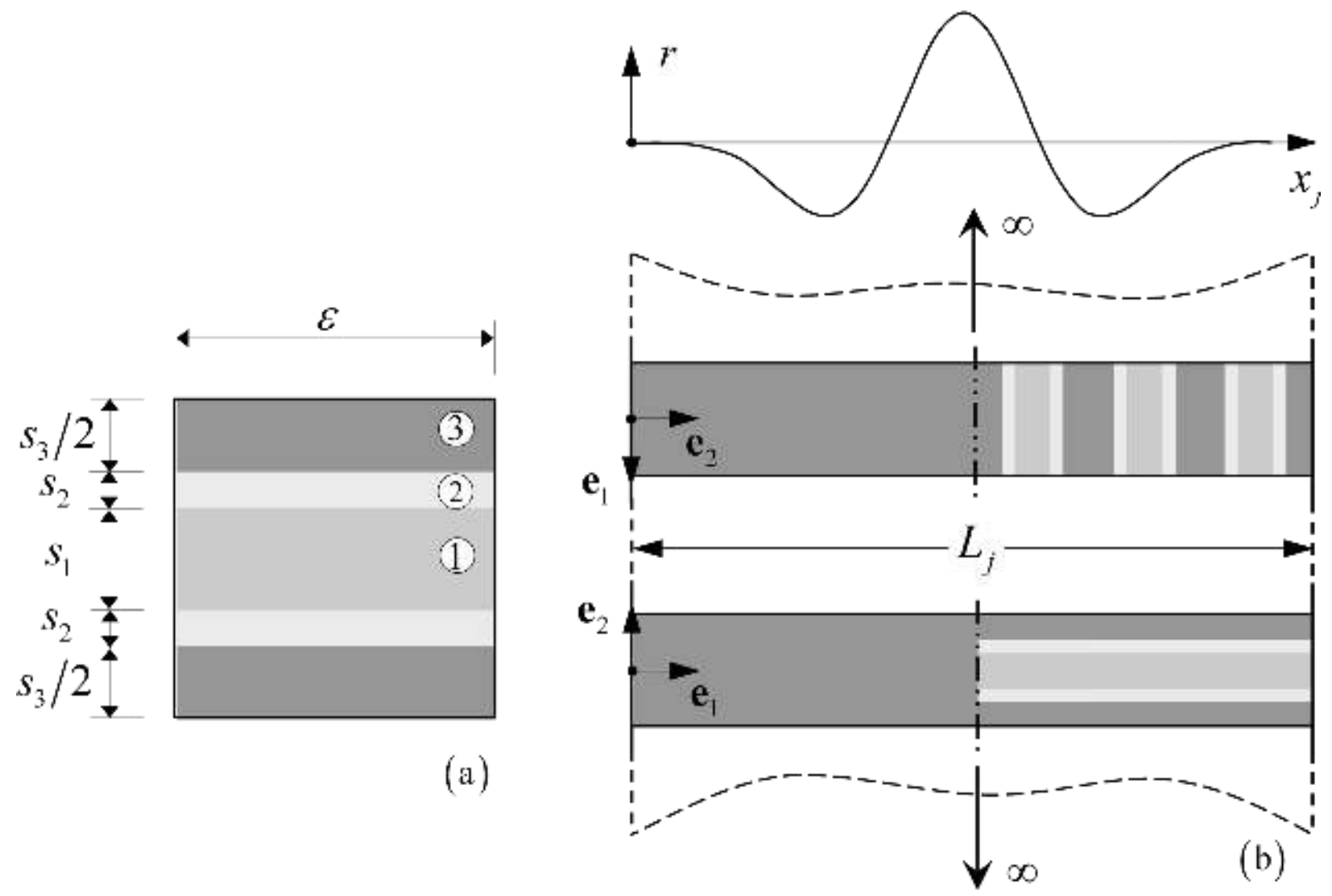

Figure 2: (a): Variation of the localized heat source $\tilde{r}$ with the normalized space variable $\tilde{x}_j$ with $j = 1, 2$ reported for $n = 1$ (red line) and $n = 2$ (black line); (b): Variation of the not localized heat source $\tilde{r}$ with the normalized space variable $\tilde{x}_j$ with $j = 1, 2$ reported for $n = 1$ (red line) and $n = 2$ (black line).

## 4.1 Perturbation functions and overall thermoelastic constants

The interconnector-electrode(anode or cathode)-electrolyte system in energy battery devices can be modelled introducing a layered material defined as an unbounded two-dimensional periodic arrangement of three different layers having thickness $s_1$, $s_2$ and $s_3$, where $d_2 = \varepsilon = s_1 + 2s_2 + s_3$, $\hat{s}_p = s_p/\varepsilon$, $p = 1, 2, 3$ and $\zeta = \hat{s}_3/2\hat{s}_2$ are defined. For simplicity, the phases are assumed homogeneous and orthotropic, with an orthotropic axis coincident with the layering direction $\mathbf{e}_1$, the geometry of the system is shown in Fig. 2a. It is important to note that, if the periodicity condition is preserved, the general asymptotic homogenization method introduced in the previous section can be applied to laminate media whose layers possess orthotropy axes of different orientation. The different orientation of the orthotropy axes brings additional complexity of the associated cell problems, whose solution may require numerical techniques.

We assume that elastic, thermoelastic and heat conduction tensors possess orthotropic symmetry. The perturbation functions $N^{(1)}_{riq_1}$, $\tilde{N}^{(1)}_k$, and $M^{(1)}_{q_1}$ are explicitly determined solving the cell problems reported in Section 3.1. Due to the particular properties of symmetry of the microstruc-ture, these functions depend only on the fast variable $\xi_2 = x_2/\varepsilon$, which is perpendicular to the layering direction, and then they are independent by $x_1$ and $\xi_1 = x_1/\varepsilon$. The non-vanishing functions $N^{(1)}_{riq_1}$ $\tilde{N}^{(1)}_k$, obtained by the solution of the cell problem of order $\varepsilon^{-1}$ (11) are given in the form:

$$N^{(1),ⓟ}_{211} = A^{(1),ⓟ}_{211}\xi^{ⓟ}_2 + B^{(1),ⓟ}_{211}, \quad N^{(1),ⓟ}_{222} = A^{(1),ⓟ}_{222}\xi^{ⓟ}_2 + B^{(1),ⓟ}_{222},$$

$$N^{(1),ⓟ}_{112} = N^{(1),ⓟ}_{121} = A^{(1),ⓟ}_{112}\xi^{ⓟ}_2 + B^{(1),ⓟ}_{112}, \tag{23}$$

and

$$\tilde{N}^{(1),ⓟ}_2 = \tilde{A}^{(1),ⓟ}_2\xi^{ⓟ}_2 + \tilde{B}^{(1),ⓟ}_2, \tag{24}$$

where $p = 1, 2, 3$ denotes respectively the phases 1,2,3 and $\xi^{①}_2 \in \left[-\frac{\hat{s}_1}{2}, \frac{\hat{s}_1}{2}\right]$, $\xi^{②}_2 \in \left[-\frac{1-\hat{s}_1}{4(1+\zeta)}, \frac{1-\hat{s}_1}{4(1+\zeta)}\right]$ and $\xi^{③}_2 \in \left[-\frac{\zeta(1-\hat{s}_1)}{2(1+\zeta)}, \frac{\zeta(1-\hat{s}_1)}{2(1+\zeta)}\right]$ are non-dimensional vertical coordinates centred in each layer. The explicit expressions for the coefficients $A^{(1),ⓟ}_{ijk}$, $B^{(1),ⓟ}_{ijk}$, $\tilde{A}^{(1),ⓟ}_{ijk}$ and $\tilde{B}^{(1),ⓟ}_{ijk}$, where $i, j, k = 1, 2$ are reported in Appendix A. At the same order $\varepsilon^{-1}$, the non-vanishing fluctuation functions associated to the heat conduction equation, derived by the solution of the cell problems of order $\varepsilon^{-1}$ (12) are given by:

$$M^{(1),ⓟ}_2 = P^{(1),ⓟ}_1\xi^{ⓟ}_2 + Q^{(1),ⓟ}_1, \quad p = 1, 2, 3, \tag{25}$$

where the coefficients $P^{(1),ⓟ}_i$ and $Q^{(1),ⓟ}_i$ are also reported in Appendix A.

Substituting the fluctuation functions (23), (24) and (25) into the expressions (22), the components of the constitutive tensors corresponding to the first order equivalent continuum are derived. The non-vanishing components of the overall elastic tensor $C_{iq_1pq_2}$ are given by:

$$C_{1111} = \frac{\mathcal{A}^1_{1111}\hat{s}^3_2 + \mathcal{A}^2_{1111}\hat{s}^3_3 + \mathcal{A}^3_{1111}\hat{s}^2_2\hat{s}_3 + \mathcal{A}^4_{1111}\hat{s}^2_3\hat{s}_2 + \mathcal{A}^5_{1111}\hat{s}^2_2 + \mathcal{A}^6_{1111}\hat{s}^2_3 + \mathcal{A}^7_{1111}\hat{s}_2\hat{s}_3}{\Delta_{1111}},$$

$$C_{2222} = \frac{\mathcal{A}^1_{2222}\hat{s}^3_2 + \mathcal{A}^2_{2222}\hat{s}^3_3 + \mathcal{A}^3_{2222}\hat{s}^2_2\hat{s}_3 + \mathcal{A}^4_{2222}\hat{s}^2_3\hat{s}_2 + \mathcal{A}^5_{2222}\hat{s}^2_2 + \mathcal{A}^6_{2222}\hat{s}^2_3 + \mathcal{A}^7_{2222}\hat{s}_2\hat{s}_3}{\Delta_{2222}},$$

$$C_{1212} = \frac{\mathcal{A}^1_{1212}\hat{s}^3_2 + \mathcal{A}^2_{1212}\hat{s}^3_3 + \mathcal{A}^3_{1212}\hat{s}^2_2\hat{s}_3 + \mathcal{A}^4_{1212}\hat{s}^2_3\hat{s}_2 + \mathcal{A}^5_{1212}\hat{s}^2_2 + \mathcal{A}^6_{1212}\hat{s}^2_3 + \mathcal{A}^7_{1212}\hat{s}_2\hat{s}_3}{\Delta_{1212}},$$

$$C_{1122} = \frac{\mathcal{A}^1_{1122}\hat{s}^3_2 + \mathcal{A}^2_{1122}\hat{s}^3_3 + \mathcal{A}^3_{1122}\hat{s}^2_2\hat{s}_3 + \mathcal{A}^4_{1122}\hat{s}^2_3\hat{s}_2 + \mathcal{A}^5_{1122}\hat{s}^2_2 + \mathcal{A}^6_{1122}\hat{s}^2_3 + \mathcal{A}^7_{1122}\hat{s}_2\hat{s}_3}{\Delta_{1122}}, \tag{26}$$

where the coefficients $\mathcal{A}^q_{ijhk}$ and $\Delta_{ijhk}$, $i, j, h, k = 1, 2$, $q = 1, \ldots, 7$ are reported in Appendix B. The non-vanishing components of the overall thermal dilatation assume the form:

$$\alpha_{11} = \frac{\mathcal{B}^1_{11}\hat{s}^2_2 + \mathcal{B}^2_{11}\hat{s}^2_3 + \mathcal{B}^3_{11}\hat{s}_2\hat{s}_3 + \mathcal{B}^4_{11}\hat{s}_2 + \mathcal{B}^5_{11}\hat{s}_2}{\Lambda_{11}},$$

$$\alpha_{22} = \frac{\mathcal{B}^1_{22}\hat{s}^2_2 + \mathcal{B}^2_{22}\hat{s}^2_3 + \mathcal{B}^3_{22}\hat{s}_2\hat{s}_3 + \mathcal{B}^4_{22}\hat{s}_2 + \mathcal{B}^5_{22}\hat{s}_2}{\Lambda_{22}}, \tag{27}$$

where the coefficients $\mathcal{B}^q_{ij}$ and $\Lambda_{ij}$, $i,j=1,2$, $q=1,\dots,5$ are reported in Appendix B. The components of the overall heat conduction tensor then become:

$$K_{11}=2(K^{②}_{11}-K^{①}_{11})\hat{s}_2+(K^{③}_{11}-K^{①}_{11})\hat{s}_3+K^{①}_{11},$$

$$K_{22}=\frac{\mathcal{D}^1_{22}\hat{s}^3_2+\mathcal{D}^2_{22}\hat{s}^3_3+\mathcal{D}^3_{22}\hat{s}^2_2\hat{s}_3+\mathcal{D}^4_{22}\hat{s}^2_3\hat{s}_2+\mathcal{D}^5_{22}\hat{s}^2_2+\mathcal{D}^6_{22}\hat{s}^2_3+\mathcal{D}^7_{22}\hat{s}_2\hat{s}_3}{\Xi_{22}}, \tag{28}$$

where the coefficients $\mathcal{D}^q_{ij}$ $i,j=1,2$, $q=1,\dots,7$ and $\Xi_{22}$ are reported in Appendix B. Note that, due to the invariance of the perturbation functions (23), (24), (25) with respect to the coordinate $x_1$, the overall constants (26), (27) and (28) do not depend on the characteristics length $L_1$, and are only functions of the phases thickness. In the case where isotropic phases are assumed, the components of the elasticity tensor become $C^{ⓟ}_{1111}=C^{ⓟ}_{2222}=\frac{\tilde{E}_p}{1-\tilde{\nu}^2_p}$, $C^{ⓟ}_{1122}=\frac{\tilde{\nu}_p\tilde{E}_p}{1-\tilde{\nu}^2_p}$, $C^{ⓟ}_{1212}=\frac{\tilde{E}_p}{2(1+\tilde{\nu}_p)}$, (with $p=1,2,3$), where for plane-strain: $\tilde{E}_p=\frac{E_p}{1-\nu^2_p}$, $\tilde{\nu}_p=\frac{\nu_p}{1-\nu_p}$, whereas for plane-stress: $\tilde{E}_p=E_p$, $\tilde{\nu}_p=\nu_p$, being $E_p$ the Young's modulus and $\nu_p$ the Poisson's ratio, respectively. The components of the thermal dilatation tensor take the forms: $\alpha^{ⓟ}_{11}=\alpha^{ⓟ}_{22}=\alpha_p$. The components of the heat conduction tensor finally become $K^{ⓟ}_{11}=K^{ⓟ}_{22}=K_p$.

For simplicity, in the illustrative examples the three phases are assumed to be isotropic. The overall thermoelastic and heat diffusion constants can be represented in the following normalized form:

$$\tilde{C}_{iq_1pq_2}\big(\hat{s}_1,\zeta,\rho^{②}_C,\rho^{③}_C,\nu_1,\nu_2,\nu_3\big)=\frac{C_{iq_1pq_2}}{\hat{C}_{iq_1pq_2}},\quad \tilde{\alpha}_{iq_1}\big(\hat{s}_1,\zeta,\rho^{②}_C,\rho^{③}_C,\nu_1,\nu_2,\nu_3,\rho^{②}_\alpha,\rho^{3}_{③}\big)=\frac{\alpha_{iq_1}}{\hat{\alpha}_{iq_1}},$$

$$\tilde{K}_{iq_1}\big(\hat{s}_1,\zeta,\rho^{②}_K,\rho^{③}_K\big)=\frac{K_{iq_1}}{\hat{K}_{iq_1}}, \tag{29}$$

where $\rho^{②}_C=E_2/E_1$, $\rho^{③}_C=E_3/E_1$, $\rho^{②}_\alpha=\alpha_2/\alpha_1$, $\rho^{③}_\alpha=\alpha_3/\alpha_1$, $\rho^{②}_K=K_2/K_1$, $\rho^{③}_K=K_3/K_1$, and $\hat{C}_{iq_1pq_2}=(C^{①}_{iq_1pq_2}+C^{②}_{iq_1pq_2}+C^{③}_{iq_1pq_2})/3$, $\hat{\alpha}_{iq_1}=(\alpha^{①}_{iq_1}+\alpha^{②}_{iq_1}+\alpha^{③}_{iq_1})/3$, $\hat{K}_{iq_1}=(K^{①}_{iq_1}+K^{②}_{iq_1}+K^{③}_{iq_1})/3$. The following values for the geometrical parameters and the Poisson's ratios have been assumed for the computations: $\hat{s}_1=5/88$, $\nu_1=0.3$, $\nu_2=0.25$, $\nu_3=0.3$ according to Bacigalupo et al. (2014). The overall thermoelastic and heat diffusion constants have been computed for the case of plane strain. Similar results can be easily obtained for the plane stress.

The variation of the normalized components of the overall elasticity tensor $\tilde{C}_{1111}$ and $\tilde{C}_{1212}$ with the ratio $\zeta$ is reported in Figs. 3 and 4. The behaviour of other elastic constants is not reported since it is qualitatively analogous to those detected for $\tilde{C}_{1111}$ and $\tilde{C}_{1212}$. The results reported in Figs. 3a and 4a have been obtained considering the value $\rho^{②}_C=5/13$, and different values for the ratio $\rho^{③}_C$: $\rho^{③}_C=31/26,1,3/2,2$, whereas the plots shown in Figs. 3b and 4b correspond to $\rho^{③}_C=5/13$ and $\rho^{②}_C=31/26,\rho^{②}_C=5/13,\rho^{②}_C=1/10,\rho^{②}_C=1/5,\rho^{②}_C=3/10$. It can be observed that as $\zeta\to 0$, and then as the thickness of the phase 3 associated to the interconnection of the battery vanishes (i. e. $\hat{s}_3=0$), the values of the overall elastic constants $\tilde{C}_{iq_1pq_2}$ tend to the overall constants of a bi-phase layered media composed only by the electrolyte-electrode system. In this case, the limit values of the constants $\tilde{C}_{iq_1pq_2}$ increase as $\rho^{③}_C$ decreases at constant $\rho^{②}_C$ (see Figs. 3a and 4a) and increase as $\rho^{②}_C$ increases at constant $\rho^{③}_C$ (see Figs. 3b and 4b). For $\zeta\to+\infty$, the thickness of the phase 2, associated to the electrodes, tends to zero (i. e. $\hat{s}_2=0$), and then the components of the overall elastic tensor assume limit values which corresponds to the overall constants of a bi-phase systems composed by the electrolyte and the interconnection. In this limit case, the value of the constants $\tilde{C}_{iq_1pq_2}$ increase as $\rho_C^{③}$ increases at constant $\rho_C^{②}$, (see Figs. 3a and 4a) and increases as $\rho_C^{②}$ decreases at constant $\rho_C^{③}$ (see Figs. 3b and 4b). The values of the ratios $\rho_C^{②}$ and $\rho_C^{③}$ considered for the analysis, associated to realistic materials commonly used in battery devices fabrication (Bacigalupo et al., 2014), imply that $E_3 > E_2$ and consequently for the values of the Poisson's ratios here assumed $\tilde{C}^{③}_{iq_1pq_2} > \tilde{C}^{②}_{iq_1pq_2}$. This means that the phase 3 is stiffer with respect the phase 2, and then the values of the constants $\tilde{C}_{iq_1pq_2}$ evaluated for $\zeta = 0$ are smaller than the values assumed for $\zeta \to \infty$. As a consequence, the components of the overall elastic tensors increase monotonically with the geometric ratio $\zeta$ (see Appendix D for the analytical expressions of the overall elastic constants in the limit cases $\zeta = 0$, $\zeta \to +\infty$).

The contour plots reported in Fig. 5, show the variation of $\tilde{C}_{1111}$ (Fig. 5a) and $\tilde{C}_{1212}$ (Fig. 5b) with $\rho_C^{②}$ and $\rho_C^{③}$ obtained for fixed values of the geometric parameters $\zeta, \hat{s}_1$ and of the Poisson's ratios. We can observe that for a constant value of the ratio $\rho_C^{②}$, as $\rho_C^{③}$ increases the overall elastic constants increases monotonically. Conversely, if the values of $\rho_C^{③}$ is fixed, $\tilde{C}_{1111}$ and $\tilde{C}_{1212}$ initially increase $\rho_C^{②}$ with and then after reaching a maximum decrease.

The variation of the normalized component of the overall thermal dilatation tensor $\tilde{\alpha}_{11}$ with the ratio $\zeta$ is reported in Fig. 6. The curves reported in Fig. 6a have been obtained assuming the values $\nu_1 = 0.3$, $\nu_2 = 0.25$, $\nu_3 = 0.25$, $\rho_C^{②} = 5/13$, $\rho_C^{③} = 31/26$, $\hat{s}_1 = 5/88$, $\rho_\alpha^{②} = 25/26$ and different values for $\rho_\alpha^{③}$: $\rho_\alpha^{③} = 109/130$, $\rho_\alpha^{③} = 2/5$, $\rho_\alpha^{③} = 6/5$, $\rho_\alpha^{③} = 8/5$, whereas the plots shown in Fig. 6b correspond to $\rho_\alpha^{③} = 109/130$ and $\rho_\alpha^{②} = 25/26$, $\rho_\alpha^{②} = 1/2$, $\rho_\alpha^{②} = 3/2$, $\rho_\alpha^{②} = 2$. The behaviour of other component $\tilde{\alpha}_{22}$ is not reported because it is qualitatively similar to those detected for $\tilde{\alpha}_{11}$. It can be observed that for $\zeta \to 0$, the limit values assumed by $\tilde{\alpha}_{11}$ become higher as $\rho_\alpha^{③}$ decreases maintaining constant $\rho_\alpha^{②}$ (see Fig. 6a), and increases as $\rho_\alpha^{②}$ increases maintaining constant $\rho_\alpha^{③}$ (see Fig. 6b). Conversely, for $\zeta \to +\infty$, the limit value of $\tilde{\alpha}_{11}$ increases as $\rho_\alpha^{③}$ increases maintaining constant $\rho_\alpha^{②}$, (see Fig. 6a) and increases as $\rho_\alpha^{②}$ decreases maintaining constant $\rho_\alpha^{③}$ (see Fig. 6b). It can be observed that in the cases where $\alpha_2 > \alpha_3$, then $\tilde{\alpha}_{iq_1}(\zeta = 0) > \tilde{\alpha}_{iq_1}(\zeta \to +\infty)$. As a consequence, the overall thermal dilatation constants increase monotonically with $\zeta$. Conversely, for $\alpha_2 < \alpha_3$ the components $\tilde{\alpha}_{iq_1}$ decreases monotonically as $\zeta$ increases (see Appendix D for the analytical expressions of the components of the overall thermal dilatation tensors in the limit cases $\zeta = 0$, $\zeta \to +\infty$).

The contour plots reported in Fig. 7, show the variation of $\tilde{\alpha}_{11}$ (Fig. 7a) and $\tilde{\alpha}_{22}$ (Fig. 7b) with $\rho_\alpha^{②}$ and $\rho_\alpha^{③}$ obtained for constant values of the geometric parameters $\zeta$, $\hat{s}_1$, $\rho_C^{②}$, $\rho_C^{③}$ and of the Poisson's ratios. We can observe that for a constant value of the ratio $\rho_\alpha^{②}$, the components of the overall thermal dilatation tensor increase monotonically as $\rho_\alpha^{③}$ increases. Conversely, for a constant value of $\rho_\alpha^{③}$, $\tilde{\alpha}_{11}$ and $\tilde{\alpha}_{22}$ decrease monotonically with $\rho_\alpha^{②}$.

The variation of the normalized components of the overall heat conductivity tensor $\tilde{K}_{11}$ and $\tilde{K}_{22}$ with the ratio $\zeta$ is reported in Figs. 8 and 9. The results reported in Figs. 8a and 9a have been obtained considering the value $\rho_K^{②} = 7/50$, and different values for the ratio $\rho_K^{③}$: $\rho_K^{③} = 53/480, 1/20, 7/50, 1/5$, whereas the plots shown in Figs. 8b and 9b correspond to $\rho_K^{③} = 53/480$ and $\rho_K^{②} = 7/50, 5/50, 53/480, 1/5$. It can be observed that for $\zeta \to 0$, the limit values assumed by $\tilde{K}_{11}$ and $\tilde{K}_{22}$ become higher as $\rho_K^{③}$ decreases maintaining constant $\rho_K^{②}$ (see Fig. 8a), and increase as $\rho_K^{②}$ increases maintaining constant $\rho_K^{③}$ (see Fig. 8b). Conversely, for $\zeta \to +\infty$, the limit values of $\tilde{K}_{11}$ and $\tilde{K}_{11}$ increase as $\rho_K^{③}$ increases maintaining constant $\rho_K^{②}$, (see Fig. 9a) and increase as $\rho_K^{②}$ decreases maintaining constant $\rho_K^{③}$ (see Fig. 9b). It can be noted that in the cases where $K_2 > K_3$, then $\tilde{K}_{iq_1}(\zeta = 0) > \tilde{K}_{iq_1}(\zeta \to +\infty)$. Consequently, the components of the overall heat conduction tensor increase monotonically with $\zeta$. Conversely, for $K_2 < K_3$ the components $\tilde{K}_{iq_1}$ decreases monotonically with $\zeta$ (see Appendix D for the analytical expressions of the components of the overall heat conduction tensor in the limit cases $\zeta = 0$, $\zeta \to +\infty$).

The contour plots reported in Fig. 10, show the variation of $\tilde{K}_{11}$ (Fig. 10a) and $\tilde{K}_{22}$ (Fig. 7b) with $\rho_K^{②}$ and $\rho_K^{③}$ obtained for constant values of the geometric parameters $\zeta$ and $\hat{s}_1$. We can observe that for a constant value of the ratio $\rho_K^{②}$, both the components of the overall heat conduction tensor increase monotonically as $\rho_K^{③}$ increases. For a constant value of $\rho_K^{③}$, $\tilde{K}_{11}$ decrease monotonically with $\rho_K^{②}$, whereas the behaviour of $\tilde{K}_{22}$ is characterized by the presence of a maximum in $\rho_K^{②} = \rho_{Kmax}^{②}$.

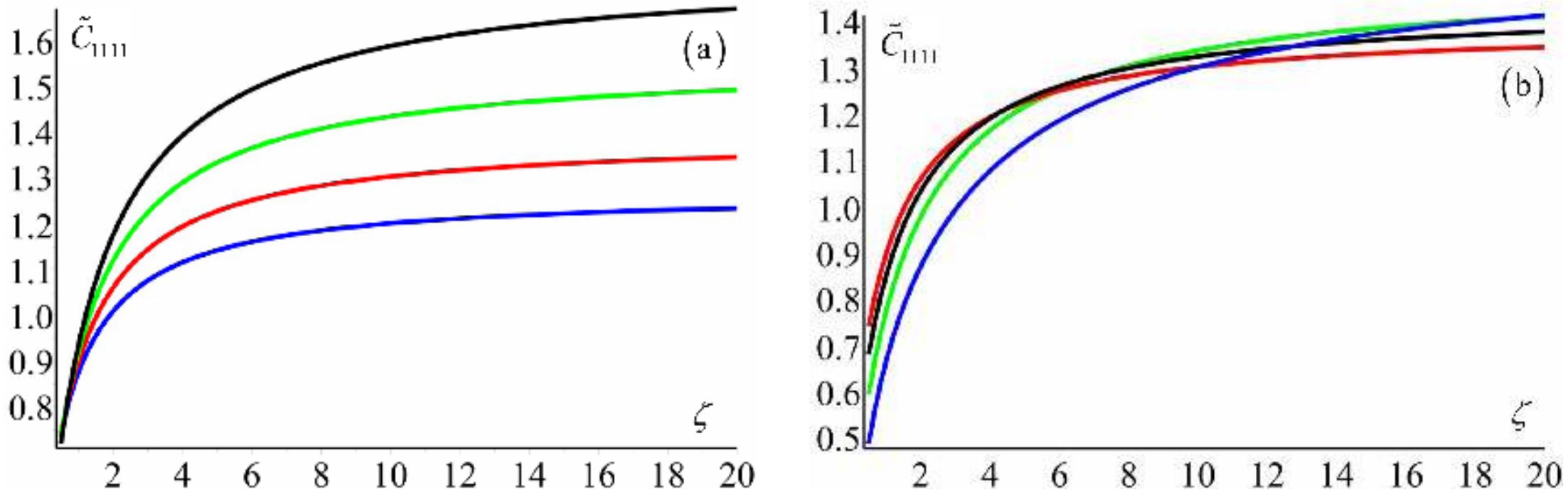

Figure 3: Dimensionless constant $\tilde{C}_{1111}$ vs. the geometric ratio $\zeta$ obtained assuming $\nu_1 = 0.3$, $\nu_2 = 0.25$, $\nu_3 = 0.25$, $\hat{s}_1 = 5/88$, $\rho_C^{②} = 5/13$ and : (a) $\rho_C^{②} = 5/13$ and different values of the ratio $\rho_C^{③}$: $\rho_C^{③} = 31/26$ red line, $\rho_C^{③} = 1$ blue line, $\rho_C^{③} = 3/2$ green line, $\rho_C^{③} = 2$ black line. (b) $\rho_C^{③} = 5/13$ and different values of the ratio $\rho_C^{②} = 31/26$: $\rho_C^{②} = 5/13$ red line, $\rho_C^{②} = 1/10$ blue line, $\rho_C^{②} = 1/5$ green line, $\rho_C^{②} = 3/10$ black line.

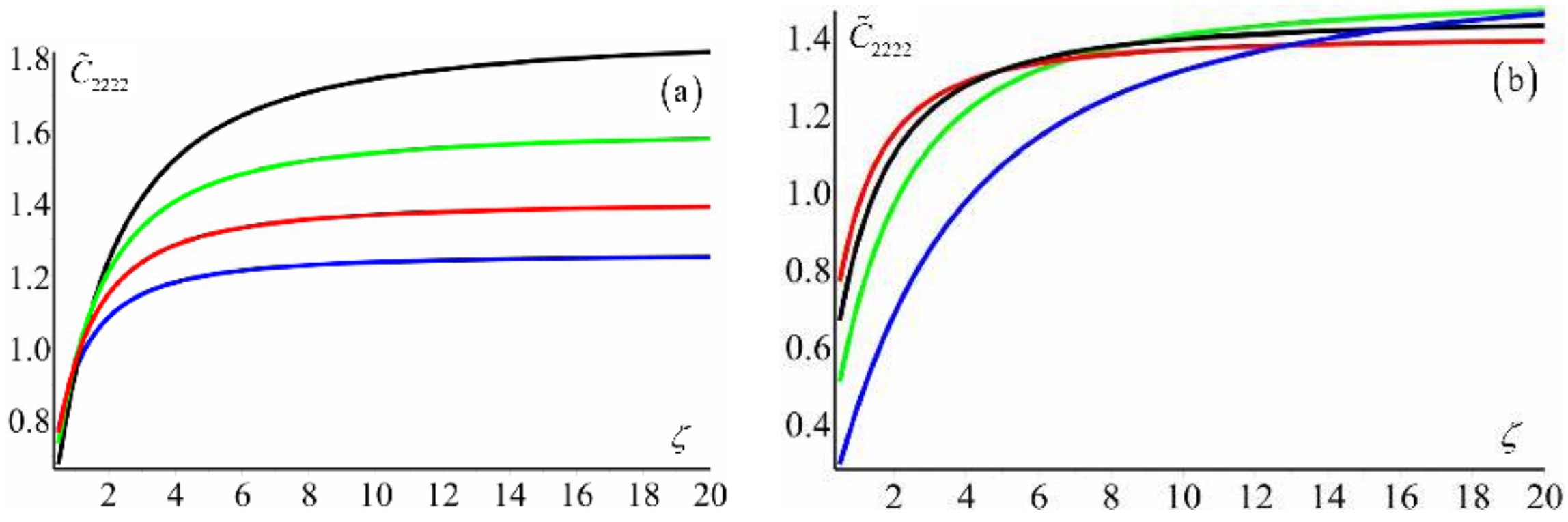

Figure 4: Dimensionless constant $\tilde{C}_{1212}$ vs. the geometric ratio $\zeta$ obtained assuming $\nu_1 = 0.3, \nu_2 = 0.25, \nu_3 = 0.25$, $\hat{s}_1 = 5/88, \rho_C^{②} = 5/13$ and : (a) $\rho_C^{②} = 5/13$ and different values of the ratio $\rho_C^{③}$: $\rho_C^{③} = 31/26$ red line, $\rho_C^{③} = 1$ blue line, $\rho_C^{③} = 3/2$ green line, $\rho_C^{③} = 2$ black line. (b) $\rho_C^{③} = 5/13$ and different values of the ratio $\rho_C^{②} = 31/26$: $\rho_C^{②} = 5/13$ red line, $\rho_C^{②} = 1/10$ blue line, $\rho_C^{②} = 1/5$ green line, $\rho_C^{②} = 3/10$ black line.

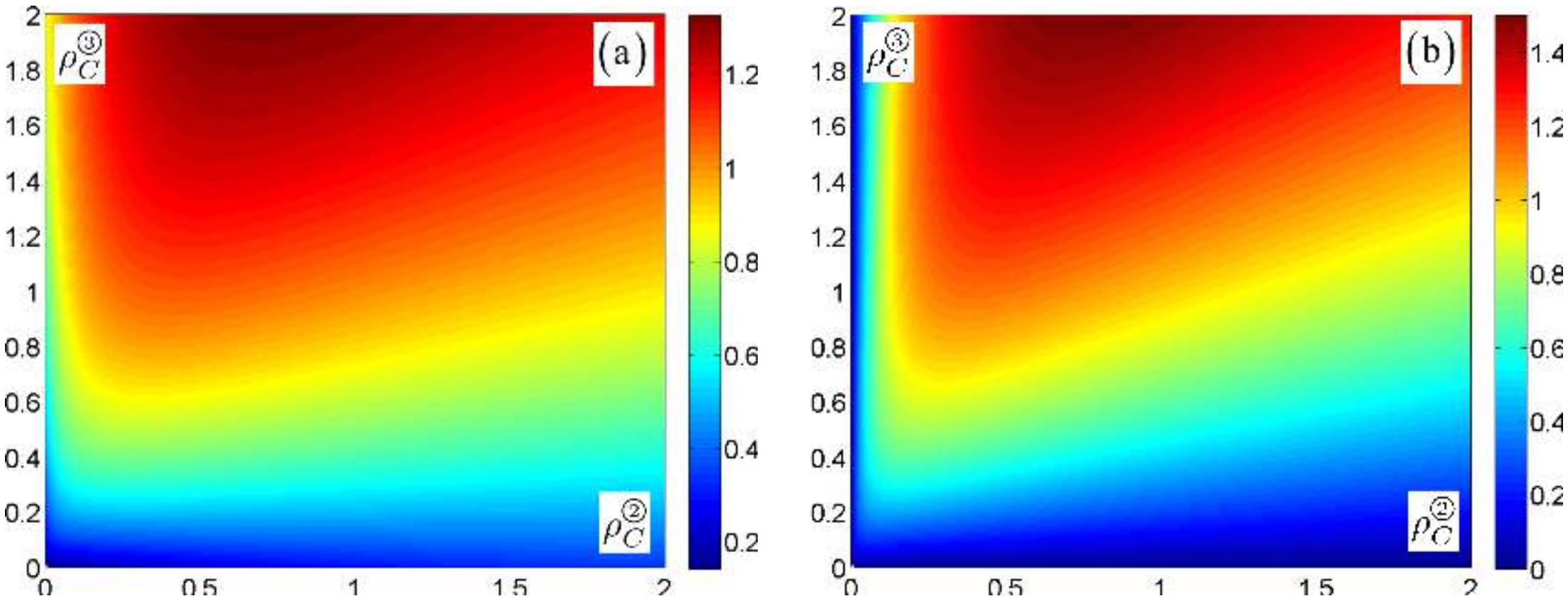

Figure 5: (a): variation of the dimensionless constant $\tilde{C}_{1111}$ with the ratios $\rho_C^{\textcircled{2}}$ and $\rho_C^{\textcircled{3}}$ obtained assuming $\nu_1 = 0.3, \nu_2 = 0.25, \nu_3 = 0.25$, $\hat{s}_1 = 5/88, \zeta = 14/5$. (b): variation of the dimensionless constant $\tilde{C}_{1212}$ with the ratios $\rho_C^{\textcircled{2}}$ and $\rho_C^{\textcircled{3}}$ obtained assuming $\nu_1 = 0.3, \nu_2 = 0.25, \nu_3 = 0.25$, $\hat{s}_1 = 5/88, \zeta = 14/5$.

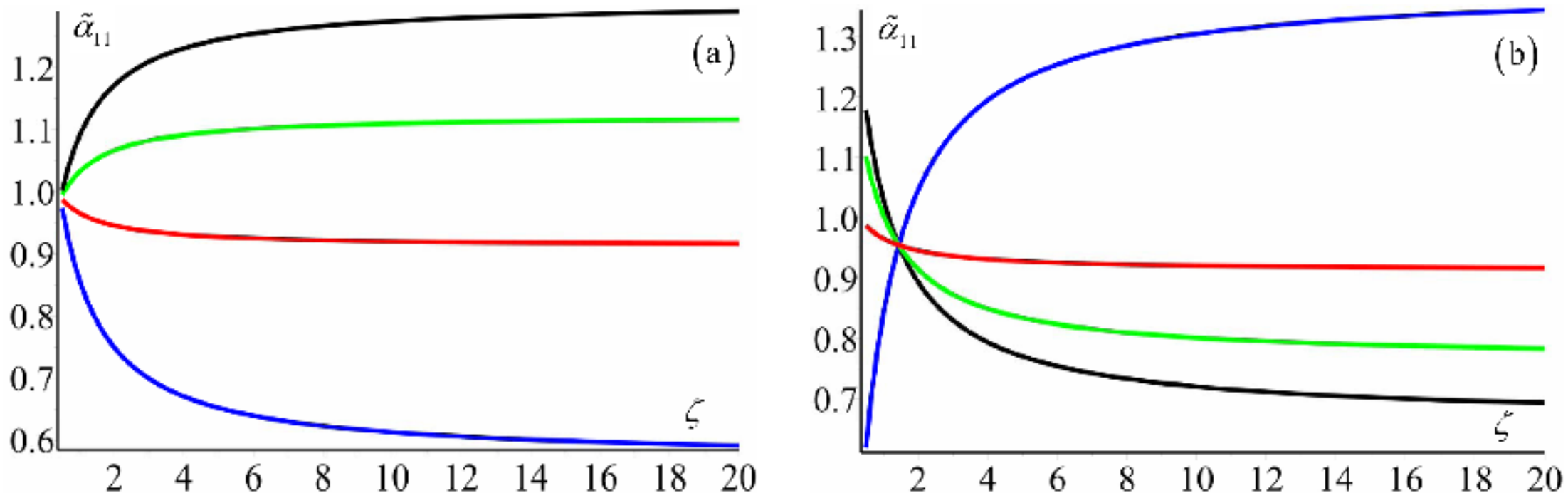

Figure 6: Dimensionless component $\tilde{\alpha}_{11}$ vs. the geometric ratio $\zeta$ obtained assuming $\nu_1 = 0.3, \nu_2 = 0.25, \nu_3 = 0.25, \rho_C^{\textcircled{2}} = 5/13, \rho_C^{\textcircled{3}} = 31/26, \hat{s}_1 = 5/88$ and: (a) $\rho_\alpha^{\textcircled{2}} = 25/26$ and different values of the ratio $\rho_\alpha^{\textcircled{3}}$: $\rho_\alpha^{\textcircled{3}} = 109/130$ red line, $\rho_\alpha^{\textcircled{3}} = 2/5$ blue line, $\rho_\alpha^{\textcircled{3}} = 6/5$ green line, $\rho_\alpha^{\textcircled{3}} = 8/5$ black line. (b) $\rho_\alpha^{\textcircled{3}} = 109/130$ and different values of the ratio $\rho_\alpha^{\textcircled{2}}$: $\rho_\alpha^{\textcircled{2}} = 25/26$ red line, $\rho_\alpha^{\textcircled{2}} = 1/2$ blue line, $\rho_\alpha^{\textcircled{2}} = 3/2$ green line, $\rho_\alpha^{\textcircled{2}} = 2$ black line.

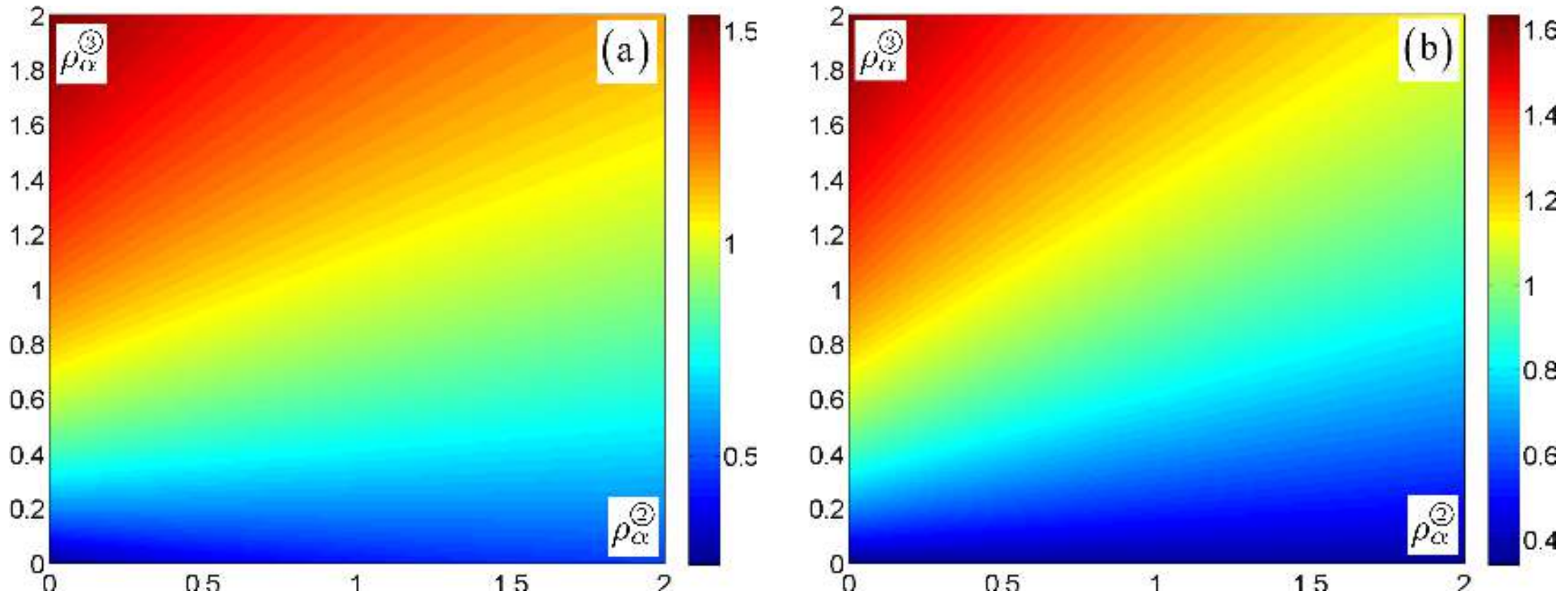

Figure 7: (a): variation of the dimensionless constant $\tilde{\alpha}_{11}$ with the ratios $\rho_{\alpha}^{②}$ and $\rho_{\alpha}^{③}$ obtained assuming $\nu_1 = 0.3, \nu_2 = 0.25, \nu_3 = 0.25,\ \hat{s}_1 = 5/88, \zeta = 14/5, \rho_{C}^{②} = 5/13, \rho_{C}^{③} = 31/26$. (b): variation of the dimensionless constant $\tilde{\alpha}_{22}$ with the ratios $\rho_{\alpha}^{②}$ and $\rho_{\alpha}^{③}$ obtained assuming $\nu_1 = 0.3, \nu_2 = 0.25, \nu_3 = 0.25,\ \hat{s}_1 = 5/88, \zeta = 14/5, \rho_{C}^{②} = 5/13, \rho_{C}^{③} = 31/26$.

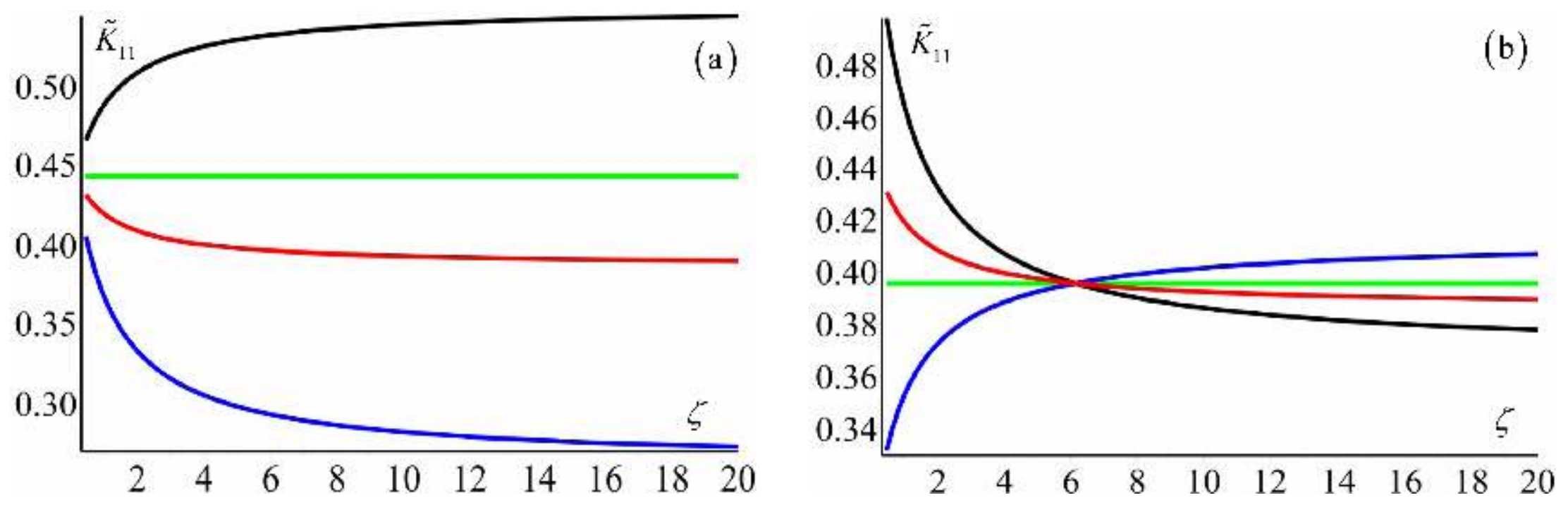

Figure 8: Dimensionless component $\tilde{K}_{11}$ vs. the geometric ratio $\zeta$ obtained assuming $\hat{s}_1 = 5/88$ and: (a) $\rho_{K}^{②} = 7/50$ and different values of the ratio $\rho_{K}^{③}$: $\rho_{K}^{③} = 53/480$ red line, $\rho_{K}^{③} = 1/20$ blue line, $\rho_{K}^{③} = 7/50$ green line, $\rho_{K}^{③} = 1/5$ black line. (b) $\rho_{K}^{③} = 53/480$ and different values of the ratio $\rho_{K}^{②}$: $\rho_{K}^{②} = 7/50$ red line, $\rho_{K}^{②} = 3/50$ blue line, $\rho_{K}^{②} = 53/480$ green line, $\rho_{K}^{②} = 1/5$ black line.

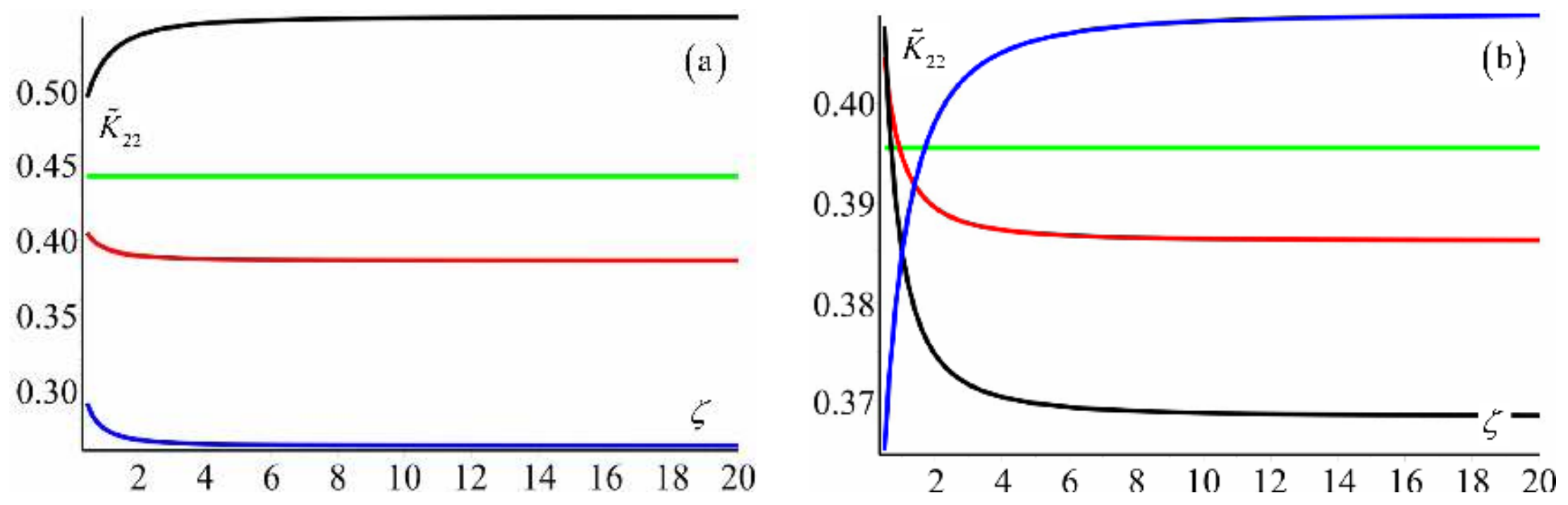

Figure 9: Dimensionless component $\tilde{K}_{22}$ vs. the geometric ratio $\zeta$ obtained assuming $\hat{s}_1 = 5/88$ and: (a) $\rho_K^{②} = 7/50$ and different values of the ratio $\rho_K^{③}$: $\rho_K^{③} = 53/480$ red line, $\rho_K^{③} = 1/20$ blue line, $\rho_K^{③} = 7/50$ green line, $\rho_K^{③} = 1/5$ black line. (b) $\rho_K^{③} = 53/480$ and different values of the ratio $\rho_K^{②}$: $\rho_K^{②} = 7/50$ red line, $\rho_K^{②} = 3/50$ blue line, $\rho_K^{②} = 53/480$ green line, $\rho_K^{②} = 1/5$ black line.

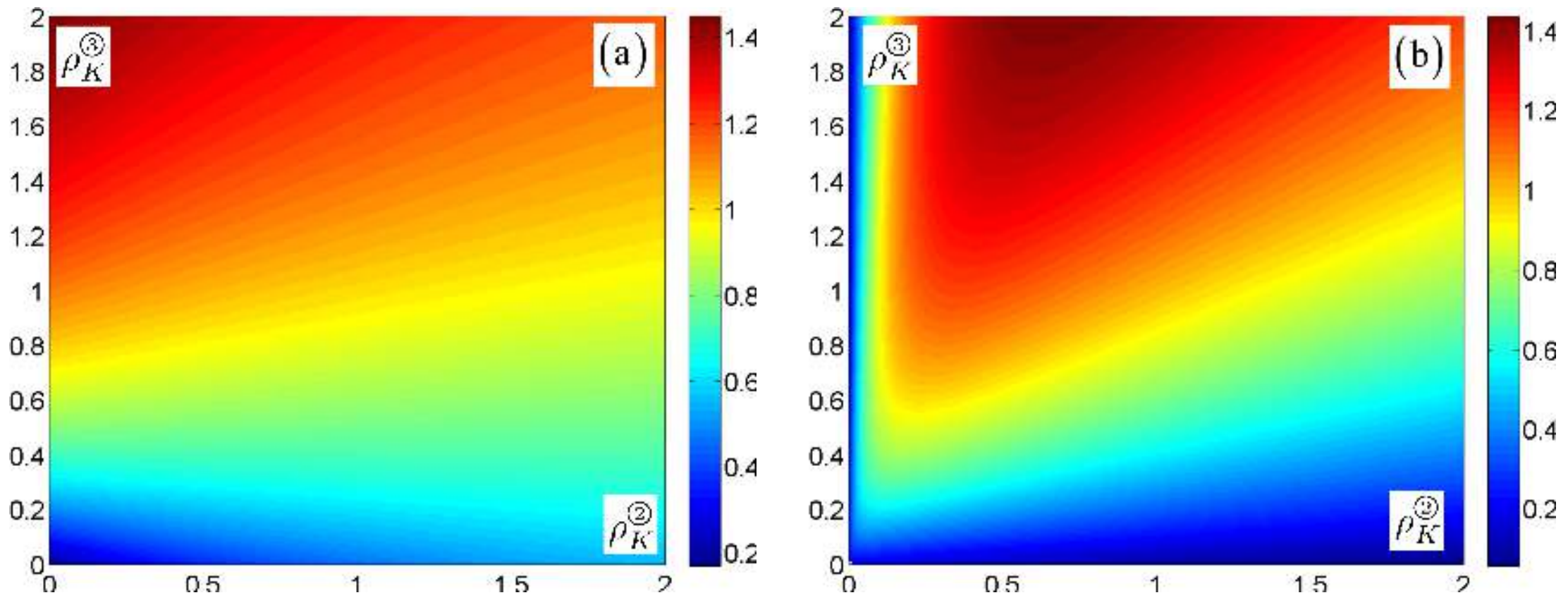

Figure 10: (a): variation of the dimensionless constant $\tilde{K}_{11}$ with the ratios $\rho_K^{②}$ and $\rho_K^{③}$ obtained assuming $\hat{s}_1 = 5/88, \zeta = 14/5$. (b): variation of the dimensionless constant $\tilde{K}_{22}$ with the ratios $\rho_K^{②}$ and $\rho_K^{③}$ obtained assuming $\hat{s}_1 = 5/88, \zeta = 14/5$.

## 4.2 Comparative analysis: homogenized model vs heterogeneous material

The two-dimensional three-phase layered material shown in Fig. 3 is assumed to be subjected to $\mathcal{L}$-periodic heat sources $r(x_j)$ (see Fig. 11). Two different distributions for these sources are introduced:

$$r(x_j) = (-1)^n R \cos\left(\frac{2\pi n x_j}{L_j}\right) e^{-[\beta(\frac{x_j}{L_j}-\frac{1}{2})]^2} + A + B\left(\frac{x_j}{L_j} - \frac{1}{2}\right)^2, \tag{30}$$

$$r^{(0)}(x_j) = (-1)^n R \cos\left(\frac{2\pi n x_j}{L_j}\right), \tag{31}$$

with $n, \beta \in \mathbb{Z}^+ - \{0\}$, constants $A$ and $B$ are reported in the Appendix C. The body forces are assumed to vanish: $b_j(x_j) = 0$.

The homogenized field equations of the first order continuum are solved analytically considering the overall thermoelastic constants (26), (27) and (28). The obtained results are then compared with those derived by means of a fully heterogeneous modelling procedure. Due to the periodicity of the heterogeneous material heat sources, only an horizontal (or vertical) characteristic portion of length $L$ of the heterogeneous model is analyzed (Fig. 3/(b)). In order to assess the reliability of the homogenized model, the macroscopic displacement, temperature fields are compared to the corresponding fields in the heterogeneous model by means of the up-scaling relations (see Bacigalupo et al. (2016)). The solution of the heterogeneous problem with $\mathcal{L}-$periodic heat sources is computed via FE analysis with periodic boundary conditions on the displacement and temperature fields. For the considered two-dimensional body subject to heat sources along the orthotropy axes, the homogenized field equations (20) and (21) take the form:

$$C_{jjjj} \frac{\partial^2 U_j}{\partial x_j^2} - \alpha_{jj} \frac{\partial \Theta}{\partial x_j} = 0, \tag{32}$$

$$K_{jj} \frac{\partial^2 \Theta}{\partial x_j^2} + r = 0, \tag{33}$$

where: $j = 1, 2,\ j \neq p$ are not summed indexes. The following conditions are imposed on the periodic domain $\mathcal{L}$ reported in Fig. 1:

$$U_j(x_j = 0) = U_j(x_j = L_j), \quad \Theta(x_j = 0) = \Theta(x_j = L_j), \tag{34}$$

$$C_{jjjj} \left.\frac{\partial U_j}{\partial x_j}\right|_{x_j=0} - \alpha_{ij}\Theta(x_j = 0) = C_{jjjj} \left.\frac{\partial U_j}{\partial x_j}\right|_{x_j=L_j} - \alpha_{ij}\Theta(x_j = L_j), \tag{35}$$

$$-K_{jj} \left.\frac{\partial \Theta}{\partial x_j}\right|_{x_j=0} = -K_{jj} \left.\frac{\partial \Theta}{\partial x_j}\right|_{x_j=L_j}. \tag{36}$$

Considering heat sources $r(x_j)$ of the form (30), the macroscopic displacements and temperature fields are given by

$$U_j(x_j) = \Omega_0(x_j) + \Omega_1(x_j)x_j + \Omega_2(x_j)x_j^2 + \Omega_3 x_j^3 + \Omega_4 x_j^4 + \Omega_5 x_j^5, \tag{37}$$

$$\Theta(x_j) = \Lambda_0(x_j) + \Lambda_1(x_j)x_j + \Lambda_2 x_j^2 + \Lambda_3 x_j^3 + \Lambda_4 x_j^4, \tag{38}$$

where explicit expressions for functions $\Omega_j$ and $\Lambda_j$ are reported in Appendix C. The non-vanishing components of the macroscopic stress fields and heat flux are given by

$$\Sigma_{jj} = C_{jjjj}E_{jj} - \alpha_{jj}\Theta, \quad \Sigma_{pp} = C_{ppjj}E_{jj} - \alpha_{pp}\Theta, \tag{39}$$

where $E_{jj} = \partial U_j / \partial x_j$ and

$$Q_j = -K_{jj} \frac{\partial \Theta}{\partial x_j}, \tag{40}$$

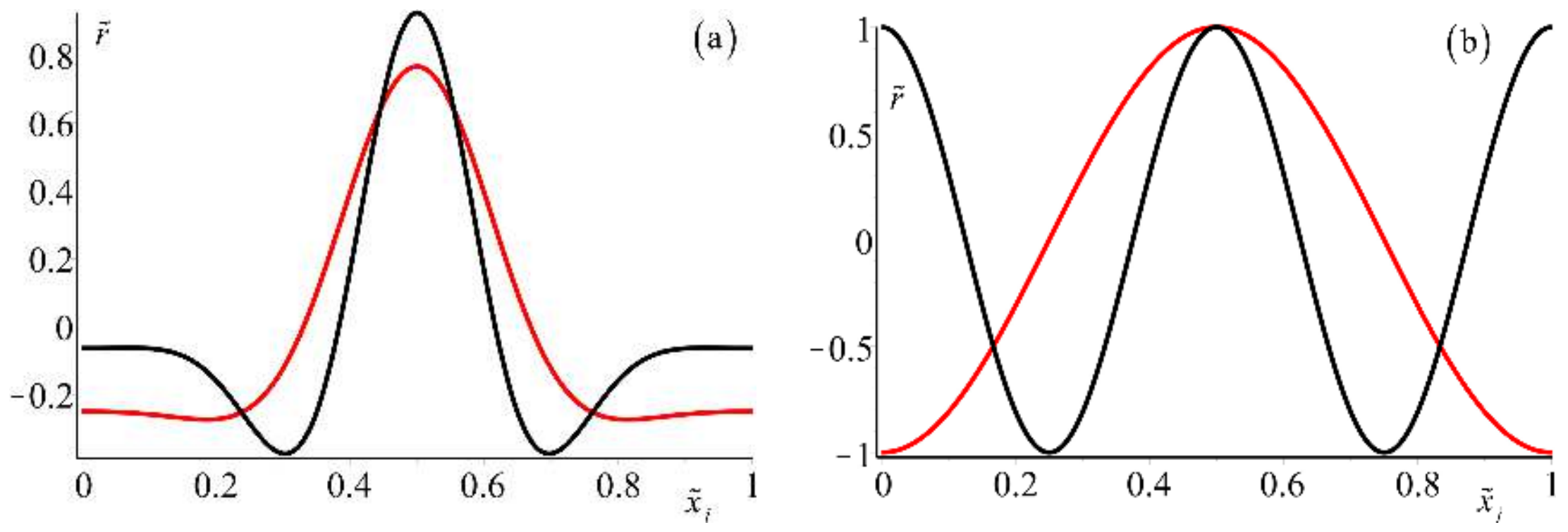

Figure 11: (a): Variation of the localized heat source $\tilde{r}$ with the normalized space variable $\tilde{x_j}$ with $j = 1, 2$ reported for $n = 1, \beta = 5$ (red line) and $n = 2, \beta = 5$ (black line); (b): Variation of the not localized heat source $\tilde{r}$ with the normalized space variable $\tilde{x_j}$ with $j = 1, 2$ reported for $n = 1, \beta = 5$ (red line) and $n = 2, \beta = 5$ (black line).

where $j = 1, 2$, $p = 1, 2$ and $j \neq p$ are not summed indexes. For $j = 1$ and $p = 2$ the source $r$ varies along $x_1-$direction, whereas for $j = 2$ and $p = 1$ $r$ depends on $x_2$.

In the case where the heat source $r_0(x_j)$ assumes the form (31), the solution becomes:

$$U_j^{(0)}(x_j) = (-1)^n \frac{RL_j^2}{4K_{jj}\pi^2 n^2} \cos\left(\frac{2\pi n x_j}{L_j}\right), \tag{41}$$

$$\Theta^{(0)} = (-1)^n \frac{\alpha_{jj} RL_j^3}{8K_{jj}\pi^3 n^3 C_{jjjj}} \sin\left(\frac{2\pi n x_j}{L_j}\right), \tag{42}$$

and $\Sigma_{jj}^0$ and $Q_j^0$ are determined by means of the relations (39) and (40). Introducing $\tilde{x}_j = x_j/L_j$ and $\tilde{r} = r/R$,

$$\tilde{U}_j = \frac{U_j K_{jj} C_{jjjj}}{L_j^3 R \alpha_{jj}}, \qquad \tilde{\Theta} = \frac{\Theta K_{jj}}{L_j^2 R}, \tag{43}$$

$$\tilde{\Sigma}_{jj} = \frac{\partial \tilde{U}_j}{\partial \tilde{x}_j} - \tilde{\Theta}, \qquad \tilde{\Sigma}_{pp} = \frac{C_{ppjj}}{C_{jjjj}} \frac{\partial \tilde{U}_j}{\partial \tilde{x}_j} - \frac{\alpha_{pp}}{\alpha_{jj}} \tilde{\Theta}, \tag{44}$$

where $\tilde{\Sigma}_{jj} = \dfrac{\Sigma_{jj} K_{jj}}{R\alpha_{jj} L_j^2}$ and $\tilde{\Sigma}_{pp} = \dfrac{\Sigma_{pp} K_{jj}}{R\alpha_{jj} L_j^2}$, and $j = 1, 2$ and $p = 1, 2$ are not summed indexes. For $j = 1$ and $p = 2$ the source $r$ varies along $x_1-$direction, whereas for $j = 2$ and $p = 1$, $r$ depends on $x_2$.

$$\tilde{Q}_j = \frac{Q_j}{L_j R}, \tag{45}$$

where $\tilde{Q}_j = -\dfrac{\partial \tilde{\Theta}}{\partial \tilde{x}_j}$ and $j = 1, 2$ is a not summed index.

The analytical solutions (37), (38), (41) and (42) derived by the sotution of the homogenized fields equations (32) and (33), is now compared with the results obtained by the finite element analysis of the heterogeneous problem corresponding to the tri-phase layered material reported in Fig. 3 subject to localized and not localized heat sources which profiles are shown in Fig.11. More precisely, finite element analysis of the heterogeneous problem, performed by means of the program COMSOL Multiphysics, provides the local fields $u_j$, $\theta$ which are used together with the *up-scaling* relations (Bacigalupo et al., 2016) for obtaining the macro-scopic fields $U_j$ and $\Theta$. These macroscopic quantities are compared with the analytical expressions (37), (38), (41) and (42). Plane strain condition has been assumed for both the solution of the homogenized equations and the heterogeneous problem, and the same values for the Poisson's coefficients and geometrical ratio $\hat{s}_1$ introduced in Section 4.1 are assumed.

In Figs. 12 and 13, the normalized macroscopic temperature field $\tilde{\Theta}$ and displacement component $\tilde{U}_1$ evaluated using analytical expressions (38), (42) (37) and (41) and considering localized and not localized heat sources varying along $x_1-$direction are reported as functions of the normalized spatial coordinate $\tilde{x}_1 = x_1/L_1$ (continuous lines in the figure) and compared with the numerical results obtained by the heterogeneous model (diamonds in the figure). The following values for the geometrical parameters, the ratios between the elastic and of thermodiffusive constants have been assumed: $L/\varepsilon = 10$, $\zeta = 7/5$ $\rho_C^{②} = 5/13$, $\rho_C^{③} = 31/26$, $\rho_\alpha^{②} = 7/52$, $\rho_\alpha^{③} = 5777/62400$, $\rho_K^{②} = 7/50$, $\rho_K^{③} = 53/480$. The macroscopic displacement and temperature fields are plotted for the characteristic portion of length $L_1 = L$, corresponding to $x_1/L = 1$ (i. e. for $0 \leq x_1/L \leq 1$), and the values for the wave number $n = 1, 2$ and $\beta = 5$ have been considered for defining the heat sources. Observing the curves, for both the quantities $\tilde{\Theta}(x_1/L)$ and $\tilde{U}_1(x_1/L)$, a good agreement is detected between the results derived by means of the first order homogenization approach and those obtained by the heterogeneous model.

The variation of the normalized macroscopic temperature field $\tilde{\Theta}$ and displacement component $\tilde{U}_2$ evaluated using analytical expressions (38), (42) (37) and (41) and considering localized and not localized heat sources varying along $x_2-$direction is reported in Figs. 14 and 15 in terms of the normalized spatial coordinate $\tilde{x}_2 = x_2/L_2$ (continuous lines in the figure) and compared with the numerical results obtained by the heterogeneous model (diamonds in the figure). Similarly to the previous case, the macroscopic displacement and temperature fields are plotted for the characteristic portion of length $L_2 = L$, corresponding to $x_2/L = 1$ (i. e. for $0 \leq x_2/L \leq 1$), and the values for the wave number $n = 1, 2$ and $\beta = 5$ have been considered for defining the heat sources. Observing the curves, for both the quantities $\tilde{\Theta}(x_2/L)$ and $\tilde{U}_2(x_2/L)$, a good agreement is detected between the results derived by the solution of the homogenized field equations and those obtained by the heterogeneous model.

In Figs. 16 and 17 the normalized components of the microscopic heat fluxes $\tilde{q}_1$ and $\tilde{q}_2$ induced respectively by localized (for $n = 1$ and $\beta = 5$) and not localized (for $n = 1$) heat sources varying along $x_1-$direction are plotted as functions of $x_1$ and $x_2$. The following intervals for the variables have been considered for these plots: $2\varepsilon < x_1 < 3\varepsilon$, $0 < x_2 < \varepsilon$. The components of the microscopic stress fields $\tilde{\sigma}_{11}$, $\tilde{\sigma}_{12}$, $\tilde{\sigma}_{22}$, generated by localized and non localized thermal sources varying along $x_1-$direction are reported as functions of $x_1$ and $x_2$ in Figs. 18, 19 and 20. The same ranges of values considered for the heat fluxes have been assumed for $x_1$ and $x_2$. In Figs. 21 and 22 the component $\tilde{q}_2$ of the microscopic heat flux and the component $\tilde{\sigma}_{22}$ of the microscopic stress fields due to localized and non localized heat sources varying along $x_2-$direction are reported as functions of $x_1$ and $x_2$. The following intervals for the variables have been considered for these plots: $-\varepsilon < x_1 < 0$, $2\varepsilon < x_2 < 3\varepsilon$.

The microscopic heat flux and stress fields illustrated in Figs. 16-22 have been evaluated by means of the down-scaling relations reported in Appendix E in terms of perturbation functions, microscopic thermoelastic constants and macroscopic fields.

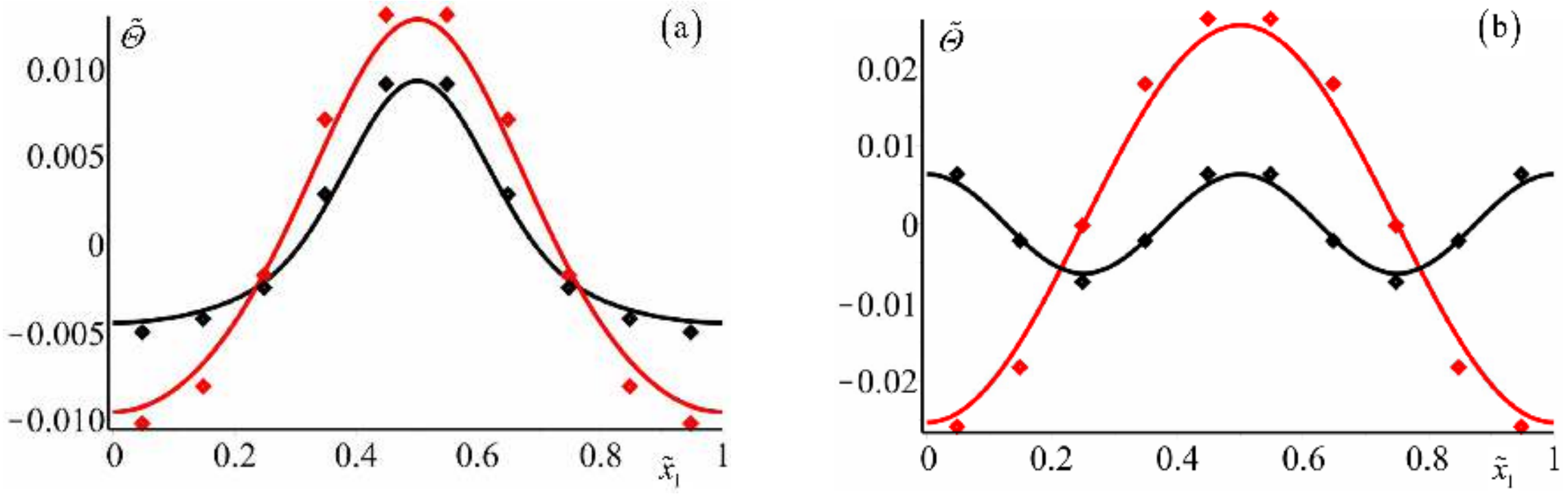

Figure 12: (a): Normalized macroscopic temperature field $\tilde{\Theta}$ due to the localized thermal source $r(x_1)$ vs. $\tilde{x}_1$ plotted for different values of wave number $n = 1, \beta = 5$ (red line), $n = 2, \beta = 5$ (black line). (b): Normalized macroscopic temperature field $\tilde{\Theta}$ due to the not localized thermal source $r(x_1)$ vs. $\tilde{x}_1$ plotted for different values of wave number $n = 1$ (red line), $n = 2$ (black line). The heterogeneous model (diamonds) is compared with the homogenized first order model.

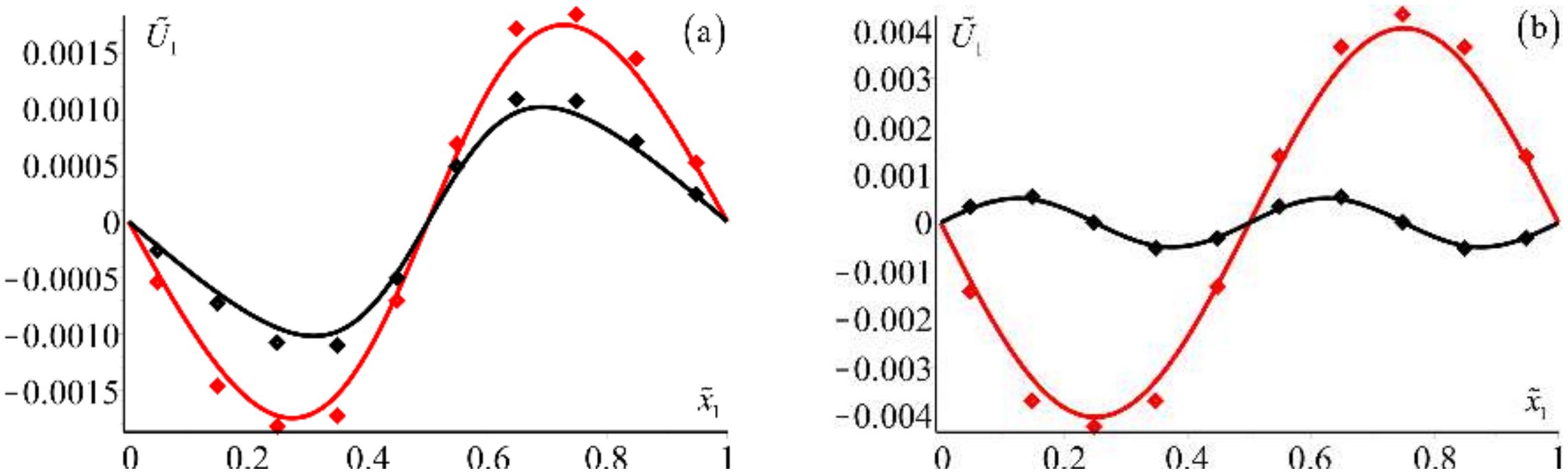

Figure 13: (a): Normalized macroscopic displacement field $\tilde{U}_1$ due to the localized thermal source $r(x_1)$ vs. $\tilde{x}_1$ plotted for different values of wave number $n = 1, \beta = 5$ (red line), $n = 2, \beta = 5$ (black line). (b): Normalized macroscopic displacement field $\tilde{U}_1$ due to the not localized thermal source $r(x_1)$ vs. $\tilde{x}_1$ plotted for different values of wave number $n = 1$ (red line), $n = 2$ (black line). The heterogeneous model (diamonds) is compared with the homogenized first order model.

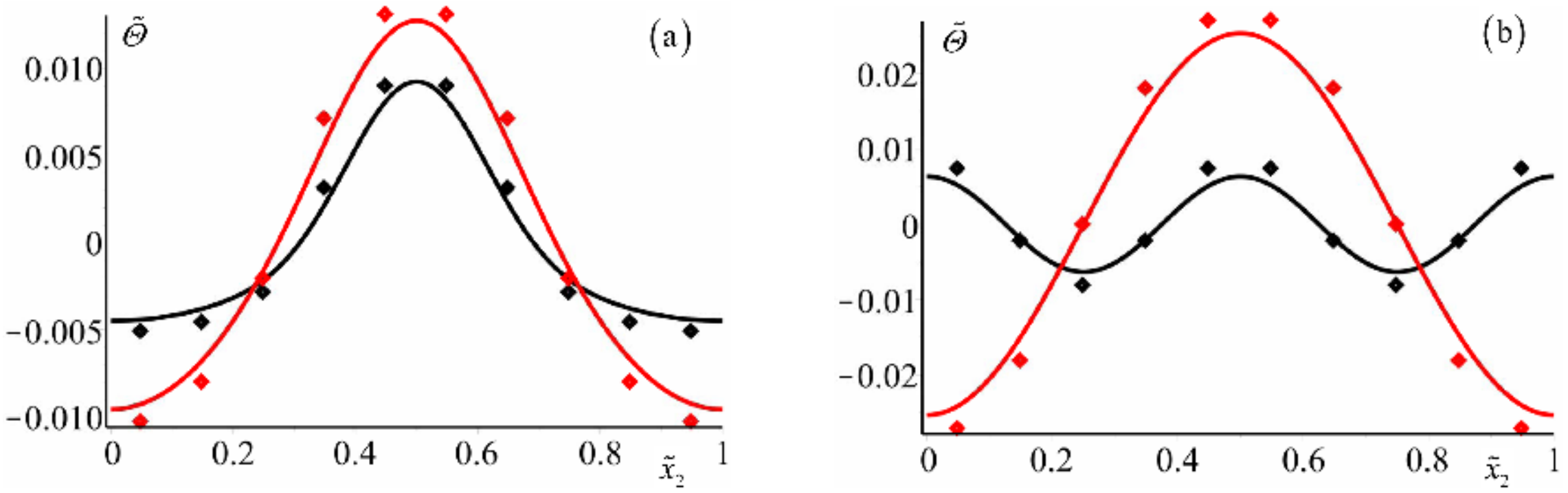

Figure 14: (a): Normalized macroscopic temperature field $\tilde{\Theta}$ due to the localized thermal source $r(x_2)$ vs. $\tilde{x}_2$ plotted for different values of wave number $n = 1, \beta = 5$ (red line), $n = 2, \beta = 5$ (black line). (b): Normalized macroscopic temperature field $\tilde{\Theta}$ due to the not localized thermal source $r(x_2)$ vs. $\tilde{x}_2$ plotted for different values of wave number $n = 1$ (red line), $n = 2$ (black line). The heterogeneous model (diamonds) is compared with the homogenized first order model.

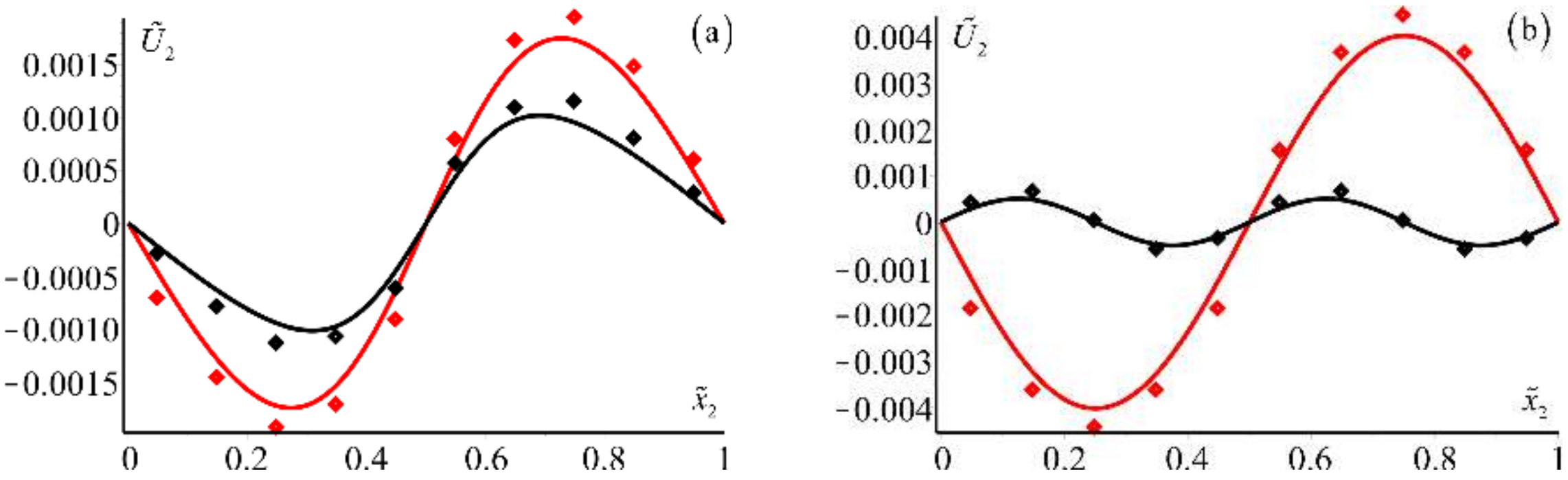

Figure 15: (a): Normalized macroscopic displacement field $\tilde{U}_2$ due to the localized thermal source $r(x_2)$ vs. $\tilde{x}_2$ plotted for different values of wave number $n = 1, \beta = 5$ (red line), $n = 2, \beta = 5$ (black line). (b): Normalized macroscopic displacement field $\tilde{U}_1$ due to the not localized thermal source $r(x_2)$ vs. $\tilde{x}_2$ plotted for different values of wave number $n = 1$ (red line), $n = 2$ (black line). The heterogeneous model (diamonds) is compared with the homogenized first order model.

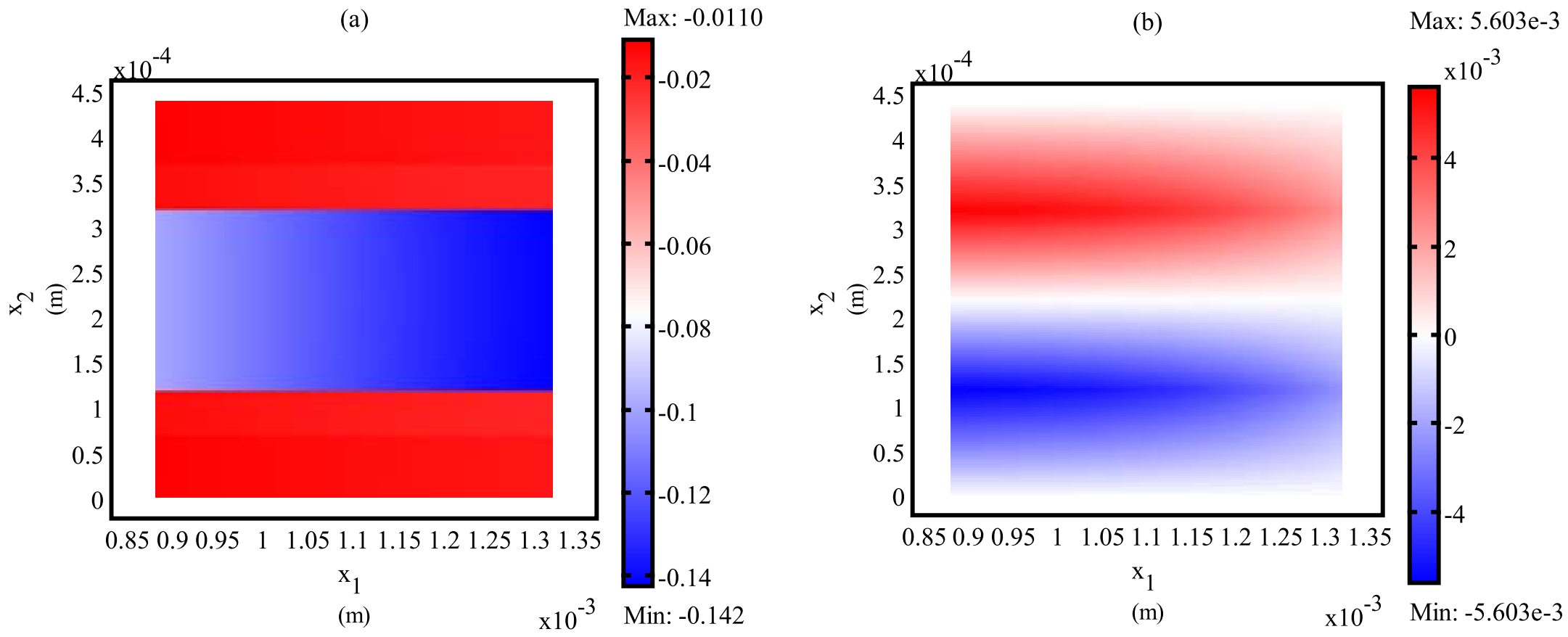

Figure 16: (a): Variation of the normalized component of the microscopic heat flux $\tilde{q}_1$ due to the localized thermal source $r(x_1)$ with $x_1$ and $x_2$ ($2\varepsilon < x_1 < 3\varepsilon$ and $0 < x_2 < \varepsilon$) plotted for $n = 1, \beta = 5$. (b): Variation of the normalized component of the microscopic heat flux $\tilde{q}_2$ due to the localized thermal source $r(x_1)$ with $x_1$ and $x_2$ ($2\varepsilon < x_1 < 3\varepsilon$ and $0 < x_2 < \varepsilon$) plotted for $n = 1, \beta = 5$.

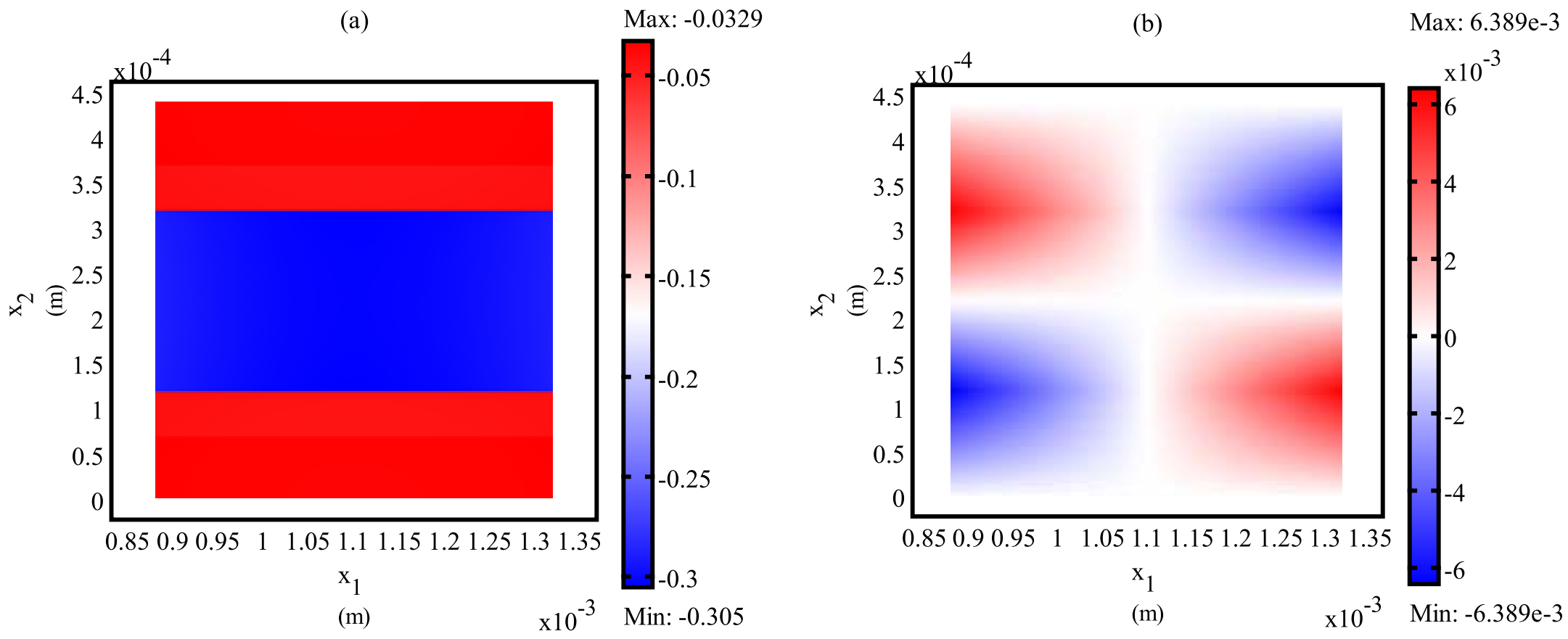

Figure 17: (a): Variation of the normalized component of the microscopic heat flux $\tilde{q}_1$ due to the not localized thermal source $r(x_1)$ with $x_1$ and $x_2$ ($2\varepsilon < x_1 < 3\varepsilon$ and $0 < x_2 < \varepsilon$) plotted for $n = 1$. (b): Variation of the normalized component of the microscopic heat flux $\tilde{q}_2$ due to the not localized thermal source $r(x_1)$ with $x_1$ and $x_2$ ($2\varepsilon < x_1 < 3\varepsilon$ and $0 < x_2 < \varepsilon$) plotted for $n = 1$.

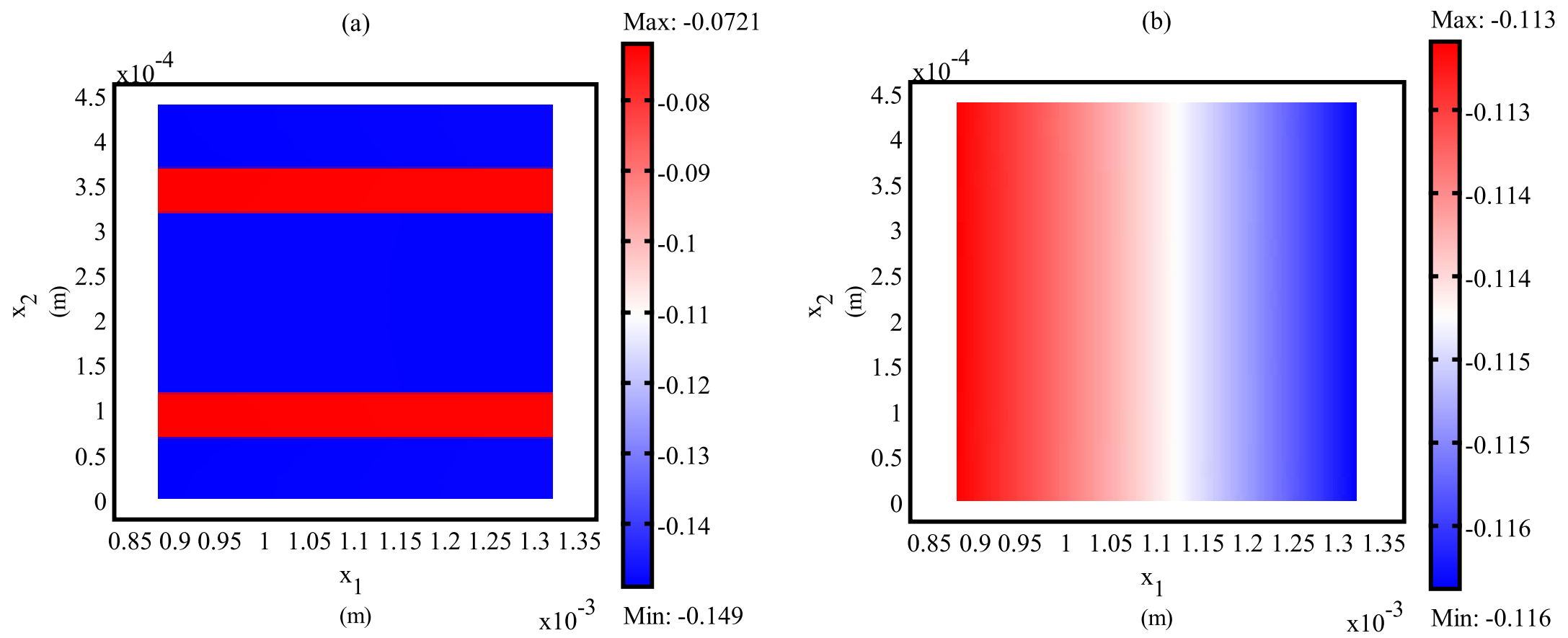

Figure 18: (a): Variation of the normalized component of the microscopic stresses $\tilde{\sigma}_{11}$ due to the localized thermal source $r(x_1)$ with $x_1$ and $x_2$ ($2\varepsilon < x_1 < 3\varepsilon$ and $0 < x_2 < \varepsilon$) plotted for $n = 1, \beta = 5$. (b): Variation of the normalized component of the microscopic stresses $\tilde{\sigma}_{22}$ due to the localized thermal source $r(x_1)$ with $x_1$ and $x_2$ ($2\varepsilon < x_1 < 3\varepsilon$ and $0 < x_2 < \varepsilon$) plotted for $n = 1, \beta = 5$.

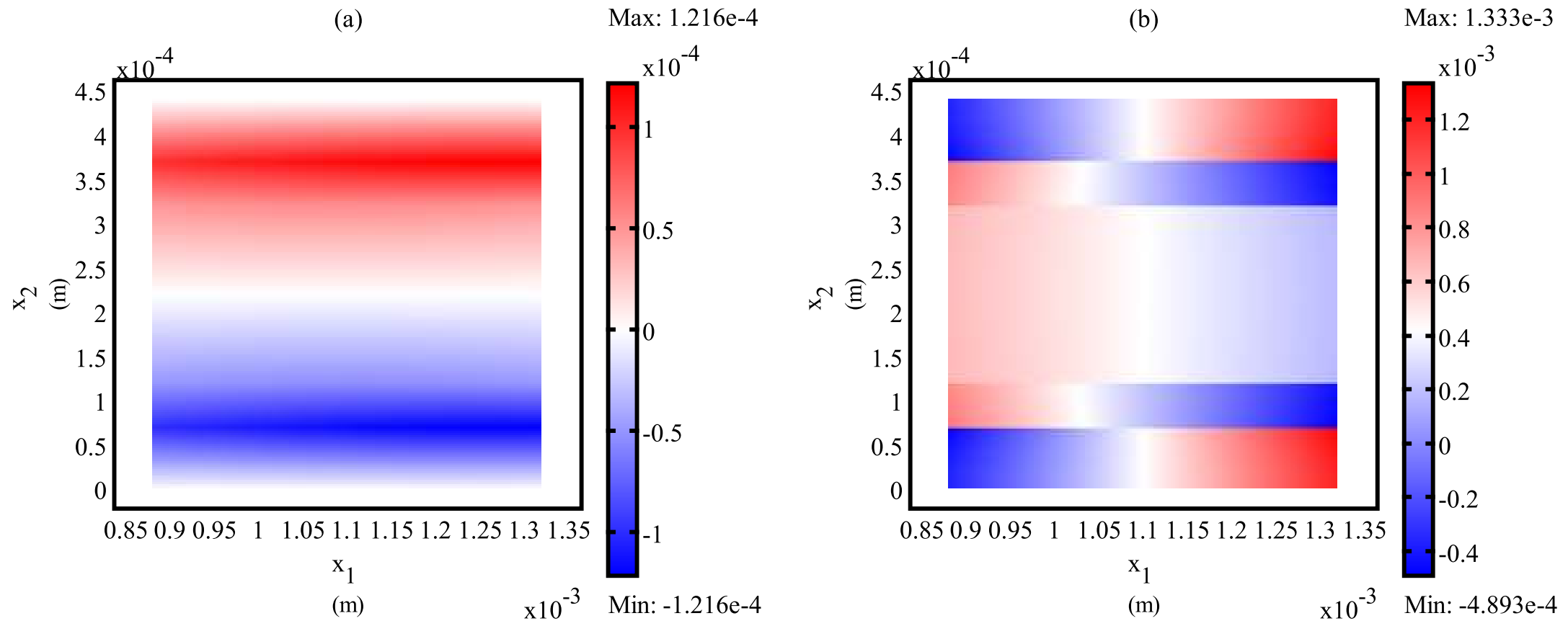

Figure 19: (a): Variation of the normalized component of the microscopic stresses $\tilde{\sigma}_{12}$ due to the localized thermal source $r(x_1)$ with $x_1$ and $x_2$ ($2\varepsilon < x_1 < 3\varepsilon$ and $0 < x_2 < \varepsilon$) plotted for $n = 1, \beta = 5$. (b): Variation of the normalized component of the microscopic stresses $\tilde{\sigma}_{11}$ due to the not localized thermal source $r(x_1)$ with $x_1$ and $x_2$ ($2\varepsilon < x_1 < 3\varepsilon$ and $0 < x_2 < \varepsilon$) plotted for $n = 1$.

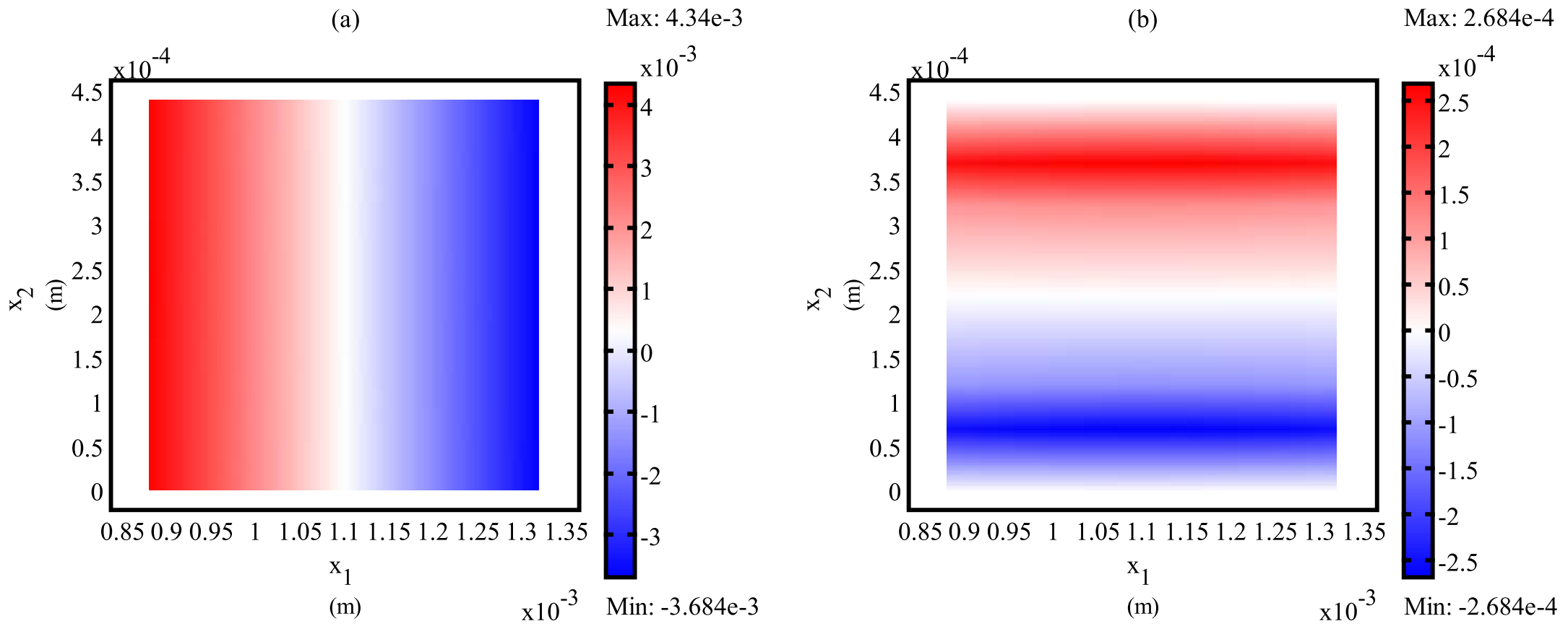

Figure 20: (a): Variation of the normalized component of the microscopic stresses $\tilde{\sigma}_{22}$ due to the not localized thermal source $r(x_1)$ with $x_1$ and $x_2$ ($2\varepsilon < x_1 < 3\varepsilon$ and $0 < x_2 < \varepsilon$) plotted for $n = 1$.. (b): Variation of the normalized component of the microscopic stresses $\tilde{\sigma}_{12}$ due to the not localized thermal source $r(x_1)$ with $x_1$ and $x_2$ ($2\varepsilon < x_1 < 3\varepsilon$ and $0 < x_2 < \varepsilon$) plotted for $n = 1$.

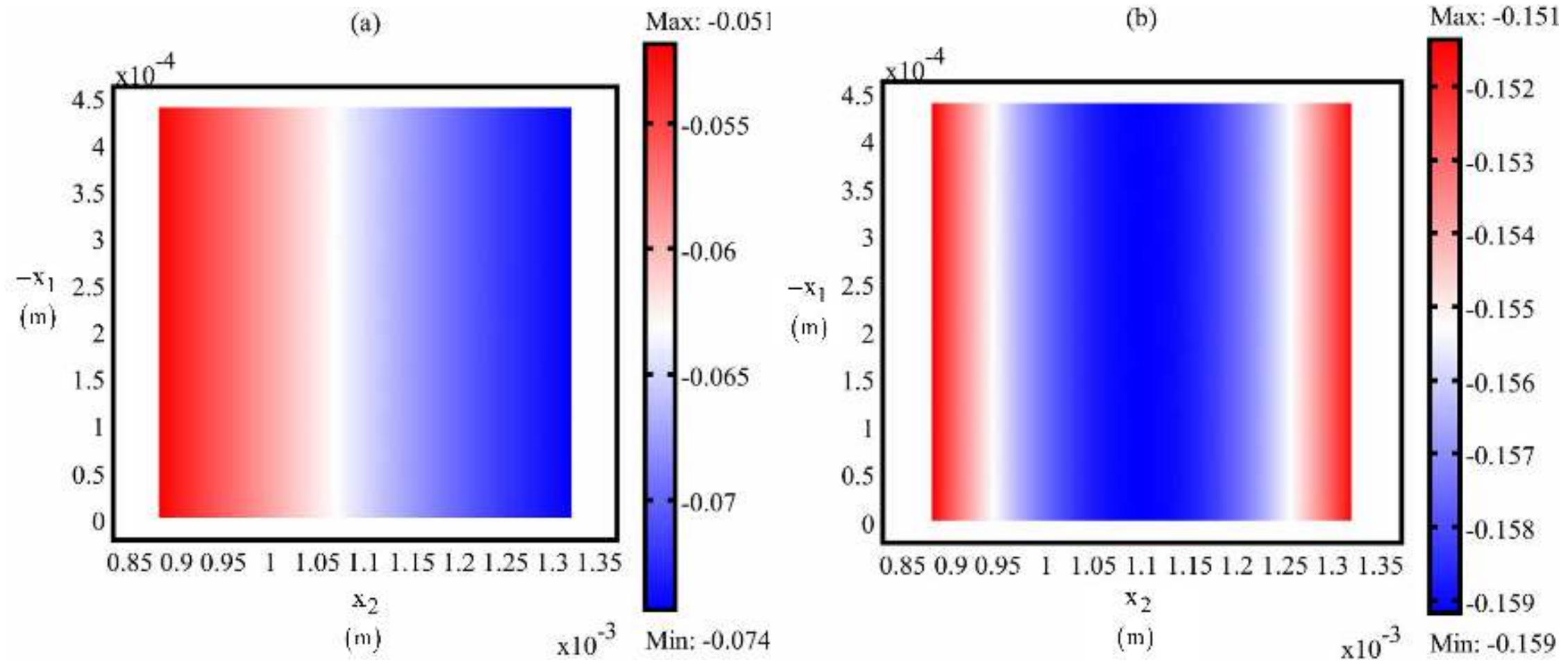

Figure 21: (a): Variation of the normalized component of the microscopic heat flux $\tilde{q}_2$ due to the localized thermal source $r(x_2)$ with $x_1$ and $x_2$ ($-\varepsilon < x_1 < 0$ and $2\varepsilon < x_2 < 3\varepsilon$) plotted for $n = 1, \beta = 5$. (b): Variation of the normalized component of the microscopic stresses $\tilde{q}_2$ due to the not localized thermal source $r(x_2)$ with $x_1$ and $x_2$ ($-\varepsilon < x_1 < 0$ and $2\varepsilon < x_2 < 3\varepsilon$) plotted for $n = 1$.

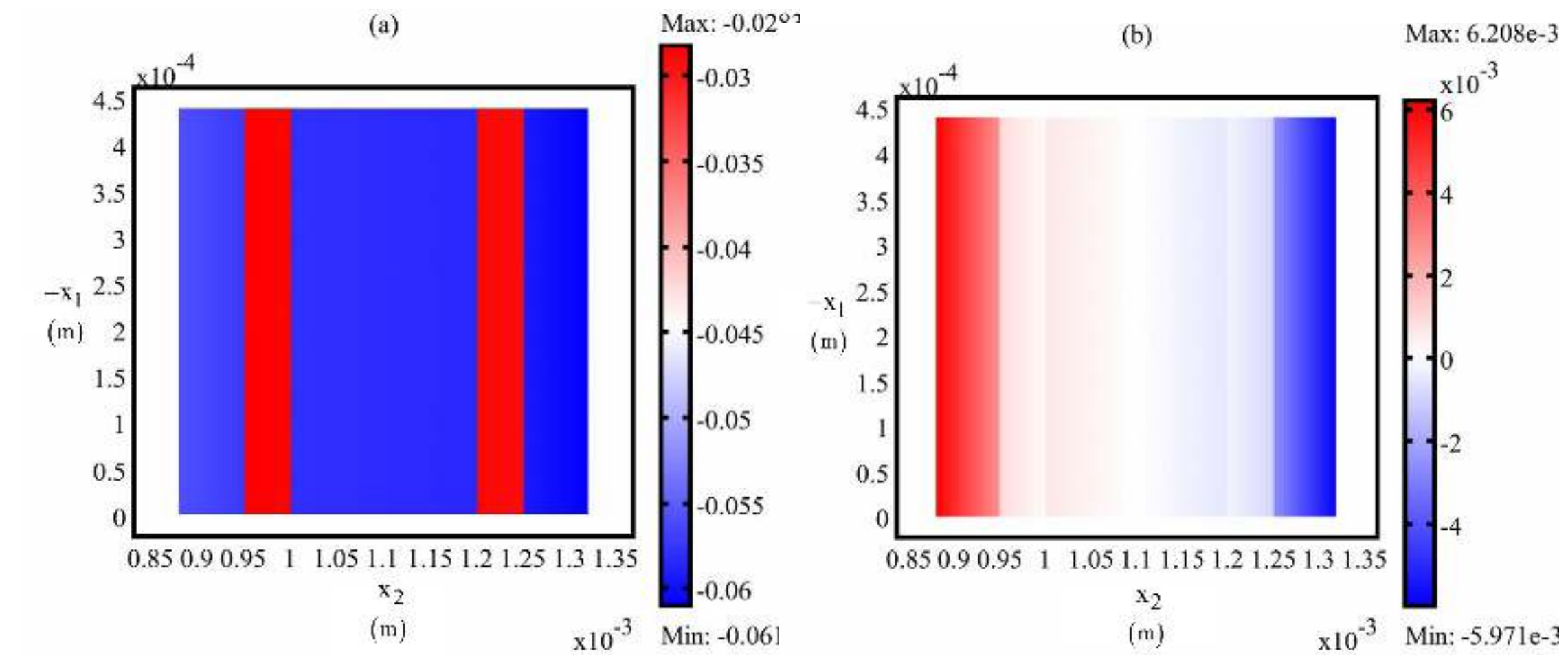

Figure 22: (a): Variation of the normalized component of the microscopic stresses $\tilde{\sigma}_{22}$ due to the localized thermal source $r(x_2)$ with $x_1$ and $x_2$ ($-\varepsilon < x_1 < 0$ and $2\varepsilon < x_2 < 3\varepsilon$) plotted for $n = 1, \beta = 5$. (b): Variation of the normalized component of the microscopic stresses $\tilde{\sigma}_{22}$ due to the not localized thermal source $r(x_2)$ with $x_1$ and $x_2$ ($-\varepsilon < x_1 < 0$ and $2\varepsilon < x_2 < 3\varepsilon$) plotted for $n = 1$.

# 5 Conclusions

Exact expressions for the components of the elastic, thermoelastic and heat conduction tensors of first order thermoelastic continuum equivalent to multi-layered battery devices have been derived by means of a general asymptotic homogenization procedure. An ideal periodic multi-layered thermoelastic composite material reproducing the planar geometry of an idealized battery device is introduced. *Down-scaling* relations associating the microscopic displacements and temperature fields to the corresponding macroscopic fields are introduced. Fluctuations functions representing the effects of the microstructures on the microscopic displacements and temperature are defined. These fluctuations functions are obtained through the solution of non-homogeneous problems on the cell defining periodic boundary conditions and normalization conditions (*up-scaling* relations). Fields equation for the homogenized thermoelastic Cauchy material are derived, and exact expressions for the overall elastic and thermodiffusive constants of the first order continuum equivalent to the periodic battery-like composite medium are obtained.

The developed general procedure is used to determine analytically the components of the overall elastic, thermoelastic and heat conduction tensors corresponding to a three-phase layered thermoelastic composite of interests for SOFCs devices fabrication. The fields equation of the first order equivalent thermoelastic media are solved considering periodic heat sources, which localized and unlocalized profiles are representative for modelling some thermal effects detected in real situations. The solution of the homogenized field equations is compared with the numerical results obtained by finite elements analysis of heterogeneous model, performed assuming periodic body force and heat sources acting on the considered three-phase layered composite. The good agreement detected between the analytical solution of the homogenized first order equations and the numerical results obtained by the heterogeneous model represents an important validation of the accuracy of the proposed asymptotic homogenization approach. The small discrepancy detected between the analytical solution of the homogenized equations and the results provided by finite element analysis of the heterogenoeus material can be further reduced considering higher order approximations of the field equations of infinite order (Bacigalupo et al., 2016), or alternatively introducing non-local elastic constitutive relations (Bacigalupo and Gambarotta, 2010, 2011, 2012).

The general method illustrated in the paper provides a synthetic description of the size effects on thermomechanical properties of both lithium ions batteries and solid oxide fuel cells devices (SOFCs), avoiding the numerical analysis of heterogeneous materials and the connected challenging computational problems. The estimation of these effective thermoelastic properties can be used in order to optimize the structural performances of the fuel cells in operative scenarios where, due to the high operational temperatures (800-1000 degrees), the components of these devices are subject to severe thermomechanical stresses which can cause damage and crack formation compromising their performances. Consequently, evaluating the overall thermoelastic properties of these battery devices through the asymptotic homogenization approach illustrated in the paper can represent an important issue in order to predict damaging phenomena and to improve the efficient design and manufacturing of these systems. The asymptotic homogenization model proposed in this work can be generalized in order to study particular interface properties which commonly characterize SOFC devices. For modelling this cases, additional thin layers with specific thermo-mechanical properties reflecting the interface characteristics can be introduced. This can be done by conveniently relaxing the interface conditions of the cell problems, which define the fluctuation functions at different orders.

# Acknowledgements

A.B. and L.M. gratefully acknowledge financial support from the Italian Ministry of Education, University and Research in the framework of the FIRB project 2010 "Structural mechanics models for renewable energy applications". AP would like to acknowledge financial support from the European Unionâs Seventh Framework Programme FP7/2007-2013/ under REA grant agreement number PCIG13-GA-2013- 618375-MeMic.

# A Explicit coefficients involved in perturbation functions

In this Appendix, the explicit expressions for coefficients $A_{ijk}^{(1),\textcircled{p}}$, $B_{ijk}^{(1),\textcircled{p}}$, $\tilde{A}_{i}^{(1),\textcircled{p}}$, $\tilde{B}_{i}^{(1),\textcircled{p}}$, $P_{i}^{(1),\textcircled{p}}$ and $Q_{i}^{(1),\textcircled{p}}$ involved in the perturbation functions (23), (24) and (25) are reported.

The coefficients $A_{ijk}^{(1),\textcircled{p}}$ and $B_{ijk}^{(1),\textcircled{p}}$, associated to the perturbation functions (23), assume the form:

$$
\begin{aligned}
A_{211}^{(1),\textcircled{1}} &= -\frac{C^{\textcircled{1}}_{1122}\,C^{\textcircled{2}}_{2222}\,\hat{s}_3 + C^{\textcircled{1}}_{1122}\,C^{\textcircled{3}}_{2222}\,\hat{s}_2 - C^{\textcircled{2}}_{1122}\,C^{\textcircled{3}}_{2222}\,\hat{s}_2 - C^{\textcircled{3}}_{1122}\,C^{\textcircled{2}}_{2222}\,\hat{s}_3}{C^{\textcircled{1}}_{2222}\left(C^{\textcircled{2}}_{2222}\,\hat{s}_3 + C^{\textcircled{3}}_{2222}\,\hat{s}_2\right)};\\
A_{211}^{(1),\textcircled{2}} &= -\frac{\left(C^{\textcircled{2}}_{1122} - C^{\textcircled{3}}_{1122}\right)\hat{s}_3}{C^{\textcircled{2}}_{2222}\,\hat{s}_3 + C^{\textcircled{3}}_{2222}\,\hat{s}_2};\\
A_{211}^{(1),\textcircled{3}} &= \frac{\left(C^{\textcircled{2}}_{1122} - C^{\textcircled{3}}_{1122}\right)\hat{s}_2}{C^{\textcircled{2}}_{2222}\,\hat{s}_3 + C^{\textcircled{3}}_{2222}\,\hat{s}_2};\\
B_{211}^{(1),\textcircled{1}} &= \frac{(2\,\hat{s}_2 + \hat{s}_3)\left(\left(\left(C^{\textcircled{2}}_{1122}\,C^{\textcircled{1}}_{2222} - C^{\textcircled{3}}_{1122}\,C^{\textcircled{1}}_{2222}\right)\hat{s}_2 + \left(C^{\textcircled{1}}_{1122}\,C^{\textcircled{2}}_{2222} - C^{\textcircled{3}}_{1122}\,C^{\textcircled{2}}_{2222}\right)\hat{s}_1\right)\hat{s}_3\right)}{2C^{\textcircled{1}}_{2222}\left(C^{\textcircled{2}}_{2222}\,\hat{s}_3 + C^{\textcircled{3}}_{2222}\,\hat{s}_2\right)(\hat{s}_1 + 2\,\hat{s}_2 + \hat{s}_3)} +\\
&\quad \frac{(2\,\hat{s}_2 + \hat{s}_3)\left(\left(C^{\textcircled{1}}_{1122}\,C^{\textcircled{3}}_{2222} - C^{\textcircled{2}}_{1122}\,C^{\textcircled{3}}_{2222}\right)\hat{s}_1\,\hat{s}_2\right)}{2C^{\textcircled{1}}_{2222}\left(C^{\textcircled{2}}_{2222}\,\hat{s}_3 + C^{\textcircled{3}}_{2222}\,\hat{s}_2\right)(\hat{s}_1 + 2\,\hat{s}_2 + \hat{s}_3)};\\
B_{211}^{(1),\textcircled{2}} &= -\frac{\hat{s}_1\,\left(\left(\left(C^{\textcircled{2}}_{1122}\,C^{\textcircled{1}}_{2222} - C^{\textcircled{3}}_{1122}\,C^{\textcircled{1}}_{2222}\right)\hat{s}_2 + \left(C^{\textcircled{1}}_{1122}\,C^{\textcircled{2}}_{2222} - C^{\textcircled{3}}_{1122}\,C^{\textcircled{2}}_{2222}\right)\hat{s}_1\right)\hat{s}_3\right)}{2C^{\textcircled{1}}_{2222}\left(C^{\textcircled{2}}_{2222}\,\hat{s}_3 + C^{\textcircled{3}}_{2222}\,\hat{s}_2\right)(\hat{s}_1 + 2\,\hat{s}_2 + \hat{s}_3)} -\\
&\quad \frac{\hat{s}_1\,\left(\left(C^{\textcircled{1}}_{1122}\,C^{\textcircled{3}}_{2222} - C^{\textcircled{2}}_{1122}\,C^{\textcircled{3}}_{2222}\right)\hat{s}_1\,\hat{s}_2\right)}{2C^{\textcircled{1}}_{2222}\left(C^{\textcircled{2}}_{2222}\,\hat{s}_3 + C^{\textcircled{3}}_{2222}\,\hat{s}_2\right)(\hat{s}_1 + 2\,\hat{s}_2 + \hat{s}_3)};\\
B_{211}^{(1),\textcircled{3}} &= -\frac{\hat{s}_1\,\left(\left(\left(C^{\textcircled{2}}_{1122}\,C^{\textcircled{1}}_{2222} - C^{\textcircled{3}}_{1122}\,C^{\textcircled{1}}_{2222}\right)\hat{s}_2 + \left(C^{\textcircled{1}}_{1122}\,C^{\textcircled{2}}_{2222} - C^{\textcircled{3}}_{1122}\,C^{\textcircled{2}}_{2222}\right)\hat{s}_1\right)\hat{s}_3\right)}{2C^{\textcircled{1}}_{2222}\left(C^{\textcircled{2}}_{2222}\,\hat{s}_3 + C^{\textcircled{3}}_{2222}\,\hat{s}_2\right)(\hat{s}_1 + 2\,\hat{s}_2 + \hat{s}_3)} -\\
&\quad \frac{\hat{s}_1\,\left(\left(C^{\textcircled{1}}_{1122}\,C^{\textcircled{3}}_{2222} - C^{\textcircled{2}}_{1122}\,C^{\textcircled{3}}_{2222}\right)\hat{s}_1\,\hat{s}_2\right)}{2C^{\textcircled{1}}_{2222}\left(C^{\textcircled{2}}_{2222}\,\hat{s}_3 + C^{\textcircled{3}}_{2222}\,\hat{s}_2\right)(\hat{s}_1 + 2\,\hat{s}_2 + \hat{s}_3)};
\end{aligned}
\tag{46}
$$

$$A_{222}^{(1),①} = -\frac{C_{2222}^{①}\,C_{2222}^{②}\,\hat{s}_3 + C_{2222}^{①}\,C_{2222}^{③}\,\hat{s}_2 - C_{2222}^{②}\,C_{2222}^{③}\,\hat{s}_2 - C_{2222}^{②}\,C_{2222}^{③}\,\hat{s}_3}{C_{2222}^{①}\,\left(C_{2222}^{②}\,\hat{s}_3 + C_{2222}^{③}\,\hat{s}_2\right)};$$

$$A_{222}^{(1),②} = -\frac{\left(C_{2222}^{②} - C_{2222}^{③}\right)\hat{s}_3}{C_{2222}^{②}\,\hat{s}_3 + C_{2222}^{③}\,\hat{s}_2};$$

$$A_{222}^{(1),③} = \frac{\left(C_{2222}^{②} - C_{2222}^{③}\right)\hat{s}_2}{C_{2222}^{②}\,\hat{s}_3 + C_{2222}^{③}\,\hat{s}_2};$$

$$B_{222}^{(1),①} = \frac{(2\,\hat{s}_2 + \hat{s}_3)\left(\left(\left(C_{2222}^{①}\,C_{2222}^{②} - C_{2222}^{①}\,C_{2222}^{③}\right)\hat{s}_2 + \left(C_{2222}^{①}\,C_{2222}^{②} - C_{2222}^{②}\,C_{2222}^{③}\right)\hat{s}_1\right)\hat{s}_3\right)}{2C_{2222}^{①}\,\left(C_{2222}^{②}\,\hat{s}_3 + C_{2222}^{③}\,\hat{s}_2\right)(\hat{s}_1 + 2\,\hat{s}_2 + \hat{s}_3)} +$$
$$\frac{(2\,\hat{s}_2 + \hat{s}_3)\left(\left(C_{2222}^{①}\,C_{2222}^{③} - C_{2222}^{②}\,C_{2222}^{③}\right)\hat{s}_1\,\hat{s}_2\right)}{2C_{2222}^{①}\,\left(C_{2222}^{②}\,\hat{s}_3 + C_{2222}^{③}\,\hat{s}_2\right)(\hat{s}_1 + 2\,\hat{s}_2 + \hat{s}_3)};$$

$$B_{222}^{(1),②} = -\frac{\hat{s}_1\,\left(\left(\left(C_{2222}^{①}\,C_{2222}^{②} - C_{2222}^{①}\,C_{2222}^{③}\right)\hat{s}_2 + \left(C_{2222}^{①}\,C_{2222}^{②} - C_{2222}^{②}\,C_{2222}^{③}\right)\hat{s}_1\right)\hat{s}_3\right)}{2C_{2222}^{①}\,\left(C_{2222}^{②}\,\hat{s}_3 + C_{2222}^{③}\,\hat{s}_2\right)(\hat{s}_1 + 2\,\hat{s}_2 + \hat{s}_3)} -$$
$$\frac{\hat{s}_1\,\left(\left(C_{2222}^{①}\,C_{2222}^{③} - C_{2222}^{②}\,C_{2222}^{③}\right)\hat{s}_1\,\hat{s}_2\right)}{2C_{2222}^{①}\,\left(C_{2222}^{②}\,\hat{s}_3 + C_{2222}^{③}\,\hat{s}_2\right)(\hat{s}_1 + 2\,\hat{s}_2 + \hat{s}_3)};$$

$$B_{222}^{(1),③} = -\frac{\hat{s}_1\,\left(\left(\left(C_{2222}^{①}\,C_{2222}^{②} - C_{2222}^{①}\,C_{2222}^{③}\right)\hat{s}_2 + \left(C_{2222}^{①}\,C_{2222}^{②} - C_{2222}^{②}\,C_{2222}^{③}\right)\hat{s}_1\right)\hat{s}_3\right)}{2C_{2222}^{①}\,\left(C_{2222}^{②}\,\hat{s}_3 + C_{2222}^{③}\,\hat{s}_2\right)(\hat{s}_1 + 2\,\hat{s}_2 + \hat{s}_3)} -$$
$$\frac{\hat{s}_1\,\left(\left(C_{2222}^{①}\,C_{2222}^{③} - C_{2222}^{②}\,C_{2222}^{③}\right)\hat{s}_1\,\hat{s}_2\right)}{2C_{2222}^{①}\,\left(C_{2222}^{②}\,\hat{s}_3 + C_{2222}^{③}\,\hat{s}_2\right)(\hat{s}_1 + 2\,\hat{s}_2 + \hat{s}_3)}; \tag{47}$$

$$A_{112}^{(1),\textcircled{1}} = -\frac{C_{1212}^{\textcircled{1}}\, C_{1212}^{\textcircled{2}}\, \hat{s}_3 + C_{1212}^{\textcircled{1}}\, C_{1212}^{\textcircled{3}}\, \hat{s}_2 - C_{1212}^{\textcircled{2}}\, C_{1212}^{\textcircled{3}}\, \hat{s}_2 - C_{1212}^{\textcircled{2}}\, C_{1212}^{\textcircled{3}}\, \hat{s}_3}{C_{1212}^{\textcircled{1}} \left(C_{1212}^{\textcircled{2}}\, \hat{s}_3 + C_{1212}^{\textcircled{3}}\, \hat{s}_2\right)};$$

$$A_{112}^{(1),\textcircled{2}} = -\frac{\left(C_{1212}^{\textcircled{2}} - C_{1212}^{\textcircled{3}}\right) \hat{s}_3}{C_{1212}^{\textcircled{2}}\, \hat{s}_3 + C_{1212}^{\textcircled{3}}\, \hat{s}_2};$$

$$A_{112}^{(1),\textcircled{3}} = \frac{\left(C_{1212}^{\textcircled{2}} - C_{1212}^{\textcircled{3}}\right) \hat{s}_2}{C_{1212}^{\textcircled{2}}\, \hat{s}_3 + C_{1212}^{\textcircled{3}}\, \hat{s}_2};$$

$$B_{112}^{(1),\textcircled{1}} = \frac{(2\,\hat{s}_2 + \hat{s}_3)\left(\left(\left(C_{1212}^{\textcircled{1}}\, C_{1212}^{\textcircled{2}} - C_{1212}^{\textcircled{1}}\, C_{1212}^{\textcircled{3}}\right) \hat{s}_2 + \left(C_{1212}^{\textcircled{1}}\, C_{1212}^{\textcircled{2}} - C_{1212}^{\textcircled{2}}\, C_{1212}^{\textcircled{3}}\right) \hat{s}_1\right) \hat{s}_3\right)}{2C_{1212}^{\textcircled{1}} \left(C_{1212}^{\textcircled{2}}\, \hat{s}_3 + C_{1212}^{\textcircled{3}}\, \hat{s}_2\right)(\hat{s}_1 + 2\,\hat{s}_2 + \hat{s}_3)} +$$

$$\frac{(2\,\hat{s}_2 + \hat{s}_3)\left(\left(C_{1212}^{\textcircled{1}}\, C_{1212}^{\textcircled{3}} - C_{1212}^{\textcircled{2}}\, C_{1212}^{\textcircled{3}}\right) \hat{s}_1\, \hat{s}_2\right)}{2C_{1212}^{\textcircled{1}} \left(C_{1212}^{\textcircled{2}}\, \hat{s}_3 + C_{1212}^{\textcircled{3}}\, \hat{s}_2\right)(\hat{s}_1 + 2\,\hat{s}_2 + \hat{s}_3)};$$

$$B_{112}^{(1),\textcircled{2}} = -\frac{\hat{s}_1 \left(\left(\left(C_{1212}^{\textcircled{1}}\, C_{1212}^{\textcircled{2}} - C_{1212}^{\textcircled{1}}\, C_{1212}^{\textcircled{3}}\right) \hat{s}_2 + \left(C_{1212}^{\textcircled{1}}\, C_{1212}^{\textcircled{2}} - C_{1212}^{\textcircled{2}}\, C_{1212}^{\textcircled{3}}\right) \hat{s}_1\right) \hat{s}_3\right)}{2C_{1212}^{\textcircled{1}} \left(C_{1212}^{\textcircled{2}}\, \hat{s}_3 + C_{1212}^{\textcircled{3}}\, \hat{s}_2\right)(\hat{s}_1 + 2\,\hat{s}_2 + \hat{s}_3)} -$$

$$\frac{\hat{s}_1 \left(\left(C_{1212}^{\textcircled{1}}\, C_{1212}^{\textcircled{3}} - C_{1212}^{\textcircled{2}}\, C_{1212}^{\textcircled{3}}\right) \hat{s}_1\, \hat{s}_2\right)}{2C_{1212}^{\textcircled{1}} \left(C_{1212}^{\textcircled{2}}\, \hat{s}_3 + C_{1212}^{\textcircled{3}}\, \hat{s}_2\right)(\hat{s}_1 + 2\,\hat{s}_2 + \hat{s}_3)};$$

$$B_{112}^{(1),\textcircled{3}} = -\frac{\hat{s}_1 \left(\left(\left(C_{1212}^{\textcircled{1}}\, C_{1212}^{\textcircled{2}} - C_{1212}^{\textcircled{1}}\, C_{1212}^{\textcircled{3}}\right) \hat{s}_2 + \left(C_{1212}^{\textcircled{1}}\, C_{1212}^{\textcircled{2}} - C_{1212}^{\textcircled{2}}\, C_{1212}^{\textcircled{3}}\right) \hat{s}_1\right) \hat{s}_3\right)}{2C_{1212}^{\textcircled{1}} \left(C_{1212}^{\textcircled{2}}\, \hat{s}_3 + C_{1212}^{\textcircled{3}}\, \hat{s}_2\right)(\hat{s}_1 + 2\,\hat{s}_2 + \hat{s}_3)} -$$

$$\frac{\hat{s}_1 \left(\left(C_{1212}^{\textcircled{1}}\, C_{1212}^{\textcircled{3}} - C_{1212}^{\textcircled{2}}\, C_{1212}^{\textcircled{3}}\right) \hat{s}_1\, \hat{s}_2\right)}{2C_{1212}^{\textcircled{1}} \left(C_{1212}^{\textcircled{2}}\, \hat{s}_3 + C_{1212}^{\textcircled{3}}\, \hat{s}_2\right)(\hat{s}_1 + 2\,\hat{s}_2 + \hat{s}_3)}. \tag{48}$$

The coefficients $\tilde{A}_i^{(1),\textcircled{p}}$, $\tilde{B}_i^{(1),\textcircled{p}}$, associated to the perturbation functions (24), take the form:

$$
\begin{aligned}
\tilde{A}_2^{(1),①} &= -\frac{C_{2222}^{②}\,\hat{s}_3\,\alpha_{22}^{①} - C_{2222}^{②}\,\hat{s}_3\,\alpha_{22}^{③} + C_{2222}^{③}\,\hat{s}_2\,\alpha_{22}^{①} - C_{2222}^{③}\,\hat{s}_2\,\alpha_{22}^{②}}{C_{2222}^{①}\left(C_{2222}^{②}\,\hat{s}_3 + C_{2222}^{③}\,\hat{s}_2\right)};\\
\tilde{A}_2^{(1),②} &= -\frac{\left(\alpha_{22}^{②} - \alpha_{22}^{③}\right)\hat{s}_3}{C_{2222}^{②}\,\hat{s}_3 + C_{2222}^{③}\,\hat{s}_2};\\
\tilde{A}_2^{(1),③} &= \frac{\hat{s}_2\left(\alpha_{22}^{②} - \alpha_{22}^{③}\right)}{C_{2222}^{②}\,\hat{s}_3 + C_{2222}^{③}\,\hat{s}_2};\\
\tilde{B}_2^{(1),①} &= \frac{(2\,\hat{s}_2 + \hat{s}_3)\left(\left(\left(C_{2222}^{①}\,\alpha_{22}^{②} - C_{2222}^{①}\,\alpha_{22}^{③}\right)\hat{s}_2 + \left(C_{2222}^{②}\,\alpha_{22}^{①} - C_{2222}^{②}\,\alpha_{22}^{③}\right)\hat{s}_1\right)\hat{s}_3 + \left(C_{2222}^{③}\,\alpha_{22}^{①} - C_{2222}^{③}\,\alpha_{22}^{②}\right)\hat{s}_1\,\hat{s}_2\right)}{2C_{2222}^{①}\left(C_{2222}^{②}\,\hat{s}_3 + C_{2222}^{③}\,\hat{s}_2\right)(\hat{s}_1 + 2\,\hat{s}_2 + \hat{s}_3)};\\
\tilde{B}_2^{(1),②} &= -\frac{\hat{s}_1\left(\left(\left(C_{2222}^{①}\,\alpha_{22}^{②} - C_{2222}^{①}\,\alpha_{22}^{③}\right)\hat{s}_2 + \left(C_{2222}^{②}\,\alpha_{22}^{①} - C_{2222}^{②}\,\alpha_{22}^{③}\right)\hat{s}_1\right)\hat{s}_3 + \left(C_{2222}^{③}\,\alpha_{22}^{①} - C_{2222}^{③}\,\alpha_{22}^{②}\right)\hat{s}_1\,\hat{s}_2\right)}{2C_{2222}^{①}\left(C_{2222}^{②}\,\hat{s}_3 + C_{2222}^{③}\,\hat{s}_2\right)(\hat{s}_1 + 2\,\hat{s}_2 + \hat{s}_3)};\\
\tilde{B}_2^{(1),③} &= -\frac{\hat{s}_1\left(\left(\left(C_{2222}^{①}\,\alpha_{22}^{②} - C_{2222}^{①}\,\alpha_{22}^{③}\right)\hat{s}_2 + \left(C_{2222}^{②}\,\alpha_{22}^{①} - C_{2222}^{②}\,\alpha_{22}^{③}\right)\hat{s}_1\right)\hat{s}_3 + \left(C_{2222}^{③}\,\alpha_{22}^{①} - C_{2222}^{③}\,\alpha_{22}^{②}\right)\hat{s}_1\,\hat{s}_2\right)}{2C_{2222}^{①}\left(C_{2222}^{②}\,\hat{s}_3 + C_{2222}^{③}\,\hat{s}_2\right)(\hat{s}_1 + 2\,\hat{s}_2 + \hat{s}_3)}.
\end{aligned}
\tag{49}
$$

Finally, the coefficients $P_i^{(1),ⓟ}$ and $Q_i^{(1),ⓟ}$ associated to the perturbation functions (25), are given by:

$$
\begin{aligned}
P_2^{(1),①} &= -\frac{K_{22}^{①} K_{22}^{②} \hat{s}_3 + K_{22}^{①} K_{22}^{③} \hat{s}_2 - K_{22}^{②} K_{22}^{③} \hat{s}_2 - K_{22}^{②} K_{22}^{③} \hat{s}_3}{K_{22}^{①} \left(K_{22}^{②} \hat{s}_3 + K_{22}^{③} \hat{s}_2\right)};\\
P_2^{(1),②} &= -\frac{\left(K_{22}^{②} - K_{22}^{③}\right) \hat{s}_3}{K_{22}^{②} \hat{s}_3 + K_{22}^{③} \hat{s}_2};\\
P_2^{(1),③} &= \frac{\hat{s}_2 \left(K_{22}^{②} - K_{22}^{③}\right)}{K_{22}^{②} \hat{s}_3 + K_{22}^{③} \hat{s}_2};\\
Q_2^{(1),①} &= \frac{(2\,\hat{s}_2 + \hat{s}_3)\left(\left(\left(K_{22}^{①} K_{22}^{②} - K_{22}^{①} K_{22}^{③}\right) \hat{s}_2 + \left(K_{22}^{①} K_{22}^{②} - K_{22}^{②} K_{22}^{③}\right) \hat{s}_1\right) \hat{s}_3 + \left(K_{22}^{①} K_{22}^{③} - K_{22}^{②} K_{22}^{③}\right) \hat{s}_1\, \hat{s}_2\right)}{2K_{22}^{①} \left(K_{22}^{②} \hat{s}_3 + K_{22}^{③} \hat{s}_2\right)(\hat{s}_1 + 2\,\hat{s}_2 + \hat{s}_3)};\\
Q_2^{(1),②} &= -\frac{\hat{s}_1 \left(\left(\left(K_{22}^{①} K_{22}^{②} - K_{22}^{①} K_{22}^{③}\right) \hat{s}_2 + \left(K_{22}^{①} K_{22}^{②} - K_{22}^{②} K_{22}^{③}\right) \hat{s}_1\right) \hat{s}_3 + \left(K_{22}^{①} K_{22}^{③} - K_{22}^{②} K_{22}^{③}\right) \hat{s}_1\, \hat{s}_2\right)}{2K_{22}^{①} \left(K_{22}^{②} \hat{s}_3 + K_{22}^{③} \hat{s}_2\right)(\hat{s}_1 + 2\,\hat{s}_2 + \hat{s}_3)};\\
Q_2^{(1),③} &= -\frac{\hat{s}_1 \left(\left(\left(K_{22}^{①} K_{22}^{②} - K_{22}^{①} K_{22}^{③}\right) \hat{s}_2 + \left(K_{22}^{①} K_{22}^{②} - K_{22}^{②} K_{22}^{③}\right) \hat{s}_1\right) \hat{s}_3 + \left(K_{22}^{①} K_{22}^{③} - K_{22}^{②} K_{22}^{③}\right) \hat{s}_1\, \hat{s}_2\right)}{2K_{22}^{①} \left(K_{22}^{②} \hat{s}_3 + K_{22}^{③} \hat{s}_2\right)(\hat{s}_1 + 2\,\hat{s}_2 + \hat{s}_3)}.
\end{aligned}
\tag{50}
$$

# B Explicit coefficients involved in overall thermoelastic constants

In this Appendix, the explicit expressions for coefficients $\mathcal{A}^p_{ijhk}$, $\Delta_{ijhk}$, $\mathcal{B}^p_{ij}$, $\Lambda_{ij}$, $\mathcal{D}^p_{ij}$ and $\Xi_{22}$ involved in the overall thermoelastic constants of the first order equivalent continuum (26), (27) and (28) are reported.

The coefficients $\mathcal{A}^p_{ijhk}$ and $\Delta_{ijhk}$, associated to the components of the overall elastic tensor (26),

assume the form:

$$
\begin{aligned}
\mathcal{A}^{1}_{1111} &= -2\, C^{\textcircled{1}}_{1111}\, C^{\textcircled{1}}_{2222}\, {C^{\textcircled{3}}_{2222}}^{2} + 2\, C^{\textcircled{2}}_{1111}\, C^{\textcircled{1}}_{2222}\, {C^{\textcircled{3}}_{2222}}^{2} + \\
& 2\, {C^{\textcircled{1}}_{1122}}^{2} {C^{\textcircled{3}}_{2222}}^{2} - 2\, {C^{\textcircled{2}}_{1122}}^{2} {C^{\textcircled{3}}_{2222}}^{2};\\[2ex]
\mathcal{A}^{2}_{1111} &= -{C^{\textcircled{2}}_{2222}}^{2} \left( C^{\textcircled{1}}_{1111}\, C^{\textcircled{1}}_{2222} - C^{\textcircled{3}}_{1111}\, C^{\textcircled{1}}_{2222} - {C^{\textcircled{1}}_{1122}}^{2} + {C^{\textcircled{3}}_{1122}}^{2} \right);\\[2ex]
\mathcal{A}^{3}_{1111} &= -C^{\textcircled{3}}_{2222} \Big( 4\, C^{\textcircled{1}}_{1111}\, C^{\textcircled{1}}_{2222}\, C^{\textcircled{2}}_{2222} + C^{\textcircled{1}}_{1111}\, C^{\textcircled{1}}_{2222}\, C^{\textcircled{3}}_{2222} - \\
& 4\, C^{\textcircled{2}}_{1111}\, C^{\textcircled{1}}_{2222}\, C^{\textcircled{2}}_{2222} - C^{\textcircled{3}}_{1111}\, C^{\textcircled{1}}_{2222}\, C^{\textcircled{3}}_{2222} - \\
& 4\, {C^{\textcircled{1}}_{1122}}^{2} C^{\textcircled{2}}_{2222} - {C^{\textcircled{1}}_{1122}}^{2} C^{\textcircled{3}}_{2222} + {C^{\textcircled{2}}_{1122}}^{2} C^{\textcircled{1}}_{2222} + \\
& {C^{\textcircled{2}}_{1122}}^{2} C^{\textcircled{3}}_{2222} - C^{\textcircled{2}}_{1122}\, C^{\textcircled{3}}_{1122}\, C^{\textcircled{1}}_{2222} + 4\, C^{\textcircled{2}}_{1122}\, C^{\textcircled{3}}_{1122}\, C^{\textcircled{2}}_{2222} + \\
& 2\, C^{\textcircled{2}}_{1122}\, C^{\textcircled{1}}_{2222}\, C^{\textcircled{2}}_{2222} - 2\, C^{\textcircled{2}}_{1122}\, C^{\textcircled{1}}_{2222}\, C^{\textcircled{3}}_{2222} - \\
& C^{\textcircled{3}}_{1122}\, C^{\textcircled{1}}_{2222}\, C^{\textcircled{2}}_{2222} + C^{\textcircled{3}}_{1122}\, C^{\textcircled{1}}_{2222}\, C^{\textcircled{3}}_{2222} \Big);\\[2ex]
\mathcal{A}^{4}_{1111} &= -C^{\textcircled{2}}_{2222} \Big( 2\, C^{\textcircled{1}}_{1111}\, C^{\textcircled{1}}_{2222}\, C^{\textcircled{2}}_{2222} + 2\, C^{\textcircled{1}}_{1111}\, C^{\textcircled{1}}_{2222}\, C^{\textcircled{3}}_{2222} - \\
& 2\, C^{\textcircled{2}}_{1111}\, C^{\textcircled{1}}_{2222}\, C^{\textcircled{2}}_{2222} - 2\, C^{\textcircled{3}}_{1111}\, C^{\textcircled{1}}_{2222}\, C^{\textcircled{3}}_{2222} - 2\, {C^{\textcircled{1}}_{1122}}^{2} C^{\textcircled{2}}_{2222} - \\
& 2\, {C^{\textcircled{1}}_{1122}}^{2} C^{\textcircled{3}}_{2222} + C^{\textcircled{2}}_{1122}\, C^{\textcircled{3}}_{1122}\, C^{\textcircled{1}}_{2222} + 2\, C^{\textcircled{2}}_{1122}\, C^{\textcircled{3}}_{1122}\, C^{\textcircled{3}}_{2222} + \\
& 2\, C^{\textcircled{2}}_{1122}\, C^{\textcircled{1}}_{2222}\, C^{\textcircled{2}}_{2222} - 2\, C^{\textcircled{2}}_{1122}\, C^{\textcircled{1}}_{2222}\, C^{\textcircled{3}}_{2222} - {C^{\textcircled{3}}_{1122}}^{2} C^{\textcircled{1}}_{2222} + \\
& 2\, {C^{\textcircled{3}}_{1122}}^{2} C^{\textcircled{2}}_{2222} - C^{\textcircled{3}}_{1122}\, C^{\textcircled{1}}_{2222}\, C^{\textcircled{2}}_{2222} + C^{\textcircled{3}}_{1122}\, C^{\textcircled{1}}_{2222}\, C^{\textcircled{3}}_{2222} \Big);\\
\mathcal{A}^{5}_{1111} &= {C^{\textcircled{3}}_{2222}}^{2} \left( C^{\textcircled{1}}_{1111}\, C^{\textcircled{1}}_{2222} - {C^{\textcircled{1}}_{1122}}^{2} + {C^{\textcircled{2}}_{1122}}^{2} \right);\\[2ex]
\mathcal{A}^{6}_{1111} &= C^{\textcircled{1}}_{1111}\, C^{\textcircled{1}}_{2222}\, {C^{\textcircled{2}}_{2222}}^{2} - {C^{\textcircled{1}}_{1122}}^{2} {C^{\textcircled{2}}_{2222}}^{2} + {C^{\textcircled{3}}_{1122}}^{2} {C^{\textcircled{2}}_{2222}}^{2};\\[2ex]
\mathcal{A}^{7}_{1111} &= 2\, C^{\textcircled{1}}_{1111}\, C^{\textcircled{1}}_{2222}\, C^{\textcircled{2}}_{2222}\, C^{\textcircled{3}}_{2222} - 2\, {C^{\textcircled{1}}_{1122}}^{2} C^{\textcircled{2}}_{2222}\, C^{\textcircled{3}}_{2222} + \\
& 2\, C^{\textcircled{2}}_{1122}\, C^{\textcircled{3}}_{1122}\, C^{\textcircled{2}}_{2222}\, C^{\textcircled{3}}_{2222};
\end{aligned}
\tag{51}
$$

$$\mathcal{A}^1_{2222} = 2\, {C^{\textcircled{3}}_{2222}}^2 C^{\textcircled{2}}_{2222} \left( C^{\textcircled{1}}_{2222} - C^{\textcircled{2}}_{2222} \right);$$

$$\mathcal{A}^2_{2222} = C^{\textcircled{3}}_{2222}\, C^{\textcircled{2}}_{2222} \left( C^{\textcircled{1}}_{2222}\, C^{\textcircled{2}}_{2222} - C^{\textcircled{2}}_{2222}\, C^{\textcircled{3}}_{2222} \right);$$

$$\mathcal{A}^3_{2222} = C^{\textcircled{2}}_{2222}\, C^{\textcircled{3}}_{2222} \left( C^{\textcircled{1}}_{2222}\, C^{\textcircled{2}}_{2222} + 4\, C^{\textcircled{1}}_{2222}\, C^{\textcircled{3}}_{2222} - 5\, C^{\textcircled{2}}_{2222}\, C^{\textcircled{3}}_{2222} \right);$$

$$\mathcal{A}^4_{2222} = 2\, C^{\textcircled{3}}_{2222}\, C^{\textcircled{2}}_{2222} \left( C^{\textcircled{1}}_{2222}\, C^{\textcircled{2}}_{2222} - C^{\textcircled{2}}_{2222}\, C^{\textcircled{3}}_{2222} \right) + \\ C^{\textcircled{3}}_{2222}\, C^{\textcircled{2}}_{2222} \left( 2\, C^{\textcircled{1}}_{2222}\, C^{\textcircled{3}}_{2222} - 2\, C^{\textcircled{2}}_{2222}\, C^{\textcircled{3}}_{2222} \right);$$

$$\mathcal{A}^5_{2222} = {C^{\textcircled{2}}_{2222}}^2 {C^{\textcircled{3}}_{2222}}^2;$$

$$\mathcal{A}^6_{2222} = {C^{\textcircled{2}}_{2222}}^2 {C^{\textcircled{3}}_{2222}}^2;$$

$$\mathcal{A}^7_{2222} = {C^{\textcircled{2}}_{2222}}^2 {C^{\textcircled{3}}_{2222}}^2;$$

$$\mathcal{A}^1_{1212} = 2\, C^{\textcircled{2}}_{1122}\, C^{\textcircled{1}}_{2222}\, {C^{\textcircled{3}}_{2222}}^2 - 2\, C^{\textcircled{2}}_{1122}\, C^{\textcircled{2}}_{2222}\, {C^{\textcircled{3}}_{2222}}^2;$$

$$\mathcal{A}^2_{1212} = C^{\textcircled{3}}_{1122}\, C^{\textcircled{1}}_{2222}\, {C^{\textcircled{2}}_{2222}}^2 - C^{\textcircled{3}}_{1122}\, {C^{\textcircled{2}}_{2222}}^2 C^{\textcircled{3}}_{2222};$$

$$\mathcal{A}^3_{1212} = C^{\textcircled{2}}_{1122}\, C^{\textcircled{1}}_{2222}\, C^{\textcircled{2}}_{2222}\, C^{\textcircled{3}}_{2222} + 2\, C^{\textcircled{2}}_{1122}\, C^{\textcircled{1}}_{2222}\, {C^{\textcircled{3}}_{2222}}^2 - \\ 3\, C^{\textcircled{2}}_{1122}\, C^{\textcircled{2}}_{2222}\, {C^{\textcircled{3}}_{2222}}^2 + 2\, C^{\textcircled{3}}_{1122}\, C^{\textcircled{1}}_{2222}\, C^{\textcircled{2}}_{2222}\, C^{\textcircled{3}}_{2222} - 2\, C^{\textcircled{3}}_{1122}\, {C^{\textcircled{2}}_{2222}}^2 C^{\textcircled{3}}_{2222};$$

$$\mathcal{A}^4_{1212} = C^{\textcircled{2}}_{1122}\, C^{\textcircled{1}}_{2222}\, C^{\textcircled{2}}_{2222}\, C^{\textcircled{3}}_{2222} - C^{\textcircled{2}}_{1122}\, C^{\textcircled{2}}_{2222}\, {C^{\textcircled{3}}_{2222}}^2 + C^{\textcircled{3}}_{1122}\, C^{\textcircled{1}}_{2222}\, {C^{\textcircled{2}}_{2222}}^2 + \\ 2\, C^{\textcircled{3}}_{1122}\, C^{\textcircled{1}}_{2222}\, C^{\textcircled{2}}_{2222}\, C^{\textcircled{3}}_{2222} - 3\, C^{\textcircled{3}}_{1122}\, {C^{\textcircled{2}}_{2222}}^2 C^{\textcircled{3}}_{2222};$$

$$\mathcal{A}^5_{1212} = C^{\textcircled{2}}_{1122}\, C^{\textcircled{2}}_{2222}\, {C^{\textcircled{3}}_{2222}}^2;$$

$$\mathcal{A}^6_{1212} = C^{\textcircled{3}}_{1122}\, {C^{\textcircled{2}}_{2222}}^2 C^{\textcircled{3}}_{2222};$$

$$\mathcal{A}^7_{1212} = C^{\textcircled{2}}_{1122}\, C^{\textcircled{2}}_{2222}\, {C^{\textcircled{3}}_{2222}}^2 + C^{\textcircled{3}}_{1122}\, {C^{\textcircled{2}}_{2222}}^2 C^{\textcircled{3}}_{2222};$$

$$\mathcal{A}^1_{1122} = 2\, {C^{\textcircled{3}}_{1212}}^2 C^{\textcircled{2}}_{1212} \left( C^{\textcircled{1}}_{1212} - C^{\textcircled{2}}_{1212} \right);$$

$$\mathcal{A}^2_{1122} = C^{③}_{1212}\, C^{②}_{1212} \left( C^{①}_{1212}\, C^{②}_{1212} - C^{②}_{1212}\, C^{③}_{1212} \right);$$

$$\mathcal{A}^3_{1122} = C^{①}_{1212}\, {C^{②}_{1212}}^2 C^{③}_{1212} + 4\, C^{①}_{1212}\, C^{②}_{1212}\, {C^{③}_{1212}}^2 - 5\, {C^{②}_{1212}}^2 {C^{③}_{1212}}^2;$$

$$\mathcal{A}^4_{1122} = 2\, C^{③}_{1212}\, C^{②}_{1212} \left( C^{①}_{1212}\, C^{②}_{1212} - C^{②}_{1212}\, C^{③}_{1212} \right) + \\ C^{③}_{1212}\, C^{②}_{1212} \left( 2\, C^{①}_{1212}\, C^{③}_{1212} - 2\, C^{②}_{1212}\, C^{③}_{1212} \right);$$

$$\mathcal{A}^5_{1122} = {C^{②}_{1212}}^2 {C^{③}_{1212}}^2;$$

$$\mathcal{A}^6_{1122} = {C^{②}_{1212}}^2 {C^{③}_{1212}}^2;$$

$$\mathcal{A}^7_{1122} = 2\, {C^{②}_{1212}}^2 {C^{③}_{1212}}^2; \tag{52}$$

$$\mathit{\Delta}_{1111} = \mathit{\Delta}_{2222} = \mathit{\Delta}_{1122} = C^{①}_{2222} \left( C^{②}_{2222}\, \hat{s}_3 + C^{③}_{2222}\, \hat{s}_2 \right)^2;$$

$$\mathit{\Delta}_{1212} = C^{①}_{1212} \left( C^{②}_{1212}\, \hat{s}_3 + C^{③}_{1212}\, \hat{s}_2 \right)^2. \tag{53}$$

The coefficients $\mathcal{B}^p_{ij}$, $\Lambda_{ij}$, associated to the components of the overall thermal dilatation tensor

(27), are given by:

$$\mathcal{B}_{11}^{1} = -2\,C_{1122}^{\textcircled{1}}\,C_{2222}^{\textcircled{3}}\,\alpha_{22}^{\textcircled{1}} + 2\,C_{1122}^{\textcircled{1}}\,C_{2222}^{\textcircled{3}}\,\alpha_{22}^{\textcircled{2}} - 2\,C_{2222}^{\textcircled{1}}\,C_{2222}^{\textcircled{3}}\,\alpha_{11}^{\textcircled{1}} + 2\,C_{2222}^{\textcircled{1}}\,C_{2222}^{\textcircled{3}}\,\alpha_{11}^{\textcircled{2}};$$

$$\mathcal{B}_{11}^{2} = -C_{1122}^{\textcircled{1}}\,C_{2222}^{\textcircled{2}}\,\alpha_{22}^{\textcircled{1}} + C_{1122}^{\textcircled{1}}\,C_{2222}^{\textcircled{2}}\,\alpha_{22}^{\textcircled{3}} - C_{2222}^{\textcircled{1}}\,C_{2222}^{\textcircled{2}}\,\alpha_{11}^{\textcircled{1}} + C_{2222}^{\textcircled{1}}\,C_{2222}^{\textcircled{2}}\,\alpha_{11}^{\textcircled{3}};$$

$$\mathcal{B}_{11}^{3} = -2\,C_{1122}^{\textcircled{1}}\,C_{2222}^{\textcircled{2}}\,\alpha_{22}^{\textcircled{1}} + 2\,C_{1122}^{\textcircled{1}}\,C_{2222}^{\textcircled{2}}\,\alpha_{22}^{\textcircled{3}} - C_{1122}^{\textcircled{1}}\,C_{2222}^{\textcircled{3}}\,\alpha_{22}^{\textcircled{1}} + C_{1122}^{\textcircled{1}}\,C_{2222}^{\textcircled{3}}\,\alpha_{22}^{\textcircled{2}} + 2\,C_{1122}^{\textcircled{2}}\,C_{2222}^{\textcircled{1}}\,\alpha_{22}^{\textcircled{3}} - C_{1122}^{\textcircled{3}}\,C_{2222}^{\textcircled{1}}\,\alpha_{22}^{\textcircled{2}} + C_{1122}^{\textcircled{3}}\,C_{2222}^{\textcircled{1}}\,\alpha_{22}^{\textcircled{3}} - 2\,C_{2222}^{\textcircled{1}}\,C_{2222}^{\textcircled{2}}\,\alpha_{11}^{\textcircled{1}} + 2\,C_{2222}^{\textcircled{1}}\,C_{2222}^{\textcircled{2}}\,\alpha_{11}^{\textcircled{2}} - C_{2222}^{\textcircled{1}}\,C_{2222}^{\textcircled{3}}\,\alpha_{11}^{1} + C_{2222}^{\textcircled{1}}\,C_{2222}^{\textcircled{3}}\,\alpha_{11}^{\textcircled{3}};$$

$$\mathcal{B}_{11}^{4} = C_{1122}^{\textcircled{1}}\,C_{2222}^{\textcircled{3}}\,\alpha_{22}^{\textcircled{1}} - C_{1122}^{\textcircled{1}}\,C_{2222}^{\textcircled{3}}\,\alpha_{22}^{\textcircled{2}} + C_{2222}^{\textcircled{1}}\,C_{2222}^{\textcircled{3}}\,\alpha_{11}^{\textcircled{1}};$$

$$\mathcal{B}_{11}^{5} = C_{1122}^{\textcircled{1}}\,C_{2222}^{\textcircled{2}}\,\alpha_{22}^{\textcircled{1}} - C_{1122}^{\textcircled{1}}\,C_{2222}^{\textcircled{2}}\,\alpha_{22}^{\textcircled{3}} + C_{2222}^{\textcircled{1}}\,C_{2222}^{\textcircled{2}}\,\alpha_{11}^{\textcircled{1}};$$

$$\mathcal{B}_{22}^{1} = -4\,C_{2222}^{\textcircled{3}}\,\alpha_{22}^{\textcircled{1}} + 4\,C_{2222}^{\textcircled{3}}\,\alpha_{22}^{\textcircled{2}};$$

$$\mathcal{B}_{22}^{2} = -2\,C_{2222}^{\textcircled{2}}\,\alpha_{22}^{\textcircled{1}} + 2\,C_{2222}^{\textcircled{2}}\,\alpha_{22}^{\textcircled{3}};$$

$$\mathcal{B}_{22}^{3} = -4\,C_{2222}^{\textcircled{2}}\,\alpha_{22}^{\textcircled{1}} + 4\,C_{2222}^{\textcircled{2}}\,\alpha_{22}^{\textcircled{2}} - 2\,C_{2222}^{\textcircled{3}}\,\alpha_{22}^{\textcircled{1}} + 2\,C_{2222}^{\textcircled{3}}\,\alpha_{22}^{\textcircled{3}};$$

$$\mathcal{B}_{22}^{4} = 2\,C_{2222}^{\textcircled{3}}\,\alpha_{22}^{\textcircled{1}} - C_{2222}^{\textcircled{3}}\,\alpha_{22}^{\textcircled{2}};$$

$$\mathcal{B}_{22}^{5} = 2\,C_{2222}^{\textcircled{2}}\,\alpha_{22}^{\textcircled{1}} - C_{2222}^{\textcircled{2}}\,\alpha_{22}^{\textcircled{3}}; \tag{54}$$

$$\Lambda_{11} = \Lambda_{22} = C_{2222}^{\textcircled{1}}\left(C_{2222}^{\textcircled{2}}\,\hat{s}_3 + C_{2222}^{\textcircled{3}}\,\hat{s}_2\right). \tag{55}$$

The coefficients $\mathcal{D}_{ij}^{p}$ and $\Xi_{22}$, associated to the components of the overall heat conduction tensor

(28), are given by:

$$\mathcal{D}^1_{22} = -2\, {K^{\textcircled{3}}_{22}}^2 \left(K^{\textcircled{1}}_{22} - K^{\textcircled{2}}_{22}\right);$$

$$\mathcal{D}^2_{22} = -K^{\textcircled{1}}_{22}\, {K^{\textcircled{2}}_{22}}^2 + {K^{\textcircled{2}}_{22}}^2 K^{\textcircled{3}}_{22};$$

$$\mathcal{D}^3_{22} = -K^{\textcircled{3}}_{22} \left(4\, K^{\textcircled{1}}_{22}\, K^{\textcircled{2}}_{22} + K^{\textcircled{1}}_{22}\, K^{\textcircled{3}}_{22} - {K^{\textcircled{2}}_{22}}^2 - 4\, K^{\textcircled{2}}_{22}\, K^{\textcircled{3}}_{22}\right);$$

$$\mathcal{D}^4_{22} = -2\, K^{\textcircled{1}}_{22}\, {K^{\textcircled{2}}_{22}}^2 - 2\, K^{\textcircled{1}}_{22}\, K^{\textcircled{2}}_{22}\, K^{\textcircled{3}}_{22} + 2\, {K^{\textcircled{2}}_{22}}^2 K^{\textcircled{3}}_{22} + 2\, K^{\textcircled{2}}_{22}\, {K^{\textcircled{3}}_{22}}^2;$$

$$\mathcal{D}^5_{22} = K^{\textcircled{1}}_{22}\, {K^{\textcircled{3}}_{22}}^2;$$

$$\mathcal{D}^6_{22} = K^{\textcircled{1}}_{22}\, {K^{\textcircled{2}}_{22}}^2;$$

$$\mathcal{D}^7_{22} = 2\, K^{\textcircled{1}}_{22}\, K^{\textcircled{2}}_{22}\, K^{\textcircled{3}}_{22};$$

$$\varXi_{22} = \left(K^{\textcircled{2}}_{22}\, \hat{s}_3 + K^{\textcircled{3}}_{22}\, \hat{s}_2\right)^2. \tag{56}$$

# C Explicit coefficients involved in the analytical solution of the homogenized field equations

The constants $A$ and $B$ involved in the heat source expressions (30) are determined by imposing the following conditions:

$$\left.\frac{\partial r}{\partial x_j}\right|_{x_j=0} = \left.\frac{\partial r}{\partial x_j}\right|_{x_j=L_j}, \qquad \int_0^{L_j} r dx_j = 0. \tag{57}$$

It is important to note that the continuity condition $r(x_j = 0) = r(x_j = L_j)$ is automatically satisfied by the structure of the function $r$. $A$ and $B$ are given by:

$$A = -\frac{R}{12\beta}\left[(-1)^n \beta^3 e^{\frac{(2\pi n-\beta^2)(2\pi n+\alpha^2)}{4\beta^2}} + 6\sqrt{\pi}\text{erf}\left(\frac{2i\pi n+\beta^2}{2\beta}\right) - 6\sqrt{\pi}\text{erf}\left(\frac{2i\pi n-\beta^2}{\beta}\right)\right] e^{\frac{2\pi n^2\beta^2}{\beta^2}}; \tag{58}$$

$$B = R(-1)^n \beta^2 e^{-\frac{\beta^2}{4}}. \tag{59}$$

Expressions (58) and (59) have been used for solving the homogenized fields equations (32) and (33) and then deriving the macroscopic fields (37) and (38). The functions $\Omega_j(x_j)$ and $\Lambda_j(x_j)$ involved respectively in the macroscopic displacement and temperature fields are here reported.

The functions $\Omega_j(x_j)$ takes the form:

$$\Omega_0(x_j) = C_1 + \frac{1}{1440 L_j^2 \beta^5 K_{jj} C_{jjjj}} \Bigg\{ 180\alpha_{jj}\pi^{\frac{5}{2}} L_j^5 R n^2 e^{-\frac{\pi^2 n^2}{\beta^2}} \operatorname{erf}\left(\frac{i(iL_j\beta^2 - 2i\beta^2 x_j + 2L_j\pi n)}{2L_j\beta}\right)$$
$$+ 180\alpha_{jj}\pi^{\frac{5}{2}} L_j^5 R n^2 e^{-\frac{\pi^2 n^2}{\beta^2}} \operatorname{erf}\left(\frac{i(iL_j\beta^2 - 2i\beta^2 x_j - 2L_j\pi n)}{2L_j\beta}\right)$$
$$+ 180i\alpha_{jj}\pi^{\frac{3}{2}} L_j^5 R n \beta^2 e^{-\frac{\pi^2 n^2}{\beta^2}} \operatorname{erf}\left(\frac{i(iL_j\beta^2 - 2i\beta^2 x_j + 2L_j\pi n)}{2L_j\beta}\right)$$
$$- 180i\alpha_{jj}\pi^{\frac{3}{2}} L_j^5 R n \beta^2 e^{-\frac{\pi^2 n^2}{\beta^2}} \operatorname{erf}\left(\frac{i(iL_j\beta^2 - 2i\beta^2 x_j - 2L_j\pi n)}{2L_j\beta}\right)$$
$$- 45\alpha_{jj}\pi^{\frac{1}{2}} L_j^5 R \beta^4 e^{-\frac{\pi^2 n^2}{\beta^2}} \operatorname{erf}\left(\frac{i(iL_j\beta^2 - 2i\beta^2 x_j + 2L_j\pi n)}{2L_j\beta}\right)$$
$$- 45\alpha_{jj}\pi^{\frac{1}{2}} L_j^5 R \beta^4 e^{-\frac{\pi^2 n^2}{\beta^2}} \operatorname{erf}\left(\frac{i(iL_j\beta^2 - 2i\beta^2 x_j - 2L_j\pi n)}{2L_j\beta}\right)$$
$$- 90\alpha_{jj}\pi^{\frac{1}{2}} L_j^5 R \beta^2 e^{-\frac{\pi^2 n^2}{\beta^2}} \operatorname{erf}\left(\frac{i(iL_j\beta^2 - 2i\beta^2 x_j + 2L_j\pi n)}{2L_j\beta}\right)$$
$$- 90\alpha_{jj}\pi^{\frac{1}{2}} L_j^5 R \beta^2 e^{-\frac{\pi^2 n^2}{\beta^2}} \operatorname{erf}\left(\frac{i(iL_j\beta^2 - 2i\beta^2 x_j - 2L_j\pi n)}{2L_j\beta}\right)$$
$$- 180i(-1)^n \alpha_{jj}\pi L_j^5 R n \beta e^{-\frac{8ix_j\pi n L_j + L_j^2\beta^2 - 4\beta^2 x_j L_j + 4\beta^2 x_j^2}{4L_j^2}}$$
$$+ 180i(-1)^n \alpha_{jj}\pi L_j^5 R n \beta e^{-\frac{-8ix_j\pi n L_j + L_j^2\beta^2 - 4\beta^2 x_j L_j + 4\beta^2 x_j^2}{4L_j^2}}$$
$$+ 90(-1)^n \alpha_{jj} L_j^3 R \beta^3 e^{-\frac{8ix_j\pi n L_j + L_j^2\beta^2 - 4\beta^2 x_j L_j + 4\beta^2 x_j^2}{4L_j^2}}$$
$$+90(-1)^n \alpha_{jj} L_j^3 R \beta^3 e^{-\frac{-8ix_j\pi n L_j + L_j^2\beta^2 - 4\beta^2 x_j L_j + 4\beta^2 x_j^2}{4L_j^2}} \Bigg\}, \tag{60}$$

$$\Omega_1(x_j) = C_2 + \frac{1}{1440 L_j^2 \beta^5 K_{jj} C_{jjjj}} \Bigg\{ -360iL_j^4\pi^{\frac{3}{2}} R\beta^2\alpha_{jj} n e^{-\frac{\pi^2 n^2}{\beta^2}} \operatorname{erf}\left(\frac{i(iL_j\beta^2 - 2i\beta^2 x_j + 2L_j\pi n)}{2L_j\beta}\right)$$
$$+ 360iL_j^4\pi^{\frac{3}{2}} R\beta^2\alpha_{jj} n e^{-\frac{\pi^2 n^2}{\beta^2}} \operatorname{erf}\left(\frac{i(iL_j\beta^2 - 2i\beta^2 x_j - 2L_j\pi n)}{2L_j\beta}\right)$$
$$+ 180L_j^4\pi^{\frac{1}{2}} R\beta^4\alpha_{jj} e^{-\frac{\pi^2 n^2}{\beta^2}} \operatorname{erf}\left(\frac{i(iL_j\beta^2 - 2i\beta^2 x_j + 2L_j\pi n)}{2L_j\beta}\right)$$
$$+ 180L_j^4\pi^{\frac{1}{2}} R\beta^4\alpha_{jj} e^{-\frac{\pi^2 n^2}{\beta^2}} \operatorname{erf}\left(\frac{i(iL_j\beta^2 - 2i\beta^2 x_j - 2L_j\pi n)}{2L_j\beta}\right)$$
$$+ 180i(-1)^{1+n} L_j^4 R\beta^2\alpha_{jj} e^{-\frac{8ix_j\pi n L_j + L_j^2\beta^2 - 4\beta^2 x_j L_j + 4\beta^2 x_j^2}{4L_j^2}}$$
$$+180i(-1)^{1+n} L_j^4 R\beta^2\alpha_{jj} e^{-\frac{-8ix_j\pi n L_j + L_j^2\beta^2 - 4\beta^2 x_j L_j + 4\beta^2 x_j^2}{4L_j^2}} \Bigg\}, \tag{61}$$

$$\Omega_2(x_j) = \frac{1}{1440L_j^2\beta^5K_{jj}C_{jjjj}}\left\{-180L_j^3\pi^{\frac{1}{2}}R\beta^4\alpha_{jj}e^{-\frac{\pi^2n^2}{\beta^2}}\mathrm{erf}\left(\frac{i(iL_j\beta^2-2i\beta^2x_j+2L_j\pi n)}{2L_j\beta}\right)\right.$$

$$-180L_j^3\pi^{\frac{1}{2}}R\beta^4\alpha_{jj}e^{-\frac{\pi^2n^2}{\beta^2}}\mathrm{erf}\left(\frac{i(iL_j\beta^2-2i\beta^2x_j-2L_j\pi n)}{2L_j\beta}\right)$$

$$+180L_j^3\pi^{\frac{1}{2}}R\beta^4\alpha_{jj}e^{-\frac{\pi^2n^2}{\beta^2}}\mathrm{erf}\left(\frac{i(i\beta^2-2\pi n}{2\beta}\right)$$

$$\left.+180L_j^3\pi^{\frac{1}{2}}R\beta^4\alpha_{jj}e^{-\frac{\pi^2n^2}{\beta^2}}\mathrm{erf}\left(\frac{i(i\beta^2+2\pi n}{2\beta}\right)\right\}, \tag{62}$$

$$\Omega_3 = \frac{1}{1440L_j^2\beta^5K_{jj}C_{jjjj}}\left\{40(-1)^{1+n}L_j^2r\beta^7\alpha_{jj}e^{-\frac{\beta^2}{4}} - 120L_j^2\pi^{\frac{1}{2}}R\beta^4\alpha_{jj}e^{-\frac{\pi^2n^2}{\beta^2}}\mathrm{erf}\left(\frac{i(i\beta^2-2\pi n}{2\beta}\right)\right.$$

$$\left.-120L_j^2\pi^{\frac{1}{2}}R\beta^4\alpha_{jj}e^{-\frac{\pi^2n^2}{\beta^2}}\mathrm{erf}\left(\frac{i(i\beta^2+2\pi n}{2\beta}\right)\right\}, \tag{63}$$

$$\Omega_4 = \frac{\alpha_{jj}\beta^2(-1)^nR}{24L_jK_{jj}C_{jjjj}}e^{-\frac{\beta^2}{4}}, \tag{64}$$

$$\Omega_5 = \frac{\alpha_{jj}\beta^2(-1)^{1+n}R}{60L_j^2K_{jj}C_{jjjj}}e^{-\frac{\beta^2}{4}}, \tag{65}$$

$$\tag{66}$$

where the constasts $C_1$ and $C_2$ assume the form:

$$C_1 = \frac{L_1^3R\alpha_{jj}}{32C_{jjjj}K_{jj}\beta^5}e^{-\frac{\pi^2n^2}{\beta^2}}\left\{4\pi^{\frac{5}{2}}n^2\mathrm{erf}\left(\frac{\beta^2+2i\pi n}{2\beta}\right) - 4\pi^{\frac{5}{2}}n^2\mathrm{erf}\left(\frac{-\beta^2+2i\pi n}{2\beta}\right)\right.$$

$$-4i\pi^{\frac{3}{2}}\beta^2n\mathrm{erf}\left(\frac{\beta^2+2i\pi n}{2\beta}\right) - 4i\pi^{\frac{3}{2}}\beta^2n\mathrm{erf}\left(\frac{-\beta^2+2i\pi n}{2\beta}\right)$$

$$-\pi^{\frac{1}{2}}\beta^4\mathrm{erf}\left(\frac{\beta^2+2i\pi n}{2\beta}\right) + \pi^{\frac{1}{2}}\beta^4\mathrm{erf}\left(\frac{-\beta^2+2i\pi n}{2\beta}\right)$$

$$-2\pi^{\frac{1}{2}}\beta^2\mathrm{erf}\left(\frac{\beta^2+2i\pi n}{2\beta}\right) + 2\pi^{\frac{1}{2}}\beta^2\mathrm{erf}\left(\frac{-\beta^2+2i\pi n}{2\beta}\right)$$

$$\left.+4(-1)^{1+n}\beta^3e^{\frac{(2\pi n-\beta^2)(2\pi n+\beta^2)}{4\beta^2}}\right\}, \tag{67}$$

$$C_2 = \frac{L_1^2R\alpha_{jj}}{720C_{jjjj}K_{jj}\beta^5}e^{-\frac{\pi^2n^2}{\beta^2}}\left\{180\pi^{\frac{5}{2}}n^2\mathrm{erf}\left(\frac{i(i\beta^2-2\pi n)}{2\beta}\right) + 180\pi^{\frac{5}{2}}n^2\mathrm{erf}\left(\frac{i(i\beta^2+2\pi n)}{2\beta}\right)\right.$$

$$-180i\pi^{\frac{3}{2}}n\beta^2\mathrm{erf}\left(\frac{i(i\beta^2-2\pi n)}{2\beta}\right) + 180i\pi^{\frac{3}{2}}n\beta^2\mathrm{erf}\left(\frac{i(i\beta^2+2\pi n)}{2\beta}\right)$$

$$-75\pi^{\frac{1}{2}}\beta^4\mathrm{erf}\left(\frac{i(i\beta^2-2\pi n)}{2\beta}\right) - 75\pi^{\frac{1}{2}}\beta^4\mathrm{erf}\left(\frac{i(i\beta^2+2\pi n)}{2\beta}\right)$$

$$-90\pi^{\frac{1}{2}}\beta^2\mathrm{erf}\left(\frac{i(i\beta^2-2\pi n)}{2\beta}\right) - 90\pi^{\frac{1}{2}}\beta^2\mathrm{erf}\left(\frac{i(i\beta^2+2\pi n)}{2\beta}\right)$$

$$+2\beta^7(-1)^ne^{-\frac{\beta^2}{4}}e^{-\frac{\pi^2n^2}{\beta^2}} + 180\beta^3(-1)^ne^{-\frac{\beta^2}{4}}e^{-\frac{\pi^2n^2}{\beta^2}}. \tag{68}$$

The functions $\Lambda_j(x_j)$ are given by:

$$\Lambda_0(x_j) = D_1 - \frac{1}{8\beta^3 K_{jj}} \Bigg\{ -2iL_j^2\pi^{\frac{3}{2}}Rne^{-\frac{\pi^2n^2}{\beta^2}}\mathrm{erf}\bigg(\frac{iL_j\beta^2 - 2i\beta^2x_j + 2L_j\pi n}{2L_j\beta}\bigg)$$
$$+ L_j^2\pi^{\frac{1}{2}}R\beta^2e^{-\frac{\pi^2n^2}{\beta^2}}\mathrm{erf}\bigg(\frac{iL_j\beta^2 - 2i\beta^2x_j + 2L_j\pi n}{2L_j\beta}\bigg)$$
$$+ 2iL_j^2\pi^{-\frac{3}{2}}Rne^{\frac{\pi n(2i\beta^2-\pi n)}{\beta^2}}\mathrm{erf}\bigg(\frac{iL_j\beta^2 - 2i\beta^2x_j - 2L_j\pi n}{2L_j\beta}\bigg)$$
$$+ L_j^2\pi^{\frac{1}{2}}R\beta^2e^{\frac{\pi n(2i\beta^2-\pi n)}{\beta^2}}\mathrm{erf}\bigg(\frac{iL_j\beta^2 - 2i\beta^2x_j - 2L_j\pi n}{2L_j\beta}\bigg) - 2L_j^2R\beta e^{\frac{i(iL_j\beta^2-2i\beta^2x_j+4L_j\pi n)(-2x_j+L_j)}{4L_j^2}}$$
$$-2L_j^2R\beta e^{\frac{i(iL_j^2\beta^2-4i\beta^2L_jx_j+4i\beta^2x_j^2+4\pi nL_j^2+8\pi nL_jx_j)}{4L_j^2}}\Bigg\}, \quad (69)$$

$$\Lambda_1(x_j) = D_2 + \frac{1}{4\beta K_{jj}} \Bigg\{ -L_j\pi^{\frac{1}{2}}Re^{-\frac{\pi^2n^2}{\beta^2}}\mathrm{erf}\bigg(\frac{iL_j\beta^2 - 2i\beta^2x_j + 2L_j\pi n}{2L_j\beta}\bigg) -$$
$$-L_j\pi^{\frac{1}{2}}e^{\frac{\pi n(2i\beta^2-\pi n)}{\beta^2}}\mathrm{erf}\bigg(\frac{iL_j\beta^2 - 2i\beta^2x_j - 2L_j\pi n}{2L_j\beta}\bigg)\Bigg\}, \quad (70)$$

$$\Lambda_2 = -\frac{R}{12\beta K_{jj}} \Bigg\{ \beta^3e^{\frac{i(\beta^2+4\pi n)}{4}} + 3\pi^{\frac{1}{2}}e^{-\frac{\pi^2n^2}{\beta^2}}\mathrm{erf}\bigg(\frac{i(i\beta^2-2\pi n)}{2\beta}\bigg)$$
$$+ 3\pi^{\frac{1}{2}}e^{-\frac{\pi^2n^2}{\beta^2}}\mathrm{erf}\bigg(\frac{i(i\beta^2+2\pi n)}{2\beta}\bigg)\Bigg\}, \quad (71)$$

$$\Lambda_3 = \frac{\beta^2R}{6L_jK_{jj}}e^{\frac{i(\eta^2+4\pi n)}{4}}, \quad (72)$$

$$\Lambda_4 = -\frac{\beta^2R}{12L_j^2K_{jj}}e^{\frac{i(\beta^2+4\pi n)}{4}}, \quad (73)$$

where the constants $D_1$ and $D_2$ assume the form:

$$D_1 = -\frac{L_j^2R}{720K_{jj}\beta^5}e^{-\frac{\pi^2n^2}{\beta^2}} \Bigg\{ 2(-1)^{1+n}\beta^7e^{\frac{(2\pi n-\beta^2)(2\pi n+\beta^2)}{4\beta^2}} + 180(-1)^{1+n}\beta^3e^{\frac{(2\pi n-\beta^2)(2\pi n+\beta^2)}{4\beta^2}}$$
$$+ 180\pi^{\frac{5}{2}}n^4\mathrm{erf}\bigg(\frac{\beta^2+2i\pi n}{2\beta}\bigg) - 180\pi^{\frac{5}{2}}n^4\mathrm{erf}\bigg(\frac{-\beta^2+2i\pi n}{2\beta}\bigg) - 180i\pi^{\frac{3}{2}}n\beta^2\mathrm{erf}\bigg(\frac{\beta^2+2i\pi n}{2\beta}\bigg)$$
$$- 180i\pi^{\frac{3}{2}}n\beta^2\mathrm{erf}\bigg(\frac{-\beta^2+2i\pi n}{2\beta}\bigg) - 75\pi^{\frac{1}{2}}\beta^4\mathrm{erf}\bigg(\frac{\beta^2+2i\pi n}{2\beta}\bigg) + 75\pi^{\frac{1}{2}}\beta^4\mathrm{erf}\bigg(\frac{-\beta^2+2i\pi n}{2\beta}\bigg)$$
$$-90\pi^{\frac{1}{2}}\beta^2\mathrm{erf}\bigg(\frac{\beta^2+2i\pi n}{2\beta}\bigg) + 90\pi^{\frac{1}{2}}\beta^2\mathrm{erf}\bigg(\frac{-\beta^2+2i\pi n}{2\beta}\bigg)\Bigg\}, \quad (74)$$

$$
\begin{aligned}
D_2 = -\frac{L_j R}{8K_{jj}\beta^3}\Bigg\{ & 2i\pi^{\frac{3}{2}} n e^{-\frac{\pi^2 n^2}{\beta^2}} \operatorname{erf}\left(\frac{i\beta^2 - 2\pi n}{2\beta}\right) - 2i\pi^{\frac{3}{2}} n e^{-\frac{\pi^2 n^2}{\beta^2}} \operatorname{erf}\left(\frac{i\beta^2 + 2\pi n}{2\beta}\right) \\
& - 2i\pi^{\frac{3}{2}} n e^{\frac{\pi n (2i\beta^2 - \pi n)}{\beta^2}} \operatorname{erf}\left(\frac{i\beta^2 - 2\pi n}{2\beta}\right) - 2i\pi^{\frac{3}{2}} n e^{\frac{\pi n (2i\beta^2 - \pi n)}{\beta^2}} \operatorname{erf}\left(\frac{i\beta^2 + 2\pi n}{2\beta}\right) \\
& - \pi^{\frac{1}{2}} \beta^2 e^{-\frac{\pi^2 n^2}{\beta^2}} \operatorname{erf}\left(\frac{i\beta^2 - 2\pi n}{2\beta}\right) - 3\pi^{\frac{1}{2}} \beta^2 e^{-\frac{\pi^2 n^2}{\beta^2}} \operatorname{erf}\left(\frac{i\beta^2 + 2\pi n}{2\beta}\right) \\
& - \pi^{\frac{1}{2}} \beta^2 e^{\frac{\pi n (2i\beta^2 - \pi n)}{\beta^2}} \operatorname{erf}\left(\frac{i\beta^2 - 2\pi n}{2\beta}\right) + \pi^{\frac{1}{2}} \beta^2 e^{\frac{\pi n (2i\beta^2 - \pi n)}{\beta^2}} \operatorname{erf}\left(\frac{i\beta^2 + 2\pi n}{2\beta}\right) \\
& + 4\beta e^{\frac{i(i\beta^2 + 4\pi n)}{4}} - 2\beta e^{\frac{i(i\beta^2 + 12\pi n)}{4}} - 2\beta e^{\frac{i(i\beta^2 - 4\pi n)}{4}}. \qquad (75)
\end{aligned}
$$

# D Three-phase vs bi-phase layered materials: limit values for the components of the overall thermoelastic tensors

In this Appendix, the explicit expressions for the components of the overall elastic, thermoelastic and heat conduction tensors in the limit cases $\zeta = 0$ and $\zeta \to +\infty$ are reported. For $\zeta = 0$ and $\zeta \to +\infty$, the components of the effective elastic tensors become

$$
\begin{aligned}
C_{1111}(\zeta = 0) &= \frac{C^{①}_{1111}C^{①}_{2222}\hat{s}_1 + 2C^{②}_{1111}C^{①}_{2222}\hat{s}_2 - (C^{①}_{1122})^2\hat{s}_2 + (C^{②}_{1122})^2\hat{s}_2}{C^{①}_{2222}};\\
C_{1111}(\zeta \to +\infty) &= \frac{C^{①}_{1111}C^{①}_{2222}\hat{s}_1 + C^{③}_{1111}C^{①}_{2222}\hat{s}_3 - (C^{①}_{1122})^2\hat{s}_1 + (C^{③}_{1122})^2\hat{s}_1}{C^{①}_{2222}};
\end{aligned} \tag{76}
$$

$$
\begin{aligned}
C_{2222}(\zeta = 0) &= \frac{C^{②}_{2222}(2C^{①}_{2222}\hat{s}_2 + C^{②}_{2222}\hat{s}_1)}{C^{①}_{2222}};\\
C_{2222}(\zeta \to +\infty) &= \frac{C^{③}_{2222}(C^{①}_{2222}\hat{s}_3 + C^{③}_{2222}\hat{s}_1)}{C^{①}_{2222}};
\end{aligned} \tag{77}
$$

$$
\begin{aligned}
C_{1122}(\zeta = 0) &= \frac{C^{②}_{1122}(2C^{①}_{2222}\hat{s}_2 + C^{②}_{2222}\hat{s}_1)}{C^{①}_{2222}};\\
C_{1122}(\zeta \to +\infty) &= \frac{C^{③}_{1122}(C^{①}_{2222}\hat{s}_3 + C^{③}_{2222}\hat{s}_1)}{C^{①}_{2222}};
\end{aligned} \tag{78}
$$

$$
\begin{aligned}
C_{1212}(\zeta = 0) &= \frac{C^{②}_{1212}(2C^{①}_{1212}\hat{s}_2 + C^{②}_{1212}\hat{s}_1)}{C^{①}_{1212}};\\
C_{1212}(\zeta \to +\infty) &= \frac{C^{③}_{1212}(C^{①}_{1212}\hat{s}_3 + C^{③}_{1212}\hat{s}_1)}{C^{①}_{1212}}.
\end{aligned} \tag{79}
$$

In the same limit, the components of the overall thermoelastic tensor assume the form:

$$
\begin{aligned}
\alpha_{11}(\zeta = 0) &= \frac{C^{①}_{1122}\alpha^{①}_{22}\hat{s}_1 - C^{①}_{1122}\alpha^{②}_{22}\hat{s}_1 + C^{①}_{2222}\alpha^{①}_{11}\hat{s}_1 + 2C^{①}_{2222}\alpha^{②}_{11}\hat{s}_2}{C^{①}_{2222}};\\
\alpha_{11}(\zeta \to +\infty) &= \frac{C^{①}_{1122}\alpha^{①}_{22}\hat{s}_1 - C^{①}_{1122}\alpha^{③}_{22}\hat{s}_1 + C^{①}_{2222}\alpha^{①}_{11}\hat{s}_1 + C^{①}_{2222}\alpha^{③}_{11}\hat{s}_3}{C^{①}_{2222}};
\end{aligned} \tag{80}
$$

$$
\begin{aligned}
\alpha_{22}(\zeta = 0) &= 2\alpha^{①}_{22}\hat{s}_1 - \alpha^{②}_{22}\hat{s}_1 + 2\alpha^{②}_{22}\hat{s}_2;\\
\alpha_{22}(\zeta \to +\infty) &= 2\alpha^{①}_{22}\hat{s}_1 - \alpha^{③}_{22}\hat{s}_1 + \alpha^{③}_{22}\hat{s}_3.
\end{aligned} \tag{81}
$$

Finally, for $\zeta = 0$ and $\zeta \to +\infty$, the components of the ffective heat conductivity tensor are given by

$$
\begin{aligned}
K_{11}(\zeta = 0) &= K^{①}_{11}\hat{s}_1 + 2K^{②}_{11}\hat{s}_2;\\
K_{11}(\zeta \to +\infty) &= K^{①}_{11}\hat{s}_1 + K^{③}_{11}\hat{s}_3;
\end{aligned} \tag{82}
$$

$$
\begin{aligned}
K_{22}(\zeta = 0) &= K^{①}_{22}\hat{s}_1 + 2K^{②}_{22}\hat{s}_2;\\
K_{22}(\zeta \to +\infty) &= K^{①}_{22}\hat{s}_1 + K^{③}_{22}\hat{s}_3;
\end{aligned} \tag{83}
$$

# E Down-scaling relations for stress fields and heat flux

In this Appendix the down-scaling relations for the stress fields and heat flux determined substituting the displacements and temperature down-scaling laws (9) and (10) into the constitutive equations (5) and (6) are reported. The down-scaling relations for $\sigma_{ij}\left(\boldsymbol{x}, \boldsymbol{\xi}=\frac{\boldsymbol{x}}{\varepsilon}\right)$ and $q_i\left(\boldsymbol{x}, \boldsymbol{\xi}=\frac{\boldsymbol{x}}{\varepsilon}\right)$ assume the following form:

$$
\begin{aligned}
\sigma_{ij}\left(\boldsymbol{x}, \boldsymbol{\xi}=\frac{\boldsymbol{x}}{\varepsilon}\right) = & \left\{ C^{\varepsilon}_{ijkr}\left[ (\delta_{kp}\delta_{rq_1} + N^{(1)}_{kpq_1,r})\frac{\partial U_p(\boldsymbol{x})}{\partial x_{q_1}} + \tilde{N}^{(1)}_{k,r}\Theta(\boldsymbol{x})\right] - \alpha^{\varepsilon}_{ij}\Theta(\boldsymbol{x})\right\}_{\boldsymbol{\xi}=\boldsymbol{x}/\varepsilon} + \\
& + \varepsilon\left\{ C^{\varepsilon}_{ijkr}\left[ (\delta_{rq_2}N^{(1)}_{kpq_1} + N^{(2)}_{kpq_1q_2,r})\frac{\partial^2 U_p(\boldsymbol{x})}{\partial x_{q_1}\partial x_{q_2}} + (\delta_{rq_1}\tilde{N}^{(1)}_{k} + \tilde{N}^{(2)}_{kq_1,r})\frac{\partial \Theta(\boldsymbol{x})}{\partial x_{q_1}}\right] - \alpha^{\varepsilon}_{ij}M^{(1)}_{q_1}\frac{\partial \Theta(\boldsymbol{x})}{\partial x_{q_1}}\right\}_{\boldsymbol{\xi}=\boldsymbol{x}/\varepsilon} \\
& + \varepsilon^2\left\{ C^{\varepsilon}_{ijkr}\left[ (\delta_{rq_3}N^{(2)}_{kpq_1q_2} + N^{(3)}_{kpq_1q_2q_3,r})\frac{\partial^3 U_p(\boldsymbol{x})}{\partial x_{q_1}\partial x_{q_2}\partial x_{q_3}} + (\delta_{rq_3}\tilde{N}^{(2)}_{kq_1} + \tilde{N}^{(3)}_{kq_1q_3,r})\frac{\partial^2 \Theta(\boldsymbol{x})}{\partial x_{q_1}\partial x_{q_2}}\right]\right. \\
& \left. -\alpha^{\varepsilon}_{ij}M^{(2)}_{q_1q_2}\frac{\partial^2 \Theta(\boldsymbol{x})}{\partial x_{q_1} x_{q_2}}\right\}_{\boldsymbol{\xi}=\boldsymbol{x}/\varepsilon} + \mathcal{O}(\varepsilon^3)
\end{aligned} \tag{84}
$$

$$
\begin{aligned}
q_i\left(\boldsymbol{x}, \boldsymbol{\xi}=\frac{\boldsymbol{x}}{\varepsilon}\right) = & -\left[K^{\varepsilon}_{ij}(\delta_{q_1j} + M^{(1)}_{q_1,j})\frac{\partial \Theta(\boldsymbol{x})}{\partial x_{q_1}}\right]_{\boldsymbol{\xi}=\boldsymbol{x}/\varepsilon} - \varepsilon\left[K^{\varepsilon}_{ij}(\delta_{q_2j}M^{(1)}_{q_1} + M^{(2)}_{q_1q_2,j})\frac{\partial^2 \Theta(\boldsymbol{x})}{\partial x_{q_1}\partial x_{q_2}}\right]_{\boldsymbol{\xi}=\boldsymbol{x}/\varepsilon} \\
& - \varepsilon^2\left[K^{\varepsilon}_{ij}(\delta_{jq_3}M^{(2)}_{q_1q_2} + M^{(3)}_{q_1q_2q_3,j})\frac{\partial^3 \Theta(\boldsymbol{x})}{\partial x_{q_1}\partial x_{q_2}\partial x_{q_3}}\right]_{\boldsymbol{\xi}=\boldsymbol{x}/\varepsilon} + \mathcal{O}(\varepsilon^3).
\end{aligned} \tag{85}
$$